\documentclass[fleqn,usenatbib]{mnras}

\usepackage{newtxtext,newtxmath}
\usepackage[normalem]{ulem}
\usepackage{units}
\usepackage{mathtools} 

\usepackage[T1]{fontenc}

\DeclareRobustCommand{\VAN}[3]{#2}
\let\VANthebibliography\thebibliography
\def\thebibliography{\DeclareRobustCommand{\VAN}[3]{##3}\VANthebibliography}

\usepackage{graphicx}	
\usepackage{amsmath}	

\usepackage{comment}

\usepackage[dvipsnames]{xcolor}

\title[Meridionally self-similar double flows]{Double flows anchored in a Kerr black hole horizon.
\\ I. Meridionally self-similar MHD models with loading terms}

   \author[L. Chantry et al.]{L. Chantry$^{1,2}$\thanks{E-mail: loic.chantry@obspm.fr},
            V. Cayatte$^{1}$,
            C. Sauty$^{1,3}$,
            N. Vlahakis$^{4}$,
            K. Tsinganos$^{4}$
            \\
            $^1$Laboratoire Univers et Th\'eories, Observatoire de Paris, Université PSL, Universit\'e Paris Cit\'e, CNRS, F-92190 Meudon, France\\
            $^2$Dipartimento di Fisica, Università degli Studi di Torino, via Pietro Giuria 1, 10125 Torino, Italy\\
            $^3$Laboratoire Univers et Particules de Montpellier (LUPM) Universit\'e Montpellier, CNRS/IN2P3, CC72, place Eug\'ene Bataillon, 34095, Montpellier Cedex 5, France \\
            $^4$Section of Astrophysics, Astronomy and Mechanics, Department of Physics, University of Athens, Panepistimiopolis Zografos, Athens 15783, Greece
          }

\date{Submitted 06/04/2022}

\pubyear{2021}

\renewcommand{\d}{\rm d}

\begin{document}
\label{firstpage}
\pagerange{\pageref{firstpage}--\pageref{lastpage}}
\maketitle

\begin{abstract}

Recent observations of supermassive black holes have brought us new information on their magnetospheres. In this study we attempt a theoretical modelling of the coupling of black holes with their jets and discs, via three innovations. First, we propose a semi-analytical MHD description of a steady relativistic inflow-outflow structure characteristic to the extraction of the hole rotational energy. The mass-loading is ensured in a thin layer, the stagnation surface, by a two-photon pair production originating to a gamma-ray emission from the surrounding disc. The double flow is described near the polar axis by an axisymmetric meridionally self-similar MHD model. Second, the inflow and outflow solutions are crossing the MHD critical points and are matched at the stagnation surface. Knowledge of the MHD field on the horizon give us the angular momentum and energy extracted from the black hole. Finally, we illustrate the model with three specific examples of double-flow solutions by varying the energetic interaction between the MHD field and the rotating black hole. When the isorotation frequency is half of the black hole one, the extracted Poynting flux is comparable to the one obtained using the force-free assumption. In two of the presented  solutions, the Penrose process dominates at large colatitudes, while the third is Poynting flux dominated at mid colatitudes. Mass injection rate estimations, from disk luminosity and inner radius, give an upper limit just above the values obtained for two solutions. This model is pertinent to describe the flows near the polar axis where pair production is more efficient. 
\end{abstract}




\section*{Introduction}

The recent images taken by the Event Horizon Telescope of the M87 black hole and also the black hole  at the center of our Galaxy (Sgr A*) have brought us new information on the black hole magnetospheres, such as the estimation of the diameter of its photon ring, or the magnitude of its surrounding magnetic field (\citealt{2019ApJ...875L...1E}, \citeyear{2021ApJ...910L..13E}). This can help us to understand the mechanisms involved in such magnetospheres, especially the origin of their observed powerful jets.

Despite recent progress on the magnetohydrodynamics (MHD) of rotating black holes {(e.g. \cite{10.1093/mnras/stab2462}, \cite{2019ApJ...880...93H}, \cite{2020ApJ...892...37P})}, several  properties of their plasma-filled magnetospheres are not completely known. In particular, the energy release at the base of jets associated with Active Galactic Nuclei (AGN) and Gamma Ray Bursts (GRB) may be explained via several mechanisms, depending on the geometry and the physical content of the black hole magnetosphere. The extraction of rotational energy from spinning black holes started to be theoretically investigated already in the 1970's \citep{1971NPhS..229..177P} and continued afterwards. 

A necessary condition for an extraction mechanism to take place is the existence of an ergo-region in the immediate vicinity of the black hole horizon. In the case where the black hole is pervaded with a magnetic field, this extraction may take place in two different ways. The extraction occurs either via the plasma itself, or via the electromagnetic field. In  particular, if the plasma inertia dominates, we have the so-called generalised Penrose mechanism. If the Poynting flux dominates, we have the so-called generalised Blandford-Znajek mechanism. Strictly speaking, the Blandford-Znajek mechanism applies in a force free magnetosphere. 
 
\cite{1969NCimR...1..252P} was the first to propose a mechanical process for extracting black hole energy, using the splitting of particles in the ergosphere. This process happens if one  of the particle resulting from the splitting falls onto the black hole with negative energy and the other one reaches infinity with more energy than the entering particle. 

On the other hand, neglecting the plasma inertia, \cite{1977MNRAS.179..433B} proposed a stationary model for energy extraction via the Poynting flux of the bulk electromagnetic field threading the black hole. In this model, where the fluid energy flux is negligible, they developed a perturbation method for the force-free  equations of electrodynamics as a function of the black hole spin parameter. Then, they applied this analytical description to a split monopole configuration modelling a black hole surrounded by an accretion disc. Thus they showed that, for a given black hole angular momentum variation, the energy extracted from the black hole can reach 50\% of the maximum extractable energy. 

These two fundamental mechanisms and their application to astrophysical phenomena are still discussed for different reasons. For example, causality agreement in the extraction of black hole rotational energy has been clearly established only for the Penrose process. The Blandford-Znajek mechanism does not explained how the electromagnetic Poynting flux is causally produced and how the black hole rotational energy is reduced (\citealt{1990ApJ...350..518P}, \citealt{2009JKPS...54.2503K}, \citealt{2014ApJ...792...88K}, \citealt{2016PTEP.2016f3E01T}). 

The expected very low rate of particle production with a relative velocity between the two fragments larger than $c/2$, seems to produce a very inefficient Penrose process (\citealt{1972ApJ...178..347B}, \citealt{1974ApJ...191..231W}). Yet, \cite{1985ApJ...290...12W} showed that the electromagnetic field may provide the required energy to put one fragment onto a negative energy orbit, without any constraint on its relative velocity (for a review of the mechanical Penrose-type processes see  \citealt{1989PhR...183..137W}). Then, a magnetic Penrose process could be extremely efficient in the entire range of expected magnetic fields (\citealt{2018MNRAS.tmpL..78D}). The plasma within the ergo-region plays the role of negative energy particles in the rotational energy extraction from the black hole. Time-dependent numerical simulations tend to support that the energy is supplied into jets from the rotational energy of the black hole (\citealt{2002Sci...295.1688K}, \citealt{2004MNRAS.350.1431K}, \citealt{2007MNRAS.377L..49K}, \citealt{2012MNRAS.423.3083M}). However, simulations  (e.g., \cite{2005MNRAS.359..801K} and \cite{2006MNRAS.368.1561M}) show that the Penrose process seems to be a transient phenomenon, which is lately replaced by a pure electromagnetic mechanism similar to the original Blandford-Znajek one.  

In this study, we consider AGN jets produced in the immediate environment of rotating supermassive black holes. Jets are multi-component outflows. Most present models are based on having a faster, mainly leptonic, core flow (the spine jet) surrounded by an outer hadronic component with mildly relativistic speeds (the sheath layer or disk wind). This two stream model was firstly proposed by \cite{1989MNRAS.237..411S} and allowed to get a unification scheme for BL Lac and radiogalaxy sources emission (\citealt{2005A&A...432..401G}). The disk wind component (proton-electron) can be modelled by radially  self-similar models including the effects of magnetic fields, gravity and enthalpy (\citealt{2003ApJ...596.1080V}, \citealt{2018MNRAS.473.4417C}). For the inner spine jet, we proposed in \cite{CC2018} an extension, in Kerr metric, of  meridionally self-similar models (\citealt{1994A&A...287..893S}, \citealt{2006A&A...447..797M},  \citealt{2014PhRvD..89l4015G}). This non-force free model is adapted to describe the spine jet close to its polar axis. The \cite{CC2018} model was built without neglecting the light cylinder radius and allowed to define a magnetic collimation criterion.

\begin{figure}       
\centering
\includegraphics[width=\linewidth]{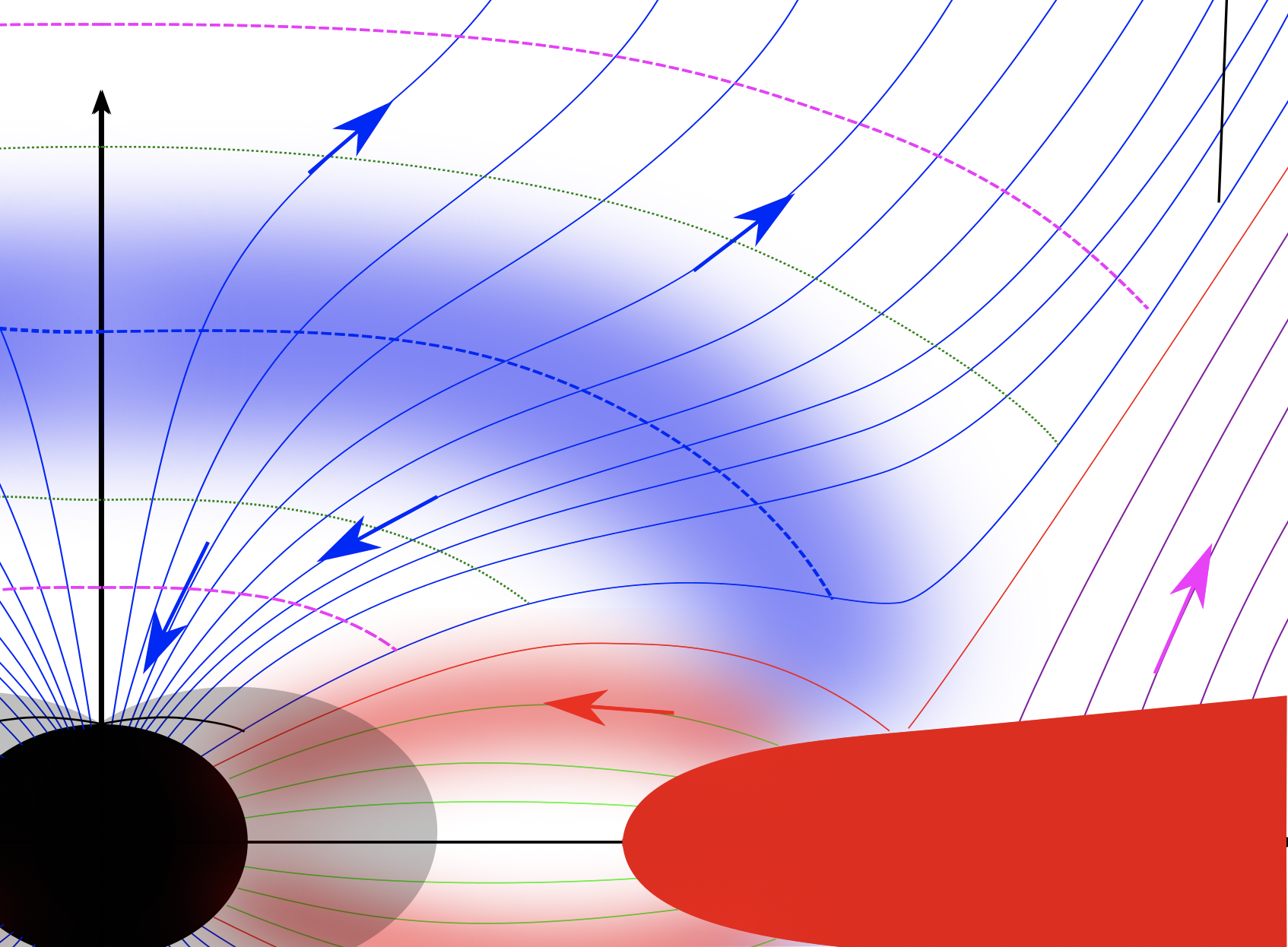}
\caption{Schematic representation of the inflow/outflow configuration in the poloidal plane. The black surface corresponds to the black hole horizon and the light grey shaded region to the ergo-region. The blue lines correspond to the magnetic field lines anchored on the black hole horizon and reaching infinity, while arrows indicate the fluid velocity direction. The green lines correspond to those lines of the black hole magnetosphere which link the disk to the black hole. The magenta lines and arrow corresponds to the magnetic field line anchored in the disk and reaching infinity. The blurred blue zone corresponds to the region where we expect that pair creation is efficient. The red zone corresponds to the disk and the blurred red zones correspond to the accretion column. The dashed blue line corresponds to the stagnation surface, and the green dotted line to the slow-magneto-sonic transition, the magenta dashed line corresponds to the Alfv\'{e}nic transition. The black line corresponds to the position of the inner and outer light cylinders.}         
\label{ShematicFlow}
\end{figure}

In steady, axisymmetric and ideal MHD, the mass flux is a conserved quantity along the magnetic flux tubes. Material slides along magnetic flux tubes. Thus to have a jet one needs to have mass injection within the flow (see the upper part of Figure \ref{ShematicFlow}).  Therefore, sufficiently close to the system axis, magnetic flux tubes should be necessarily anchored onto the black hole horizon. Therefore, the only way to obtain an outflow on such tubes is to inject mass at some location above the horizon.
The main process capable of mass loading is the creation of electron/positron pairs. \cite{Levinson2011VARIABLETE} estimate the amount of pairs produced from the hard photons emitted by a radiatively inefficient accretion disk. The authors conclude that the disk could not produce enough pairs to obtain the necessary Goldreich-Julian charge density in the black hole magnetosphere (\citealt{2016ApJ...818...50H} and references therein). Additionally, because of this low charge density, an electric gap forms, accelerating the particles along the flow. Due to an inverse Compton mechanism, the acceleration goes together with an increase of hard photon production. This induces the creation of additional pairs.  These gap models have been studied extensively in the literature  \citep{1992SvA....36..642B,1998ApJ...497..563H,2016ApJ...833..142H,2016ApJ...818...50H,2017PhRvD..96l3006L}. In particle-in-cell simulations \citep{refId0}, the magnetic reconnection on the equatorial plane and the formation of an intermittent spark gap lead to bursts of pair creation near the inner light cylinder. During these bursts, the density of pairs can reach values more than a thousand times the Goldreich-Julian density.

The extraction of rotational energy from a central supermassive black hole is a mechanism suspected to play a dominant role in the formation of flows around it.  Those energy extraction mechanisms are not limited {\it a priori} to  the Penrose, or the Blandford-Znajek processes. One can generalise the problem of energy extraction by studying directly the Noether currents associated to the energy and angular momentum, as explained in  \cite{2014PhRvD..89b4041L}. It allows us to study the interactions between the black hole and the surrounding hydromagnetic field. We recall that the spine jet is launched from the vicinity of the black hole magnetosphere around the rotation axis. The source of the spine jet energy can be the injection of matter/energy (pair creation, flow of energy in the magnetosphere or, any other mechanism), or the extraction of rotational energy from the black hole. The outflow starts at some stagnation radius where the poloidal velocity is zero, to reach a large distance and the inflow starts also at the same stagnation radius to fall into the black hole horizon (see Figure \ref{ShematicFlow}).

\cite{1986A&A...162...32C}, \cite{1990ApJ...363..206T}, and \cite{1992ApJ...386..455H} developed a general formalism allowing to solve the longitudinal fluid motion in a Kerr metric for a steady, axisymmetric, and magnetised flow in ideal MHD. This method requires that the geometry of the poloidal field lines is given, in order to solve the Bernoulli equation along the poloidal magnetic lines. Another way to deduce the field line shape is to solve the transverse Grad-Shafranov equation (see. \citealt{PhysRevD.44.2295}, \citealt{Beskin1993} and \citealt{2011PhRvD..83j4007G}). For a fixed geometry of the magnetic field lines, several approaches have been used to match the inflow and outflow solutions with loading terms localised on the stagnation surface (\citealt{2013PhRvD..88h4046G}, \citealt{2015ApJ...801...56P}, \citealt{2020ApJ...892...37P}). For example,  \cite{2019ApJ...880...93H} pursue this inflow/outflow matching approach by using the numerical methods introduced by \cite{2014ApJ...788..186N} to solve the Grad-Shafranov equation in the force-free approximation. \cite{2014ApJ...796...26G} used the \cite{1986A&A...162...32C} formalism with loading terms, controlling the magnitude of the bulk mass flux.

In this paper, we use the meridionally self-similar model developed in \cite{CC2018} to create inflow/outflow solutions. By using a non force-free model, we aim at producing a spine jet with a non  zero density on the polar axis. At the interface between the inflow and the outflow we need to include loading terms and are able to produce inflow solutions by reversing the flow direction. Our goal is to build complete solutions of inflow/outflow to reproduce a spine-jet with a given isorotation frequency for an AGN with a given black hole mass, spin, and magnetic field at a few gravitational radii above the black hole. In Sec. \ref{sec-sec1}, we summarize the \cite{CC2018} model and realize an extension of the MHD equations for an ideal plasma in Kerr metric adding mass loading terms. This extension is written using the 3+1 formalism and some general results of axisymmetric and steady configuration are deduced. In Sec.\ref{sec-sec2} we solve the MHD equations for the inflow and outflow parts, satisfying the matching conditions at the stagnation surface, where the loading of matter occurs in a thin layer. This allows us to correctly quantify the energy and angular momentum exchanges between the rotating black hole and the magnetised flow surrounding it, without assuming force-free condition. Hence, we may deduce the mass, angular momentum and energy injected at the stagnation surface. In Sec.\ref{sec-sec3} we analyse three inflow/outflow solutions. In particular, we discuss the role of the magnitude of the mass injection rate, the kinetic and dynamical behaviour of the flows, as well as  the interaction between the black hole and the MHD fields.

\section{Inflow and outflow via pair injection}\label{sec-sec1}

We model the problem under the assumptions of stationarity and axisymmetry. This means that all physical quantities in our study are invariant with time and along the azimuthal coordinate. We also consider an ideal relativistic plasma in which takes place pair creation.

	\subsection{Space-time geometry} 

The Kerr space-time is the simplest geometrical frame allowing to study energetic and angular momentum exchanges between black holes and magnetized fluids. Choosing a Kerr space-time implies that we  neglect the self-gravitation of the energy-momentum tensor field (plasma + radiation + electromagnetic fields). It is a reasonable assumption, because a perturbation to the Kerr metric due to the self-gravitation of the energy-momentum field is negligible compared to the Kerr geometry.

Let $(\mathcal{M};\mathbf{g})$ be the Kerr manifold using the usual Boyer-Lindquist map coordinates. Its line element is,
\begin{equation}
\d s^2 = -\dfrac{\rho^2 \Delta}{\Sigma^2} c^2{\d t}^2 +\varpi^2 \,  \left({\d \varphi}-\frac{\omega}{c}c{\d t}\right)^2 +\dfrac{\rho^2}{\Delta}{\d r}^2 +\rho^2{\d \theta}^2\,,
\end{equation}	
where,
\begin{equation*}
    \begin{array}{cc}
\Delta=r^2+r_g^2 a^2-2r_g r \,, & \Sigma^2=(r^2+r_g^2 a^2)^2-r_g^2 a^2\Delta \sin^2\theta \,, \\
 \rho^2=r^2+r_g^2 a^2\cos^2\theta\,, & \varpi = \dfrac{\Sigma}{\rho}\sin\theta\,, \quad \omega =\dfrac{2 r_g^2 a c r}{\Sigma^2}\,, 
    \end{array}
\end{equation*} 
with $a=\dfrac{\mathcal{J} c }{ {\rm M}^2 \mathcal{G}}\,$ and $r_g=\dfrac{ \mathcal{G}{\rm M}}{c^2}$
the dimensionless black hole spin ($0\leq a <1$) and the gravitational radius, respectively. We note that $\mathcal{J}$ is the angular momentum of the massive central object and ${\rm M}$ its mass. Also note that $\varpi$ corresponds to the usual Boyer-Lindquist cylindrical coordinate, i.e. $2\pi\varpi$ is the perimeter of the circle centered on the axis at constant $t$, $\theta$, and $r$.

A Kerr space-time has two Killing vectors, $\boldsymbol{\eta}=(1/c){\boldsymbol{\partial_t}}$ and $\boldsymbol{\xi}=\boldsymbol{\partial_\varphi}$ associated to stationarity and axisymmetry of this space-time. For a circular space-time as the Kerr space-time [\cite{2010arXiv1003.5015G}], the fiducial observer is called the zero-angular-momentum observer (ZAMO). The ZAMO 4-velocity of the Kerr space-time is $\mathbf{n}=\dfrac{1}{h}\left(\boldsymbol{\eta}-\boldsymbol{\beta}\right)\,$ where $h=\dfrac{\d \tau}{{\rm d}t}$ is the lapse function converting  Boyer-Lindquist time to ZAMO proper time, 
\begin{equation}
h=\left(1-\dfrac{2 r_g r}{\rho^2}+\beta^{\varphi}\beta_{\varphi}\right)^{1/2}=\dfrac{\rho}{\Sigma}\sqrt{\Delta} \,,
\end{equation}
and $\boldsymbol{\beta}=-\dfrac{\omega}{c}\boldsymbol{\partial}_\varphi$ is the shift vector of the ZAMO, where $\omega$ is the ZAMO shift pulsation. The ZAMO frame, the conventions and notations are introduced in \cite{CC2018}.

\subsection{Overview of the self-similar MHD model in Kerr metric}
   
We built a semi-analytical magnetohydrodynamic model (MHD) based on a self-similar separation of the variables, in Kerr metric. This model was already presented in \cite{CC2018} and is an extension of a similar one developed in Schwarzschild metric  (\citealt{2006A&A...447..797M}). After recalling the assumptions, we summarize briefly what has been presented in \cite{CC2018}. 

The Alfv\'en surface is spherical and we note with the subscript $\star$ the value of quantities at the intersection of the polar axis and the Alfv\'en surface. On the polar axis, the Alfv\'en radius is noted by $r_\star$, while we fix the typical magnetic flux scale $B_\star r_\star^2/2$ where $B_\star$ is the radial magnetic field value. We assume that the dimensionless magnetic flux is equal to $\alpha = f(r)\times \sin^2\theta$, where $f$ is an unknown function to be determined. The cylindrical radius on a flux tube is then equal to $\varpi(\alpha,r) = G(r) \sqrt{\alpha} =  \sqrt{{\alpha/\alpha_0}} \varpi(\alpha_0,r)$. The shape of the flux tube set is simply deduced from a radial function or from another flux tube. Another consequence is that the magnetic flux can be used as the second variable in our self-similar approach instead of the colatitude. Then we assume a variable separation for the poloidal Alfv\'en Mach number $M_{\rm alf} = M(r)m_1(\alpha)$ and for the pressure $P=P_0+({B_\star^2/8\pi})\Pi(r)\pi_1(\alpha)$ ($P_0$ being a constant value). The continuity conditions on the Alfv\'en surface lead to determine the expansion of $m_1(\alpha)$ to first order and of the isorotation frequency $\Omega$ to zeroth order. Then, it is possible to expand at order two in $\sin\theta$ the Euler equation.

Two parameters ($\mu$ and $l$) are derived from the physical properties of the Kerr black hole, the gravitational radius $r_g$ and its spin $a$. These parameters $\mu$ and $l$ are normalised relatively to the radius $r_\star$ of the Alfv\'en surface,
\begin{equation}
    \mu = \dfrac{2 r_g}{r_\star} \quad l = \dfrac{a r_g}{r_\star} = a \dfrac{\mu}{2}\,. 
\end{equation}
Another parameter ($\nu$) allows to link the gravitational potential to the flow kinetics, and is defined as the ratio of the escape speed to the flow speed at the Alfv\'en surface on the polar axis,
\begin{equation}
    \nu = \dfrac{1}{V_\star} \sqrt{\dfrac{2\mathcal{G}{\rm M}}{r_\star}}
\end{equation}
$V_\star$, the radial component of the flow on the polar axis at the Alfv\'en surface, will be negative for an inflow and positive for an outflow.  

Some functions have to be expanded in a limited $\alpha$-based development and other parameters are used for it. Without matter injection, the model assumptions lead to a conservation of mass flux par unit of magnetic flux, of angular momentum flux per unit of magnetic flux and energy integrals. Those quantities depends only on the dimensionless magnetic flux. The Euler equation has to be expanded to the second order in colatitude. It requires the following form for the square of the mass to magnetic flux, $\Psi_A^2 \propto 1 + \delta \alpha $. $L\Psi_A$ is the angular momentum per unit of magnetic flux and is also conserved along the flow. Then we introduce the parameter $\lambda$ to write $L\Psi_A \propto \lambda \alpha$. Note that the toroidal velocity and magnetic field are written in the form $\lambda g(r) \sin\theta$. The last motion integral gives the total specific energy $\mathcal{E}$ (see Eqs.(24) and (60) in \cite{CC2018}). We need to take only the first order term in an expansion of $\mathcal{E}$ on $\alpha$, $\mathcal{E}\propto (1+e_1 \alpha)$ with $e_1$ a parameter describing deviations from a spherical symmetry energy in this first order scheme. Similarly to this last constant of motion, we applied the same choice for the pressure. In the $\pi_1$ function we introduced a parameter $\kappa$ measuring how the pressure evolves with the  magnetic flux, i.e. deviates from a spherically symmetric value. $\kappa>0$ means an under-pressurized jet while $\kappa<0$ an over pressurized jet at the launching region. The four final ordinary differential equations are presented in Appendix C of \cite{CC2018}. A solution of these equations which crosses the different critical surfaces is fully determined by an additional parameter $\Pi_\star$ quantifying the dimensionless pressure at the Alfv\'en surface and on the polar axis. 

Finally, the MHD field solutions of this model are determined from eight model parameters $\lambda,\kappa,\delta,\mu,\nu,l,e_1$ and $\Pi_\star$.

For the Newtonian case ($\mu=0$ and $l=0$), or in a non rotating black hole $l=0$, in the ordinary differential equations system $\nu$ and $\lambda$ appear only with a square, and thus the system is invariant under the transformation $\nu \longleftrightarrow -\nu$ or $\lambda \longleftrightarrow -\lambda$. This property is due to the symmetry of ideal, axisymmetric and stationary MHD equations in Newtonian gravity or around a Schwarzschild black hole. Nevertheless, in Kerr configuration the Lense-Thirring term breaks the previous symmetry. The system of equations for our model (Eqs (C.3) and (C.4) of \citealt{CC2018}) is invariant under sign change preserving the product $\lambda \nu l$. Then the model system of equations is invariant under the following transformations,
\begin{equation}
    \left\{\begin{array}{ccc}
         \lambda &\longleftrightarrow & -\lambda  \\
         \nu &\longleftrightarrow & -\nu\\
         l &\longleftrightarrow & l
    \end{array}\right. \quad, \quad \left\{\begin{array}{ccc}
         l &\longleftrightarrow & -l  \\
         \nu &\longleftrightarrow & -\nu\\
         \lambda &\longleftrightarrow & \lambda
    \end{array}\right.\quad \text{or} \quad \left\{\begin{array}{ccc}
         l &\longleftrightarrow & -l  \\
         \lambda &\longleftrightarrow & -\lambda\\
         \nu &\longleftrightarrow & \nu
    \end{array}\right.\,.
\end{equation}
This discussion on the invariance of ideal, axisymmetric and stationary MHD equations around Kerr black hole is postponed to a future work.

A careful look of the self-similar model developed in \cite{CC2018} shows that it produces inflow and outflow  solutions, starting from the radius where the poloidal velocity of the flow is zero. This radius defines the stagnation radius and the stagnation sphere. 
	
We build inflow solution by taking negative values of $\nu$. Taking $\nu$ negative means that we allow the value of $V_\star$ to be negative. Thus it becomes the radial component of velocity at the intersection of the Alfv\'en surface and the axis. However $\nu$ negative is not sufficient to distinguish an inflow solution from an outflow one. A solution can be interpreted as a physical inflow only if it accelerates from the stagnation surface towards the black hole horizon, crossing first the slow magnetosonic surface and then the Alfv\'{e}n/fast magnetosonic surface before reaching the horizon. 

In this framework we can build a combination of outflow and inflow solutions in order to describe the jet from the black hole up to large distances. For that purpose we need to consider the physics of magnetized fluid with particle injection.

	\subsection{Particle number continuity}
	
Here, we consider a scenario where highly energetic photons or relativistic neutrinos, which are in the very close black hole vicinity, load the magnetosphere with electron-positron plasma via the mechanism of pair creation,
\begin{align}
\label{CreationPairsPhoton}
\gamma + \gamma \rightleftharpoons e_+ + e_- \, ,\\
\label{CreationPairsNeutrino}
\nu + \bar{\nu} \rightleftharpoons e_+ + e_- \,.
\end{align}
	
In the following, we refer to neutrinos or photons as the radiative component
and index quantities linked to these with the subscript $_r$. 
In a medium composed of radiation and leptonic plasma, the mechanism of pair creation implies a modified  expression of the particle number continuity equation. For electrons and positrons, Eqs. (\ref{CreationPairsPhoton}) and (\ref{CreationPairsNeutrino}) we get, respectively,
\begin{equation}
\label{CreationElectronPositron}
    \begin{array}{cc}
\nabla \cdot n_{+} \mathbf{u}_{+}=\dfrac{1}{c}\dfrac{\delta^4 n_{+}^{\rm created}}{\delta^3 \mathcal{V}_{\mathbf{u}_{+}}\delta \tau_+}\,, & \nabla \cdot n_{-} \mathbf{u}_{-}=\dfrac{1}{c}\dfrac{\delta^4 n_{-}^{\rm created}}{\delta^3 \mathcal{V}_{\mathbf{u}_{-}}\delta\tau_-}\,,
    \end{array}
\end{equation}
where $n_-$ ($n_+$) is the electron (positron) number density, $\mathbf{u}_-$ ($\mathbf{u}_+$) the 4-velocity of the electron (positron) fluid, $\delta^3 \mathcal{V}_{\mathbf{u}_{-}}$ (or $\delta^3 \mathcal{V}_{\mathbf{u}_{+}}$) is the elementary volume in the reference frame of the electron (positron) fluid, $\delta \tau_+$ ($\delta \tau_-$) is the  positron (electron) fluid proper time and $\delta^4 n_{-}^{\rm created}$ ($\delta^4 n_{+}^{\rm created}$) is the total injection of electrons (positrons) due to photon or neutrino annihilation in the respective elementary time and volume. 

For each process, Eq. (\ref{CreationPairsPhoton}) or  (Eq. \ref{CreationPairsNeutrino}), the number of created electrons is equal to the number of created positrons, which is also equal to the number of disappearing photons for the first process, Eq. (\ref{CreationPairsPhoton}) or neutrinos for the second one, Eq. (\ref{CreationPairsNeutrino}). This exchange of different components implies that the electron-positron fluid component $\rho_0\mathbf{u}\hat{=}m_{e}\left(n_{+} \mathbf{u}_{+}+n_{-} \mathbf{u}_{-}\right)$ is no longer conserved,
\begin{equation}
    \begin{array}{lcc}
        \! \nabla \cdot \left(\rho_0 \mathbf{u}+m_{e}\mathbf{N}_r\right)=0  \! \!& \overset{\text{3+1 formalism}}{\Longrightarrow} \! \! & \boldsymbol{\nabla} \cdot \rho_0 h \gamma \mathbf{V}_{\rm p}=c h k_m\,,
    \end{array}
    \label{CreationPlasma}
\end{equation}
where $\boldsymbol{\nabla}$ indicates the covariant derivative on the spatial $\Sigma_t$ manifold, $\mathbf{N}_r$ is the Feynman number 4-current of radiative component and $\rho_0$ is the mass density  in the electron-positron fluid reference frame. The second equation derives from the steadiness and axisymmetry assumptions.

The term $c k_m = \dfrac{\delta^4 m^{\rm created}}{\delta^3 \mathcal{V}_{\mathbf{u}}\delta\tau_{\mathbf{u}}} $ in 
Eq. (\ref{CreationElectronPositron}) and Eq.(\ref{CreationPlasma})
corresponds to the rate of the created electron-positron mass per unit volume measured in the fluid reference frame and per  fluid proper time unit. $\delta^3\mathcal{V}_{\mathbf{u}}$ is the elementary volume in the fluid reference frame and $\delta \tau_{\mathbf{u}}$ is the elementary proper time. 

Then, $c h k_m $ is the rate of created electron-positron mass per unit volume measured by the ZAMO and per unit time in the Boyer-Lindquist coordinates. We have,
\begin{equation}
    chk_m = - c m_e h \, \nabla \cdot \mathbf{N}_r = \dfrac{\delta^4 m^{\rm created}}{\delta^3 \mathcal{V}_{\rm ZAMO}\delta t} \,.
\end{equation}
We remind that $\delta^3\mathcal{V}_{\mathbf{u}}=\gamma \delta^3 \mathcal{V}_{\rm ZAMO}$, $\delta \tau=\gamma \delta\tau_{\mathbf{u}}$ and $\delta \tau = h \delta t$. 

	\subsection{4-current density and Maxwell's equations}

For both pair creation processes the initial particles are not charged, so the source terms of the electromagnetic field $\mathbf{j}$ are due only to the electron-positron plasma,  $\mathbf{j}=e\left(n_+ \mathbf{u}_+ - n_- \mathbf{u}_-\right)$. Hence, the 3+1 decomposition of Maxwell's equations in a Kerr space-time maintains their ordinary expressions,
\begin{align}
\boldsymbol{\nabla}\cdot \mathbf{E} & =  4\pi \rho_e\,,\label{Maxwell-Gauss}\\
\boldsymbol{\nabla}\cdot \mathbf{B} & =  0\,,\label{Maxwell-Flux}\\
\boldsymbol{\nabla}\times(h\mathbf{E}) & =  \left( \mathbf{B}\cdot\boldsymbol{\nabla}\dfrac{\omega}{c} \right) \varpi\mathbf{e}_{\varphi}\,,\label{Maxwell-Faraday}\\
\boldsymbol{\nabla}\times(h\mathbf{B}) & =  \dfrac{4\pi h}{c}\mathbf{J}-\left(\mathbf{E}\cdot\boldsymbol{\nabla}\dfrac{\omega}{c}\right)\varpi\mathbf{e}_{\varphi}
\label{Maxwell-Ampere}\,,
\end{align}
where $\rho_e$ is the electric charge density and $\mathbf{J}$ the charged current measure by the ZAMO, given by the 3+1 decomposition of the 4-current density $\mathbf{j}=\rho_e \mathbf{n}+\mathbf{J}$. We also assume infinite electrical conductivity,
\begin{equation}\label{idealconduction}
\mathbf{E}+\dfrac{\mathbf{V}\times \mathbf{B}}{c}=0 \,.
\end{equation}     
 	
	\subsection{Euler's equation and effective enthalpy equation}\label{subsec-1.4}

The momentum and the energy equations, respectively the Euler's equation and the first law of thermodynamics, are obtained using the 3+1 decomposition of the energy-momentum conservation. Here our energy-momentum tensor is composed of the electro-magnetic part $\mathbf{T}_{\rm EM}$, the electron-positron part $\mathbf{T}_{\rm FL}$ and the radiative part $\mathbf{T}_{r}$. We note that $\mathbf{T}_{\rm MHD}=\mathbf{T}_{\rm FL}+\mathbf{T}_{\rm EM}$ is the MHD part of the energy-momentum tensor and $\mathbf{k}=-\nabla\cdot\mathbf{T}_{r}=\nabla\cdot\mathbf{T}_{\rm MHD}$ is the 4-force exerted by radiation on the fluid of pairs.  It may also include Compton or Inverse-Compton forces due to pair creation. We will focus our attention on the motion and dynamics of the electron-positron fluid. We make the additional assumption that the distribution function of the considered electron-positron plasma $(m_+ + m_-)f=m_+ f_+ + m_- f_-$ is isotropic for the particle 4-velocities in the fluid reference frame. With these assumptions the energy-momentum of the pair plasma is,
\begin{equation}
\label{def-EnergyMmentumTensorPairs}
\mathbf{T}_{\rm FL}=\rho_0 \xi c^2 \mathbf{u}\otimes\mathbf{u}+P\mathbf{g} \, ,
\end{equation}
where $P$ is an effective pressure. Here, the effective enthalpy per unit mass in the fluid frame $\xi$, which plays a role in the fluid inertia, differs from that obtained for an ideal fluid in thermodynamical equilibrium at a relativistic temperature. We have $\xi_{\rm eq}(\Theta)$ where $\Theta={P/\rho_0 c^2}$ the dimensionless temperature, given by Synge law (see \citealt{synge1957relativistic}, \citealt{osti_4772082}, \citealt{Marle1}) or the Taub-Matthews equation of state (see \citealt{1971ApJ...165..147M}). The difference ${\delta e}/{c^2}=\xi-\xi_{\rm eq}(\Theta)$ is positive where the wings of the distribution function are larger than those of thermodynamics equilibrium fluid and negative in the reverse case. For more details see part 3.3.2 of \cite{chant2018}. 

From the previous equations, we deduce the 3+1 decomposition,
\begin{equation}\label{eulereq}
\begin{array}{cc}
    \rho_0 \gamma \left(\mathbf{V}\cdot \boldsymbol{\nabla} \right)\left(\gamma \xi \mathbf{V}\right)+& \rho_0\xi\gamma^2\left[c^2 {\boldsymbol{\nabla} \ln h}+\dfrac{\varpi\omega V^{\hat{\varphi}}}{h}{\boldsymbol{\nabla}\ln\omega}\right] +\boldsymbol{\nabla} P \\
    & = \rho_e \mathbf{E}+\dfrac{\mathbf{J}\times\mathbf{B}}{c}+\left(\mathbf{k}\cdot\mathbf{n}\right)\mathbf{n}+\mathbf{k}-\gamma\xi c k_m \mathbf{V}\,,
\end{array}
\end{equation}
where $\mathbf{V}$ is the speed of the electron-positron fluid measured by the ZAMO. Note that the mechanism of pair creation produces forces
on the electron-positron fluid. One force comes from the direct effect of the radiative components, the second one comes from the variation of the inertia due to the transformation. In \cite{2019ApJ...880...93H} one assumption is added that in our notations is written as $\mathbf{k}=c^2\left(\nabla\cdot{\rho_0 \mathbf{u}}\right)\mathbf{u}$. This assumption leads to some differences in the treatment of the matching conditions compared to ours, as we will see below.
    
The projection of the energy-momentum conservation equation along the electron-positron fluid 4-velocity in the comoving frame gives,
\begin{equation}
{\rho_0 } \left(\mathbf{V}_{\rm p}\cdot \boldsymbol{\nabla}\right) (\xi c^2)=  \left(\mathbf{V}_{\rm p}\cdot \boldsymbol{\nabla}\right)P-\dfrac{c}{\gamma}\mathbf{k}\cdot\mathbf{u}\,,
\end{equation}
where the ${\rm p}$ subscript means poloidal projection of a vector field. The additional term	$\dfrac{c}{\gamma} \mathbf{k}\cdot\mathbf{u}$ corresponds to the way the injection of pairs contributes to the effective internal enthalpy.
	
	\subsection{Integrals of motion}
	
The first integrals derived 
from Maxwell equations in \cite{CC2018} are still valid here  because those equations are not modified by the introduction of pair production. Therefore, due to axisymmetry, magnetic flux conservation can be written,
 \begin{equation}
\mathbf{B}_{\rm p}=\frac{\boldsymbol{\nabla}A\times \mathbf{e}_\varphi}{\varpi} \, ,
\label{champsMagneicPoloidal}
\end{equation}
with $A$ the magnetic flux function and $\partial_\varphi A =0$. Faraday's  induction law (Eq. \ref{Maxwell-Faraday}) leads immediately to the existence of an electrical potential $\Phi$, such that the electric field is written as $h\mathbf{E}=\dfrac{\omega}{c}\nabla A - \nabla \Phi$.  Axisymmetry also implies that $\partial_\varphi \Phi= 0$. From the symmetries and Ohm's law for infinite electrical  conductivity, Eq. (\ref{idealconduction}), we deduce that the poloidal velocity and the poloidal magnetic field are aligned. Hence, there is a function $\Psi_A$ such that,
\begin{equation}
\label{def-massflux}
\Psi_A \mathbf{B}_{\rm p}=4\pi\rho_0h\gamma \mathbf{V}_{\rm p} \, .
\end{equation}
Inserting this result into  Eq. (\ref{CreationPlasma}) we can interpret $\Psi_A$ as the mass flux per unit magnetic flux on a given  magnetic flux tube. Combining the continuity equation, Eq. (\ref{CreationPlasma}), the divergence-free property of the magnetic field , Eq. ({\ref{Maxwell-Flux}) together with the previous equation we get},
\begin{equation}
    \label{VariationMassFlux}
    \mathbf{B}_{\rm p}\cdot\mathbf{\nabla}\Psi_A = 4\pi c h k_m \, ,
\end{equation}
which governs how the mass flux per unit of magnetic flux $\Psi_A$ varies along a poloidal field line. Together with Ohm's law, Eq. (\ref{idealconduction}), we get,
\begin{equation*}
    c\nabla \Phi = \left[\omega+\dfrac{1}{\varpi}\left(h V^{\hat{\varphi}} - \dfrac{\Psi_A B^{\hat{\varphi}}}{4\pi\rho_0\gamma}\right)\right]\nabla A \, .
\end{equation*}
The electrical potential $\Phi$ is then a function of $A$, $\Phi=\Phi(A)$. We may introduce the so-called  frequency of isorotation $\Omega$, which is a function of the magnetic flux,
\begin{equation}\label{def-isorotation+conservation}
\Omega(A) \equiv c\dfrac{d\Phi}{dA} \,\text{ with }\, \mathbf{E}=-\dfrac{\Omega-\omega}{hc}\nabla A \, .
\end{equation}
We introduce the poloidal Mach number,
\begin{equation}\label{def-MachAlfvenique}
  M_{\rm Alf}^2=h^2\dfrac{{V_{\rm p}}^2}{V_{\rm Alf}^2}=\dfrac{4{\pi}h^2\rho_0\xi\gamma^2{V_{\rm p}}^2}{{B_{\rm p}}^2}=\dfrac{\xi {\Psi_A}^2}{4\pi\rho_0} \, .
\end{equation}

Following \cite{2014PhRvD..89b4041L}, 
we calculate Noether's current densities associated to the two space-time Killing vectors,
\begin{align}
    \label{NoetherAngularMomentumFlux}
    \mathbf{M}_{\rm MHD} &= \;\;\mathbf{T}_{\rm MHD}(\,.\,,\boldsymbol{\xi}) \, , \\
    \label{NoetherEnergyFlux}
    \mathbf{P}_{\rm MHD} &= -\mathbf{T}_{\rm MHD}(\,.\,,\boldsymbol{\eta}) \, .
\end{align}
$\mathbf{M}_{\rm MHD}$ ($\mathbf{P}_{\rm MHD}$) corresponds respectively to the Noether's current density, also called flux, associated with the symmetry generator $\boldsymbol{\xi}$ ($\boldsymbol{\eta}$). In what follows, we will refer to these quantities as the Noether's angular momentum and energy current or flux of the MHD part of the energy-momentum tensor.

From Eq. (\ref{def-EnergyMmentumTensorPairs}), the previous equation and the usual 3+1 decomposition of the electromagnetic energy-momentum tensor allow to get the poloidal component of these Noether's current densities,
\begin{align}
h\mathbf{M}_{\rm MHD,p}&=\frac{\Psi_A L}{4\pi c}\mathbf{B}_{\rm p} \,,\\ \label{polnoetheren}
h\mathbf{P}_{\rm MHD,p}&=\frac{\Psi_A\mathcal{E}}{4\pi c}\mathbf{B}_{\rm p} \,,
\end{align}
where,
\begin{align} 
\label{def-AngularMomentum}
L&= \varpi \left( \gamma \xi V^{\hat{\varphi}}-\dfrac{h B^{\hat{\varphi}}}{\Psi_A}\right)\,,\\ 
\label{def-Energy}
\mathcal{E}&={\gamma\xi hc^2\left(1+\frac{\varpi \omega V^{\hat{\varphi}}}{hc^2}\right)}-\frac{h\varpi \Omega}{\Psi_A} B^{\hat{\varphi}}\, ,
\end{align}
are the usual specific angular momentum and specific energy. Using $\mathbf{k}=\nabla\cdot\mathbf{T}_{\rm MHD}$ and the Killing vector definition we get,
\begin{align}
\label{NoetherConservationAngularMomentum}
\mathbf{\nabla}\cdot \left(h\mathbf{M}_{\rm MHD,p}\right) &= \;\; h \mathbf{k}\cdot\boldsymbol{\xi} \, ,\\
\label{NoetherConservationEnergy}
\mathbf{\nabla}\cdot \left(h\mathbf{P}_{\rm MHD,p}\right) &= - h \mathbf{k}\cdot\boldsymbol{\eta} \, ,
\end{align}
which leads to,
\begin{align}
\label{VariationAngularFlux}
\mathbf{B}_{\rm p}\cdot\boldsymbol{\nabla} \left(\Psi_A L\right)&=4\pi h(\mathbf{k}\cdot\boldsymbol{\xi}) \, ,\\
\label{VariationEnergyFlux}
\mathbf{B}_{\rm p}\cdot\boldsymbol{\nabla} \left(\Psi_A\mathcal{E}\right)&=-4\pi ch(\mathbf{k}\cdot\boldsymbol{\eta}) \, .
\end{align}	
Those equations describe how the angular momentum and energy fluxes of the MHD fields evolve along a poloidal field line. Thus the mass, the angular momentum and the energy of the fluid plus the electromagnetic fields are loaded by the mechanism of pair creation. Therefore the isorotational function remains a constant along a poloidal field line.

	\subsection{Energetic balance on the black hole horizon}
	
It is well known that a Kerr black hole can transfer part of its rotational energy to its environment. The processes of \cite{1971NPhS..229..177P} and \cite{1977MNRAS.179..433B} explain how a Kerr black hole may transfer part of its rotational energy to  particles, or the electromagnetic field. Penrose's process involves particles and Blandford-Znajek's process force-free MHD fields obtained by a perturbation method, expanding the spin for radial or paraboloidal fields. In fact, the black hole may transfer angular momentum and rotational energy to the plasma, as explained and discussed in \cite{2014PhRvD..89b4041L}.  
\begin{figure} 
\centering
\def\svgwidth{0.95\linewidth}{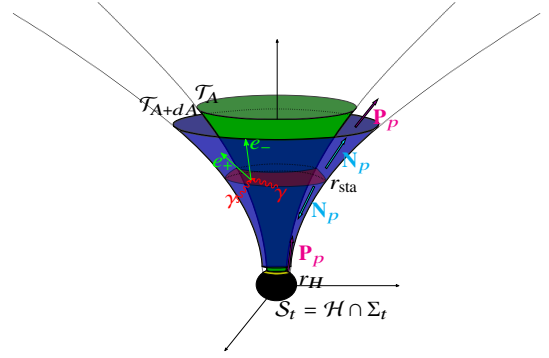}
\caption{Schematic representation of two neighbouring flux tubes $\mathcal{T}_A$ and $\mathcal{T}_{A+dA}$ anchored in the black hole horizon at the $r_H$ radius.  The stagnation surface is defined as the surface with null poloidal speed ($\mathbf{V}_{\rm p}=0$). It is the red disk labelled with $r_{\rm sta}$. $\mathbf{N}_{\rm p}$ and $\mathbf{P}_{\rm p}$ are the poloidal flux of particle and the poloidal Noether's energy flux in the outflow and inflow.}
\label{Fig-IntegrationFigure}
\end{figure}

In our configuration, we consider the interaction between the non force-free magnetised fluid and the black hole. We neglect how the radiation exchanges energy and angular momentum with the Kerr black hole. The exchange of rotational energy of angular momentum is determined by the value of the $\mathbf{T}_{\rm MHD}$ tensor at the horizon of the black hole.  As in \cite{2014PhRvD..89b4041L}, we can integrate Noether's flux conservation equations Eqs.  (\ref{NoetherConservationAngularMomentum})  - (\ref{NoetherConservationEnergy}), or the equivalent system, but this time in the volume of space-time between two 
neighbouring flux tubes $\mathcal{T}_A$ and $\mathcal{T}_{A+{\rm d}A}$ anchored on the horizon of the black hole (see Fig. \ref{Fig-IntegrationFigure}). This integration is performed for radii between $r_H$ and $r>r_H$ and for a time interval between $t$ and $t+{\rm d}t$. In fact,  we could directly integrate Eq. (\ref{VariationMassFlux}) and Eqs. (\ref{VariationAngularFlux})-(\ref{VariationEnergyFlux}) along a poloidal field line anchored into the black hole (see Fig.  \ref{Fig-IntegrationFigure}),
\begin{align}
    \label{MassIntegratedFlux}
    \frac{{\rm d}\dot{M}}{{\rm d}A}&=\Psi_{A,H}\left(A\right)+\int_0^\ell \left.\frac{4\pi h c k_m}{\mid\mid\mathbf{B}_{\rm p}\mid\mid}\right| _{A={\rm cst}}{\rm d}\ell\,,\\
    \label{AngularIntegratedFlux}
    \Psi_{A} L&=\Psi_{A,H} L_H\left(A\right)+\int_0^\ell \left.\frac{4\pi h \mathbf{k}\cdot\boldsymbol{\xi}}{\mid\mid\mathbf{B}_{\rm p}\mid\mid}\right| _{A={\rm cst}} {\rm d}\ell\,,\\
    \label{EnergyIntegratedFlux}
    \Psi_{\it A} \mathcal{E}&=\Psi_{A,H} \mathcal{E}_H\left(A\right)-\int_0^\ell \left.\frac{4\pi hc \mathbf{k}\cdot\boldsymbol{\eta}}{\mid\mid\mathbf{B}_{\rm p}\mid\mid}\right| _{A={\rm cst}} {\rm d}\ell \,.
\end{align}
These equations are easily interpreted, since the mass, the angular momentum and the energy fluxes, per unit magnetic flux, at a given point of a poloidal field line are composed of the contribution of the black hole and the injection of mass, angular momentum and energy, respectively, due to the radiation.

Horizon is a one-way hypersurface, then $\Psi_{A,H}<0$. Energy is extracted from the black hole if $\Psi_{A,H}\mathcal{E}_H\left(A\right)>0$, which is equivalent to $\mathcal{E}_H<0$. The black horizon absorbs negative energy per unit mass $\mathcal{E}_H$ (see also \citealt{2016PTEP.2016f3E01T}).

Then, following \cite{PhysRevLett.25.1596} and \cite{Thorne1987} the black hole physical parameters evolve in time, according to,
\begin{align}
\label{MassEvolution-Hole}
\dfrac{r_H-r_g}{2r_H r_g} \dfrac{d^2 {\rm M}_{\rm irr} c^2 }{ dt dA}&=-\left(\Psi_{A,H}\mathcal{E}_{H}-\omega_H \Psi_{A,H} L_{H}\right)  \, ,\\
\label{AngularMomentumEvolution-Hole}
\dfrac{d^2 \mathcal{J}}{dt dA}&=-\Psi_{A,H} L_{H} \,,\\
\label{EnergyEvolution-Hole}
\dfrac{d^2 {\rm M} c^2}{dt dA}&=-\Psi_{A,H} \mathcal{E}_{H} \,.
\end{align}
where $M_{\rm irr}$ is the irreducible mass of the black hole, 
\begin{equation}
    M_{\rm irr}=M\sqrt{\dfrac{1+\sqrt{1-a^2}}{2}}\,.
\end{equation}
Here we apply the formulation given by \cite{Thorne1987} to link the Noether's energy and angular momentum fluxes on the horizon to the black hole parameters.

In order to interpret physically the energy, it is useful to decompose the energy flux in its different physical contributions, 
\begin{equation}\label{decompofluxenergy}
\Phi_{\mathcal{E}}=\Psi_A \mathcal{E}=\Psi_A\gamma\xi hc^2\left(1+\frac{\varpi \omega V^{\hat{\phi}}}{hc^2}\right)-{h\varpi \Omega} B^{\hat{\varphi}}=\Phi_{\rm FL}+\Phi_{\rm EM} \, .
\end{equation}
 
The first term is the Noether energy flux of the fluid $\Phi_{\rm FL}$. The Noether Poynting flux $\Phi_{\rm EM}$ corresponds to the electromagnetic energy flux. The fluid term $\Phi_{\rm FL}$ is composed of two terms. The first term $\Phi_{\rm M}=\Psi_A\gamma\xi hc^2$ (negative on the black hole horizon) contains the absorption by the black hole of pure massive energy, plus the internal and the kinetic energy of the fluid. We must have $\gamma\xi{\longrightarrow}+\infty$ on the horizon in order to let $h\gamma\xi$ finite and non-zero. The second term  is $\Phi_{\rm LT}=\Psi_A\gamma\xi\varpi \omega V^{\hat{\varphi}}$ that we call the Lense-Thirring term. It can be positive. Its sign will depend on the sign of $V^{\hat{\varphi}}$. Note also that the pair fluid contribution $\Phi_{\rm FL}$ can be written as,
\begin{equation}
\Phi_{\rm FL}=-\Psi_A\xi c^2\left(\mathbf{u}\cdot\boldsymbol{\eta}\right)\,.
\label{PhiPF-Killing}
\end{equation}
 As a consequence, $\Phi_{\rm FL}$ can be positive in an inflow ($\Psi_A < 0$), only if $\boldsymbol{\eta}$ is not a time-like future oriented vector, that is to say in the ergo-region.  

\cite{1977MNRAS.179..433B} show that under certain conditions the Poynting flux can be transported to infinity, meaning that the electromagnetic field is fed by the rotation energy of the Kerr black hole. We say that the \emph{electromagnetic extraction process is active where $\left.\Phi_{\rm EM}\right|_{H}>0$. We also say that the pair fluid process is active where} $\left.\Phi_{\rm FL}\right|_{H}=\left.\Phi_{\rm M}\right|_{H}+\left.\Phi_{\rm LT}\right|_{H}>0$. 

The null energy condition applying to $\mathbf{T}_{\rm MHD}$ on the horizon (see \citealt{2014PhRvD..89b4041L}) writes,
\begin{equation}
    \label{nullenergycondition}
    \left.\Phi_{\mathcal{E}}\right|_{H}\le\left.\Psi_A {L}\right|_{H}\omega_{H} \, ,
\end{equation} 
which implies the impossibility of \emph{generalised energy extraction} ($\left.\Phi_{\mathcal{E}}\right|_{H} > 0$) for non rotating black hole. With regard to Eq.(\ref{MassEvolution-Hole}), this condition implies an increase of the irreducible mass, which also implies an increase of the black hole entropy.	

We can pursue the calculation of this decomposition, by inverting the motion integrals system on the black hole horizon. For the details of the  inversion procedure, see \cite{CC2018}. The result on the horizon is,
\begin{align}
\left.\Phi_{\rm M}^{} \right\rvert_H&=-\:\frac{\left.M^2_{\rm Alf}\right\rvert_H}{\left.M^2_{\rm Alf}\right\rvert_H+{\varpi_H^2(\Omega-\omega_H)^2}/{c^2}}\left(\Psi_A L\omega_H-\Psi_A \mathcal{E}\right) \, ,\\
\left.\Phi_{\rm LT}^{} \right\rvert_H&=\Psi_A {L}\omega_H+\frac{{\varpi_H^2\omega_H(\Omega-\omega_H)}/{c^2}}{\left.M^2_{\rm Alf}\right\rvert_H+{\varpi_H^2(\Omega-\omega_H)^2}/{c^2}}\left(\Psi_A L\omega_H-\Psi_A \mathcal{E}\right) \, ,\\
\label{fluxemhole}
\left.\Phi_{\rm EM}^{}\right\rvert_H&=\frac{{\varpi_H^2\Omega(\omega_H-\Omega)}/{c^2}}{\left.M^2_{\rm Alf}\right\rvert_H+{\varpi_H^2(\Omega-\omega_H)^2}/{c^2}}\left(\Psi_A L\omega_H-\Psi_A \mathcal{E}\right) \, ,
\end{align}
where each quantity is evaluated on the black hole horizon. 

\begin{figure}      
\centering
\def\svgwidth{0.95\linewidth}
\begingroup%
  \makeatletter%
  \providecommand\color[2][]{%
    \errmessage{(Inkscape) Color is used for the text in Inkscape, but the package 'color.sty' is not loaded}%
    \renewcommand\color[2][]{}%
  }%
  \providecommand\transparent[1]{%
    \errmessage{(Inkscape) Transparency is used (non-zero) for the text in Inkscape, but the package 'transparent.sty' is not loaded}%
    \renewcommand\transparent[1]{}%
  }%
  \providecommand\rotatebox[2]{#2}%
  \newcommand*\fsize{\dimexpr\f@size pt\relax}%
  \newcommand*\lineheight[1]{\fontsize{\fsize}{#1\fsize}\selectfont}%
  \ifx\svgwidth\undefined%
    \setlength{\unitlength}{841.88976378bp}%
    \ifx\svgscale\undefined%
      \relax%
    \else%
      \setlength{\unitlength}{\unitlength * \real{\svgscale}}%
    \fi%
  \else%
    \setlength{\unitlength}{\svgwidth}%
  \fi%
  \global\let\svgwidth\undefined%
  \global\let\svgscale\undefined%
  \makeatother%
  \begin{picture}(1,0.70707071)%
    \lineheight{1}%
    \setlength\tabcolsep{0pt}%
    \put(0,0){\includegraphics[width=\unitlength,page=1]{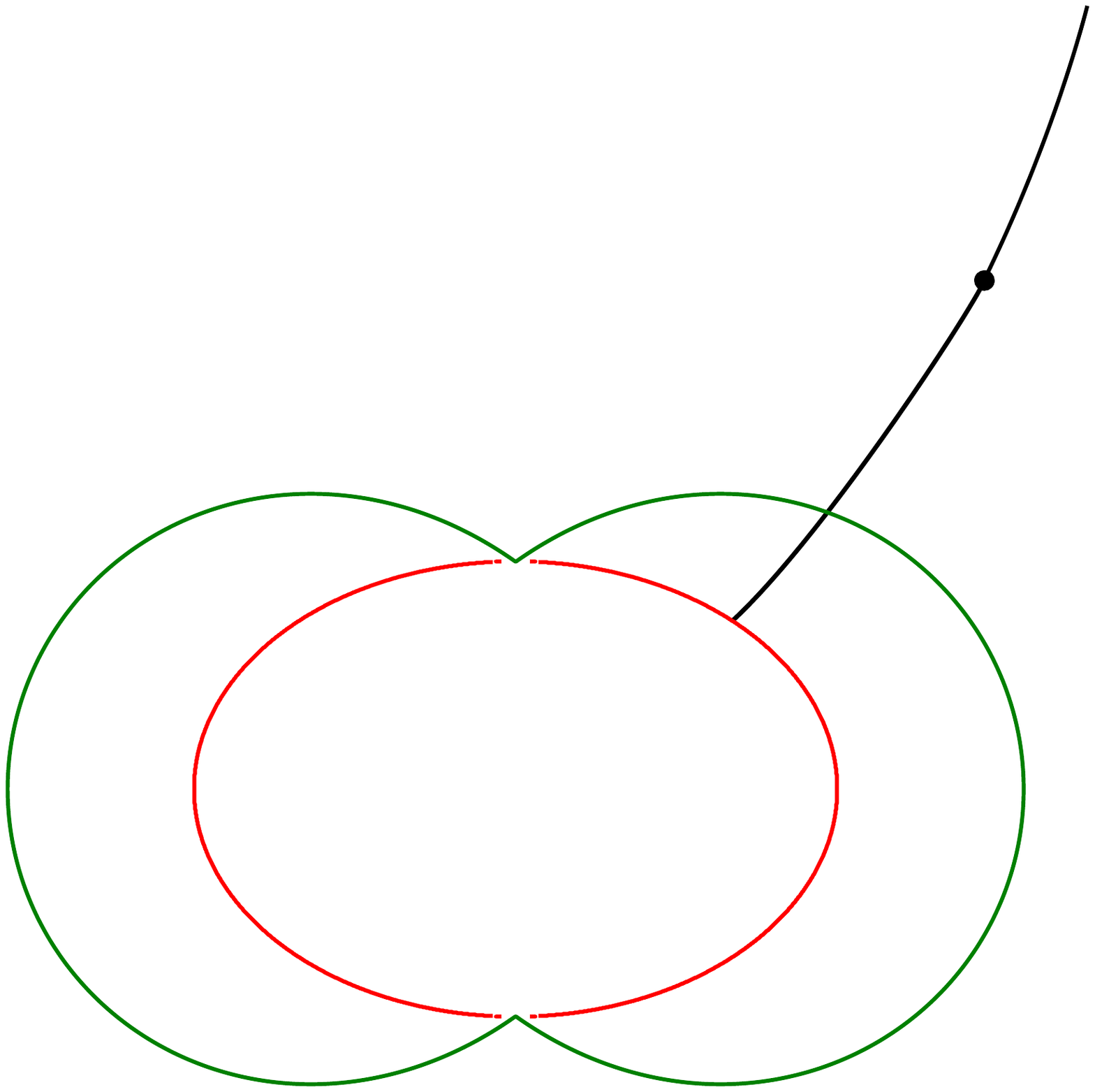}}%
    \put(0.69,0.49){\makebox(0,0)[lt]{\lineheight{1.25}\smash{\begin{tabular}[t]{l}$r_I$\end{tabular}}}}%
    \put(0.595,0.36){\makebox(0,0)[lt]{\lineheight{1.25}\smash{\begin{tabular}[t]{l}{\color{green}$r_E$}\end{tabular}}}}%
    \put(0.73,0.66){\makebox(0,0)[lt]{\lineheight{1.25}\smash{\begin{tabular}[t]{l}$A$\end{tabular}}}}%
    \put(0.53,0.31){\makebox(0,0)[lt]{\lineheight{1.25}\smash{\begin{tabular}[t]{l} {\color{red}$r_H$} \end{tabular}}}}%
    \put(0.3,0.2){\makebox(0,0)[lt]{\lineheight{1.25}\smash{\begin{tabular}[t]{l}\color{red}{$\mathcal{H}$}\end{tabular}}}}%
  \end{picture}%
\endgroup%

\caption{Representation of a typical poloidal velocity and magnetic field line. The injection of pairs is supposed to be located on the black dot outside the ergosphere, i.e. $r_I>r_E$. Without a magnetic field, energy extraction is impossible, if the injection point is outside the ergo-region.} 
\label{Fig-InjectionPosition}
\end{figure}

We focus our attention on a generic field line. Let us now consider the case where injection on this line is entirely located on a point $r=r_I>r_H$. We also suppose that injection on this line is sufficient to have both an inflow ($\Psi_A\le 0$ for $r<r_I$), and an outflow ($\Psi_A> 0$ for $r\ge r_I$). On this line there is no injection $k_m,\mathbf{k}=0$ except at the injection point (black dot in the right part of Fig.\ref{Fig-InjectionPosition}). We should now look at the conditions that the extraction process imposes on the position of the starting point, which is the injection point at $r_{\rm I}$. 

In the case of a pure pair fluid, without electro-magnetic field, $\Phi_{\rm EM}=0$, on the section of the poloidal field line going from the black hole to the location of injection ($r_H\le r \le r_{\rm I}$) (see Fig. \ref{Fig-InjectionPosition}), we have $\Psi_{A}={\rm Cst}<0$. In addition Eqs.(\ref{EnergyIntegratedFlux} and \ref{decompofluxenergy}) imply that for all $r\in]r_H;r_I[$ we have $\Psi_A\mathcal{E}(r)=\Phi_{\rm FL}(r)={\rm Cst}$. In this case, the fluid extraction process cannot be active if the injection is located outside of the ergosphere. Indeed, if the injection starts outside of the ergosphere ($r_{\rm I}\ge r_{\rm E}$), then for a point outside of the ergosphere and below the injection point ($r_E\le r \le r_{\rm I}$)  $\Phi_{\rm FL,H}=\Phi_{\rm FL}(r)=-\Psi_A\xi c^2\left(\mathbf{u}\cdot\boldsymbol{\eta}\right)<0$. In this zone, $-\Psi_A(r)>0$ ($r<r_I$) and $\left(\mathbf{u}\cdot\boldsymbol{\eta}\right)<0$ because outside of the ergosphere $\boldsymbol{\eta}$ is time-like and future oriented. This is equivalent to the necessity in the Penrose process to get the fission of the particles inside the ergosphere.

In the MHD case, we can have extraction via a process of fluid extraction and have a point of injection outside of the ergosphere because we can have exchange between the ideal fluid energy flux and the Poynting flux. Thus, in this case, the Poynting flux increases as one moves out of the ergosphere while the ideal fluid flux became positive inside the ergosphere, being negative outside of the ergosphere. Eq.(\ref{fluxemhole}) implies that the electromagnetic process is active where $0<\Omega<\omega_{H}$, which is a result already obtained by \cite{1977MNRAS.179..433B}.
 
\section{The model on the black hole horizon and on the pair creation layer} \label{sec-sec2}

\subsection{Inflow/outflow model with a thin layer}

In order to describe the MHD field, from horizon to infinite, we need to use source of material to match under some continuity conditions (matching conditions) an inflow to an outflow. We propose here to link an outflow and an inflow solutions of the self-similar model, which have the same stagnation radius, with a thin injection (pair creation or other processes) layer at the level of the stagnation surface. 

Indeed this means that the pair creation terms $k_m$ and $\mathbf{k}$ are null \emph{except at the stagnation surface of the solution} (a sphere). Similar kind of double flow are exposed in \cite{2013PhRvD..88h4046G} or \cite{2019ApJ...880...93H}. In the following we will use the notations adopted by \cite{CC2018}. We also use $_{\rm in}$ and $_{\rm out}$ subscript to refer to a quantity calculated just down the stagnation radius (for inflow) and just up the stagnation radius (for outflow).

The electro-magnetic field source is only due to the electron-positron 4-current.
Nevertheless the creation of pairs or other processes on this thin layer can be at the origin of current and charge density in the layer, which will imply the discontinuity of some components of the electromagnetic fields. The electromagnetic surface sources are then located on the stagnation sphere. 

    \subsection{Matching conditions for both flows}

First, let us consider some thin layer at the stagnation surface position ($r\in\left[r_{\rm sta}-\dfrac{\Delta r}{2};r_{\rm sta}+\dfrac{\Delta r}{2}\right]$) where the pairs are created. The Maxwell-Flux Eq.(\ref{Maxwell-Flux}), the Maxwell-Faraday Eq.(\ref{Maxwell-Faraday}) and the assumptions of axisymmetry and ideality (Eq.\ref{idealconduction}) ensure the continuity of the magnetic flux and the isorotation function along the field line of magnetic field,
\begin{equation}\label{CoFluxIsor}
  \left\{
      \begin{aligned}
\mathbf{B}_{\rm p}\cdot\boldsymbol{\nabla} A &=& 0\\
\mathbf{B}_{\rm p}\cdot\boldsymbol{\nabla} \Omega &=& 0
      \end{aligned}
    \right.
\quad\underset{\Delta r \rightarrow 0}{\Longrightarrow}\quad    
  \left\{
      \begin{aligned}
\left[A\left(r_{\rm sta},\theta\right)\right]_{\rm in}^{\rm out}=0\\
\left[\Omega\left(r_{\rm sta},\theta\right)\right]_{\rm in}^{\rm out}=0
      \end{aligned}
    \right.    \,.
\end{equation}
We require the same stagnation radius for both the inflow and the outflow, assuming $\Delta r\rightarrow 0$. Thus the two flows coincide on the stagnation surface, 
\begin{equation}\label{eq-rsta}
r_{\rm sta}^{\rm out}=r_{\rm sta}^{\rm in}=r_{\rm sta}
\,.
\end{equation}

The stagnation surface is spherical. This induces a continuity of the radial magnetic field component $B^{\hat{r}}$ and, as a consequence, of $\partial_\theta A$.
Thus, the integration of the Maxwell-Flux Eq. (\ref{Maxwell-Flux}) on a infinitesimal volume around the stagnation surface leads that the magnetic field component $B^{\hat{r}}$, perpendicular to this surface, has to be continuous. It is equivalent to the continuity of $\partial_\theta A$, 
\begin{equation}\label{CoBr}
\left[\partial_\theta A\left(\,r_{\rm sta}\,,\theta\right)\right]_{\rm in}^{\rm out}=0 \, .
\end{equation}
Matching inflow and outflow solutions of the meridional self-similar model (\citealt{CC2018}), the continuity of $\partial_\theta A$ is directly obtained from the continuity of the magnetic flux (first line of Eq.\ref{CoFluxIsor}) and from, 
\begin{equation} 
A_{\rm in/out} = A_{\star {\rm in/out}}f_{\rm in/out}(R)\sin^2 \theta\,.
\end{equation}
Here $R$ corresponds to the dimensionless radius $R=r/r_\star$, where $r_\star$ is the Alfv\'en radius. Finally, Eq.(\ref{CoBr}) does not add a matching constraint since it is equivalent to the continuity of the magnetic field component $B^{\hat{r}}$ across the stagnation layer.

In the same way, the integration of Maxwell-Faraday Eq.(\ref{Maxwell-Faraday}) on a small surface delimited by a small loop, using the continuity of the magnetic flux, induces that the latitudinal electric field component, $E^{\hat{\theta}}$, is also continuous across the stagnation layer. 
We may have a discontinuity of the derivative $\partial_r A$ across the stagnation layer, and this jump is due to the toroidal surface current flux, $J_\sigma^{\hat{\varphi}}$, and the surface charge density, $\sigma_e$. After integrating the Maxwell-Gauss Eq.(\ref{Maxwell-Gauss}) and the Maxwell-Ampere Eq.(\ref{Maxwell-Ampere}), we get,
\begin{equation}
\label{Surfacic-DetermineByAngle}
  \left\{
      \begin{aligned}
-\frac{\Omega-\omega}{h h_r c}\left[\partial_r A\left(\,r_{\rm sta}\,,\theta\right)\right]^{\rm out}_{\rm in} &=& 4\pi\sigma_e \, ,\\
\frac{1}{\varpi h_r}\left[\partial_r A\left(\,r_{\rm sta}\,,\theta\right)\right]^{\rm out}_{\rm in} &=& \frac{4\pi}{c} J_{\sigma}^{\hat{\varphi}} \, ,
      \end{aligned}
    \right.  
\end{equation}
where the surface current is defined by $J_\sigma^{\hat{k}}=\underset{\Delta r \rightarrow 0}{\lim}\int_{r_{\rm sta}-\Delta r/2}^{r_{\rm sta}+\Delta r/2}J^{\hat{k}}{\rm{d}}r$, with $k=\theta$ or $\varphi$.

As in \cite{CC2018}, the equations (Eqs.\ref{def-isorotation+conservation},\ref{def-MachAlfvenique},\ref{def-AngularMomentum},\ref{def-Energy}) can be reversed in order to get the expression of the $B^{\hat{\varphi}}$ component as a function of the different physical quantities. The toroidal magnetic field is also linked to the intensity of charges, which cross the surface inside a circle $\mathcal{C}_{r,\theta}=\left\{M\in\Sigma\mid\theta(M)=\theta,r(M)=r\right\}$. In order to calculate the intensity of charges across this surface per unit Boyer-Lindquist time $t$, we need to calculate the flux of $h\mathbf{J}$ across it. We use $h\mathbf{J}$ instead of $\mathbf{J}$ because the current flux $\mathbf{J}$ is calculated in the ZAMO proper time. Thus we are able to show, in the frame of our assumptions, that $I(r,\theta)=\int_{\mathcal{D}_{r,\theta}}h\mathbf{J}\cdot{\rm \mathbf{dS}}=\frac{h c \varpi B^{\hat{\varphi}}}{2}$. Under axisymmetric and stationarity condition the poloidal electromagnetic Noether's flux can be expressed as $\mathbf{P}_{\rm EM,p}=-\frac{ \varpi\Omega B^{\hat{\varphi}}}{4\pi c}\mathbf{B}_{\rm p}=-\frac{I\Omega}{2\pi h c^2}\mathbf{B}_{\rm p}$.

The matching conditions induced by possible electromagnetic surface sources do not impose the  continuity of the global current intensity $I$ across the stagnation surface. The current is also proportional to the Poynting flux per unit of magnetic flux $\Phi_{\rm EM}=-2\frac{\Omega I}{c}$. The global current can be expressed from the inversion of the motion integrals as it has been done in \cite{CC2018}, using the cylindrical radius per unit of light cylinder radius $x=\frac{\varpi\left(\Omega-\omega\right)}{hc}$. For the current we get,
\begin{equation}
\dfrac{2I}{c}=-\Psi_A L \left[1-\dfrac{M_{\rm Alf}^2+\frac{\varpi^2\Omega\left(\Omega-\omega\right)}{c^2}}{M_{\rm Alf}^2-h^2\left(1-x^2\right)}\right]
-\dfrac{\Psi_A \mathcal{E}}{c^2}\dfrac{\varpi^2\left(\Omega-\omega\right)}{M_{\rm Alf}^2-h^2\left(1-x^2\right)}\, .
\label{Integrated-Current}
\end{equation}
It implies that the continuity of the global current requires another constraint on how the  injected angular momentum $\Delta \Psi_A L$ is related to the injected energy $\Delta \Psi_A \mathcal{E}$. Since there is no physical reason to have this requirement, our  way of matching can support to have a discontinuity in the global current.

Eqs.(\ref{VariationMassFlux},\ref{VariationAngularFlux},\ref{VariationEnergyFlux}) imply the discontinuity of $\Psi_A$, $\Psi_A L$ and $\Psi_A \mathcal{E}$ across the stagnation layer. In other words the r.h.s of those three equations are zero except onto the stagnation surface itself where they are equal to a Dirac delta function. Thus from Eq.(\ref{Integrated-Current}) we also get a discontinuity of the current intensity $I$, which is equivalent to a discontinuity of $B^{\hat{\varphi}}$, becoming a discontinuity of the Poynting flux. It implies the presence of some meridional surface current $J_{\sigma}^{\hat{\theta}}$. 

\begin{figure}     
\centering
\includegraphics[width=\linewidth]{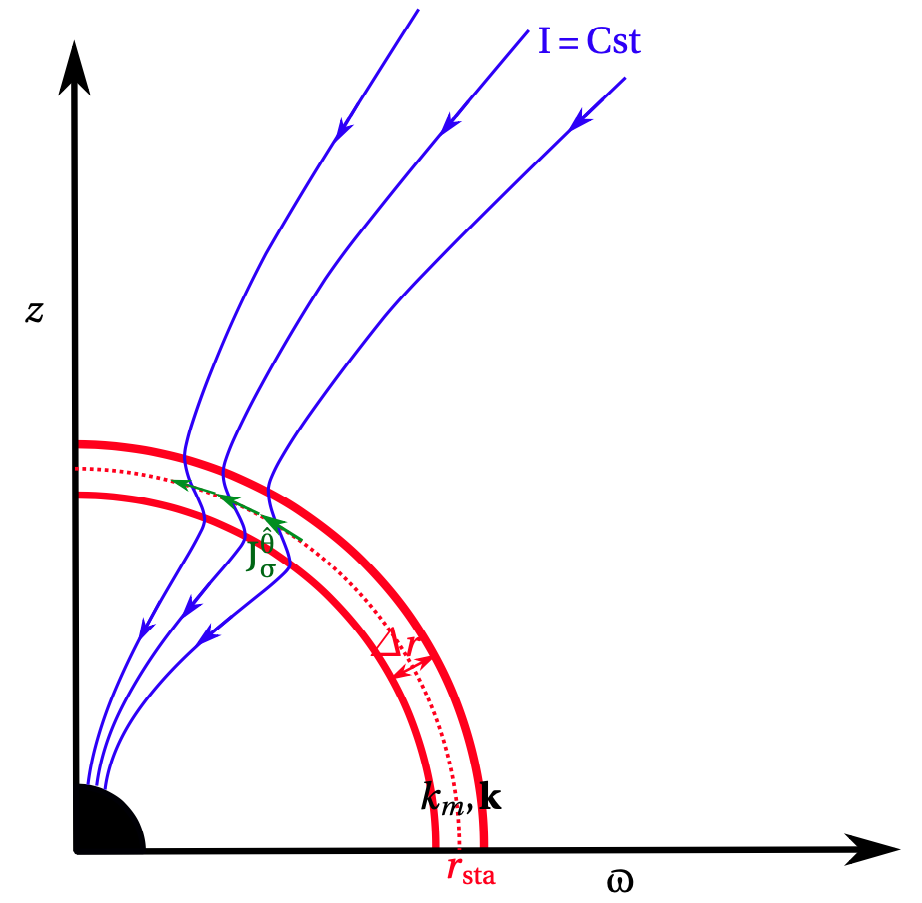}
\caption{The meridional surface current $j_\sigma^{\hat{\theta}}$ in the stagnation layer of thickness $\Delta r$ creates a discontinuity of the global current function $I$. }      
\label{CurrentIntensity}
\end{figure}

A paradox seems to appear because charges may accumulate somewhere on the stagnation surface, due to the existence of non null $J_\sigma^{\hat{\theta}}$. To solve it, we use a  schematic view of constant intensity $I$ tube in the poloidal surface (see Fig. \ref{CurrentIntensity}). 

In our assumptions $\Delta r \rightarrow 0$ and we get a discontinuity of the intensity function, which implies a discontinuity of $hJ^{\hat{r}}$. It is linked to the variation with $\theta$ of the surface current $J_\sigma^{\hat{\theta}}$. Using the charge conservation, we can calculate the jump of the radial current at a co-latitude $\theta$,
\begin{equation}
hJ^{\hat{r}}_{\rm in}\left(r_{\rm sta}\right)=hJ^{\hat{r}}_{\rm out}\left(r_{\rm sta}\right)+\frac{1}{\Sigma\sin\theta}\frac{\partial}{\partial \theta}\left(\rho\sin\theta J_{\sigma}^{\hat{\theta}}\right) \, .
\end{equation}

Integrating this equation in order to let appear the discontinuity of the current function, we get,
\begin{equation}\label{eq-increascurrent}
I_{\rm in}=I_{\rm out}+ 2\pi\rho(r_{\rm sta},\theta)\sin\theta J_{\sigma}^{\hat{\theta}} \, .
\end{equation}

For each component of the flow, the parameters of the physical quantities are normalised to their own Alfv\'en radius. Thus, we need to adjust the two components of the flows, such that they correspond to a black hole of a given mass and spin. From the previous discussion, we also get a continuity of the isorotation function and the radial magnetic field component. Then, the matching conditions write, 
\begin{equation}\label{eq-mmc-1}
\begin{array}{lllll}
    \left[r_g\right]^{\rm out}_{\rm in}=0,& 
    \left[a\right]^{\rm out}_{\rm in}=0,&
    \left[\Omega\right]^{\rm out}_{\rm in}=0,&
    \left[B^{\hat{r}}\right]^{\rm out}_{\rm in}=0 .
\end{array}
\end{equation}

The first three conditions combined with the assumption written in Eq.(\ref{eq-rsta}) lead to the three following jump conditions for the parameters,
\begin{equation}\label{eq-mmc}
  \left\{
      \begin{aligned}
\left[\frac{R_{\rm sta}}{\mu}\right]^{\rm out}_{\rm in}&=0\,, \\
\left[\frac{l}{\mu}\right]^{\rm out}_{\rm in}&=0 \, ,\\
\left[\dfrac{\mu^3}{2\left(1+l^2\right)^2}+\dfrac{ \lambda\mu^{3/2} }{\nu}\sqrt{1-\dfrac{\mu}{1+l^2}}\right]^{\rm out}_{\rm in}&=0 \, .
      \end{aligned}
    \right.  
\end{equation}
We use those three conditions for numerical matching. Note that the third condition ensures the continuity of the isorotation function across the stagnation surface. 

The last equation of Eqs. (\ref{eq-mmc-1}) set and the Eq. (\ref{eq-rsta}) provide for the two flows the ratio  between the Alfv\'en radii and the one between the magnetic field values on the axis at the Alfv\'en surfaces,
\begin{equation}\label{eq-mmc-2}
  \left\{
      \begin{aligned}
\dfrac{r_\star^{\rm in}}{r_\star^{\rm out}}&=\dfrac{\mu_{\rm out}}{\mu_{\rm in}}\\
\dfrac{B_\star^{\rm in}}{B_\star^{\rm out}}&=\left(\dfrac{\mu_{\rm in}}{\mu_{\rm out}}\right)^2\dfrac{f_{\rm out}}{f_{\rm in}}
      \end{aligned}
    \right.  
\end{equation}
The second equation ensures the continuity of the magnetic flux. It imposes the ratio between $B_\star^{\rm in}$ and $B_\star^{\rm out}$ without bringing any extra constraint. As already seen above, in our model the continuity of the radial magnetic field component written in Eq.(\ref{CoBr}) is directly derived from the continuity of the magnetic flux. 

Here we do not impose continuity of the Poynting flux between the inflow and the outflow because pair creation can induce the production of Poynting flux. Discontinuity of the Poynting flux is equivalent to discontinuity of the total current intensity $I$ (see Fig. \ref{CurrentIntensity}).

	\subsection{Energetic balance at the stagnation surface}
	\label{subsec-theoryloadingoninterface}
	
Once, we have the inflow and outflow solutions from our semi-analytical model, which satisfy the system of Eqs.(\ref{eq-mmc}), we deduce the mass injection rate $k_m$ of the pair creation and the 4-force $\mathbf{k}$ of the radiation field on the fluid of pairs. This includes the Compton, inverse Compton and pair creation. These terms take the following forms, 
\begin{align}
\label{SourceTerme-MaterialCreationRate}
k_m&=k_{m,{\rm sta}}\left(\theta\right)\delta\left(r-r_{\rm sta}\right) \, , &\\ \label{SourceTerme-ForceCreationpairs}
\mathbf{k}&=\mathbf{k}_{\rm sta}\left(\theta\right)\delta\left(r-r_{\rm sta}\right)\, .&
\end{align}

The integration on the stagnation surface of Eqs.\ref{VariationMassFlux}, \ref{VariationAngularFlux} and \ref{VariationEnergyFlux} gives the variation of the mass, angular momentum and energy fluxes. On the stagnation surface we get the following system,
\begin{align}
\label{Transition-MassFlux}
\Psi_A^{\rm out}\left(A\right)&=\Psi_A^{\rm in}\left(A\right)+\frac{4\pi c h h_r}{B^{\hat{r}}}k_{m,{\rm sta}}\left(\theta_{\rm sta}\left(A\right)\right)\, , &   \\ \label{Transition-AngularMomentumFlux}
\left(\Psi_A L\right)^{\rm out}\left(A\right)&=\left(\Psi_A L\right)^{\rm in}\left(A\right)+\frac{4\pi h h_r}{B^{\hat{r}}}\boldsymbol{\xi}\cdot \mathbf{k}_{\rm sta}\left(\theta_{\rm sta}\left(A\right)\right) \, ,&  \\
\label{Transition-EnergyFlux}
\left(\Psi_A \mathcal{E}\right)^{\rm out}\left(A\right)&=\left(\Psi_A \mathcal{E}\right)^{\rm in}\left(A\right)-\frac{4\pi c h h_r}{B^{\hat{r}}}\boldsymbol{\eta}\cdot \mathbf{k}_{\rm sta}\left(\theta_{\rm sta}\left(A\right)\right) \, .&
\end{align}

In the outflow, the mass flux is positive, i.e. directed outwards, whereas in the inflow, it is negative, i.e. directed inwards. Applying the first condition to Eq.\ref{Transition-MassFlux} implies that for each colatitude we have $\frac{4\pi c h}{B^{\hat{r}}}k_{m,{\rm sta}}\left(\theta\left(A\right)\right)\ge\left(-\Psi_A^{\rm in}\left(A\right)\right)$, which means that the rate of pair creation needs to be sufficient to reverse the mass flux. The mass, angular momentum and energy injected per unit time and per unit dimensionless magnetic flux $\alpha$, evaluated for the inflow solution, are given by the following expressions,
\begin{align}
\label{InjectedMassPerAlpha}
\dfrac{{\rm d}^2 M_{\rm Inj}}{{\rm d}t {\rm d}\alpha}&=\dot{M}_{\rm Inj}^\star\frac{4}{\mu_{\rm out}^2}\left(\frac{G_{\rm out}}{G_{\rm in}}\right)^4\left[\dfrac{c\Psi_A^{\rm out}}{B_\star^{\rm out}}-\left(\dfrac{G^2_{\rm in}}{G^2_{\rm out}}\right)\dfrac{c\Psi_A^{\rm in}}{B_\star^{\rm in}}\right]\,,\\
\label{InjectedAngularMomentumPerAlpha}
\dfrac{{\rm d}^2 J_{\rm Inj}}{{\rm d}t {\rm d}\alpha}&=\dot{J}_{\rm Inj}^\star\frac{8}{\mu_{\rm out}^3}\left(\frac{G_{\rm out}}{G_{\rm in}}\right)^4\left[\dfrac{\Psi_A^{\rm out}L_{\rm out}}{B_\star^{\rm out}r_\star^{\rm out}}-\left(\dfrac{\mu_{\rm out} G^2_{\rm in}}{\mu_{\rm in} G^2_{\rm out}}\right)\dfrac{\Psi_A^{\rm in}L_{\rm in}}{B_\star^{\rm in}r_\star^{\rm in}}\right]\,,\\ 
\label{InjectedEnergyPerAlpha}
\dfrac{{\rm d}^2 E_{\rm Inj}}{{\rm d}t {\rm d}\alpha}&=\dot{E}_{\rm Inj}^\star\frac{4}{\mu_{\rm out}^2}\left(\frac{G_{\rm out}}{G_{\rm in}}\right)^4\left[\dfrac{\Psi_A^{\rm out}\mathcal{E}_{\rm out}}{B_\star^{\rm out}c}-\left(\dfrac{G^2_{\rm in}}{G^2_{\rm out}}\right)\dfrac{\Psi_A^{\rm in}\mathcal{E}_{\rm in}}{B_\star^{\rm in}c}\right]\,,
\end{align}
with the following constants for the injection,
\begin{equation}\label{eq-typicalscale}
    \begin{array}{ccc}
    {\dot{M}_{\rm Inj}}^\star = \dfrac{{r_g}^2 {B_\star^{\rm in}}^2}{2 c}\,, &  {\dot{J}_{\rm Inj}}^\star = \dfrac{{r_g}^3 {B_\star^{\rm in}}^2}{2}\,, & {\dot{E}_{\rm Inj}}^\star = \dfrac{c {r_g}^2 {B_\star^{\rm in}}^2}{2}\,.
    \end{array}
\end{equation}
To get an order of magnitude for these quantities, we need the black hole mass and the value of the magnetic field at the Alfv\'{e}n surface on the axis for the outflow $B_\star^{\rm out}$. For M87 the mass of the supermassive black hole is ${M_H}\approx\left(6.6\pm0.4\right) \times10^9 M_{\odot}$ \citep{2011ApJ...729..119G}. On the axis of the M87 jet, \cite{kino2014relativistic} give at the distance of $20r_g$ a magnetic field of the order of few Gauss, $B(20r_g)\approx \unit[5\pm4]{G}$. In our model, along the axis $B^r=B_\star/G^2$ (see \citealt{CC2018}). We restrain ourselves to solutions with $r_{\rm sta} < 20 r_g$.  In a recent publication, the \citeauthor{2021ApJ...910L..13E} uses polarised emission imaging to estimate the magnetic field. They obtain a typical value of $B\approx \unit[1-30]{G}$ in the region near the horizon. They also use a one-zone isothermal sphere model to estimate the magnitude of the magnetic field and get $\approx \unit[5]{G}$ at $5r_g$. For solutions with $r_{\rm sta} < 5 r_g$ we prefer to use the last observational constrain to fix the value of $B_\star^{\rm out}$. Then $B_\star^{\rm in}$ is calculated using the last equation of Eq.(\ref{eq-mmc-2}) set.

Since the fluxes are conserved in the inflow along a given magnetic field line, which crosses the horizon of the black hole, and using Eqs.\ref{AngularMomentumEvolution-Hole} and \ref{EnergyEvolution-Hole}, the fluxes can be related to the variations of the black hole mass, angular momentum and energy. It writes,
\begin{align}
\left(\Psi_A L\right)^{\rm out}\left(A\right)&=-\frac{d^2 J_{\rm H}}{dtdA}\left(\theta_{\rm {H}}\left(A\right)\right)+\frac{4\pi  h}{B^{\hat{r}}}\mathbf{\xi}\cdot \mathbf{k}_{\rm sta}\left(\theta_{\rm sta}\left(A\right)\right) \, , \\
\label{HoleInfinite-EnergyFlux}
\left(\Psi_A \mathcal{E}\right)^{\rm out}\left(A\right)&=-\frac{d^2 M_{\rm H}c^2}{dtdA}\left(\theta_{\rm {H}}\left(A\right)\right)-\frac{4\pi c h}{B^{\hat{r}}}\mathbf{\eta}\cdot \mathbf{k}_{\rm sta}\left(\theta_{\rm sta}\left(A\right)\right) \, .
\end{align}

Thus for a line, which crosses the black hole horizon, the flux at infinity is constituted from the flux given by extraction from the black hole and the flux given by the photons which are transformed into pairs.

In the frame of the model proposed by \cite{CC2018}, 
the rate at which energy is extracted from the rotating black hole is,
\begin{equation}
-\frac{dE_{H}}{dt}=\int_0^{A_H}\Psi_A\mathcal{E}dA=B_{\star,{\rm in}}^2 r_s^2 c\frac{h_\star^2 \nu}{2\mu^{3/2}}\dot{w}\left(\alpha_{H},\delta,e_1\right) \, ,
\end{equation}
with,
\begin{equation*}
\begin{aligned}
\dot{w}\left(\alpha_{H},\delta,e_1\right)&=\frac{2}{3\delta}\left\{\left(1+\delta\alpha_{H}\right)^{3/2}\left(1+\frac{3e_1\alpha_{H}}{5}-\frac{2e_1}{5\delta}\right)-1+\frac{2e_1}{5\delta}\right\}\\
&\underset{\delta\rightarrow 0}{\approx} \alpha_{H}+\frac{e_1\alpha_{H}^2}{2}+\frac{\delta\alpha_{H}^2}{4}\left(1+\frac{2e_1\alpha_{H}}{3}\right)\, .
\end{aligned}
\end{equation*}


\section{Double flow solutions}\label{sec-sec3}
    \subsection{Parameters of three matching solutions}
    
\begin{table*}
\centering
\begin{tabular}{*{10}{c}}
   \hline
    \multicolumn{2}{c}{\rm Solution} & $\lambda$ & $\kappa$ & $\delta$ & $\nu$ & $l$ & $\mu$ & $\Pi_\star$ & $e_1$ \\
    \hline\hline
    {M1} & I1 & $0.036$ & $0.468$ & $0.075$ & $-1.79$ & $0.12$ & $0.442$ & $1.4$ & $-0.21$ \\
    & O1 & $0.985$ & $0.230$ & $1.328$ & $0.386$ & $1.016 \times 10^{-2}$ & $3.758 \times 10^{-2}$ &  & $6.892\times 10^{-3}$ \\ \hline  
    {M2} & I2 & $0.392$ & $1.341$ & $0.355$ & $-1.562$ & $0.17$ & $0.807$ & $0.859 $ & $-0.349$ \\ 
    & {O2} & $0.998$ & $0.280$ & $1.296$ & $0.234$ & $6.502\times10^{-3}$ & $3.012\times 10^{-2}$ & & $6.892\times 10^{-3}$ \\
    \hline
   {M3} & {I3} & $0.388$ & $5.898$ & $0.259$ & $-1.443$ & $0.25$ & $0.978$ & $0.275$ & $-0.555$\\
   & {O3} & $1.171$ & $0.291$ & $1.319$ & $0.600$ & $4.767\times 10^{-2}$ & $0.184$ & & $-6,268\times 10^{-2}$\\
    \hline 
\end{tabular}
\caption{Input parameters for the three solutions. For each of them, the first line presents the parameters for the inflow solution and the second line the parameters for the outflow solutions. The parameter $\lambda$ is the dimensionless ratio of angular momentum flux per unit of magnetic flux. The parameter $\kappa$ is the deviation from spherical symmetry of the pressure, while $\delta$ is the deviation from spherical symmetry of the number density/enthalpy ratio. $\nu$ is the escape speed per unit velocity of the fluid at the Alfv\'{e}n point, along the polar axis. $l$ is the dimensionless black hole spin  and $\mu$ the Schwarzschild radius per unit Alfv\'{e}n radius. $\Pi_\star$ is the dimensionless pressure at the Alfv\'{e}n point along the polar axis and $e_1$ the deviation from spherical symmetry of the total energy. \label{Tab-Param-Input}}

\end{table*}

As explained above, a solution of inflow or outflow is fully determined by eight parameters called the input parameters ($\lambda,\kappa,\delta,\nu,l,\mu,\Pi_\star$ and $e_1$). They must be fixed in order to solve the ordinary differential equations system (see Appendix C of \citealt{CC2018}). For the outflow solution, the $\Pi_\star$ value is automatically adapted by lowering its value to the limiting value to avoid oscillations in the jet.  Thus the non-oscillating outflows are determined by seven input parameters. Conversely, inflows are determined by eight parameters, as $\Pi_\star$ remains free. 

A necessary condition, to extract energy at some colatitude from the black hole, is to choose negative values of $e_1$ for the inflow solution. In this case, it is possible to inject negative energy (see Eq.(60) of \citealt{CC2018}), as long as the black hole accretes enough magnetic flux. The magnetic flux on the equator of the black hole horizon must be higher than a minimum threshold value. In dimensionless form, it writes as $\alpha_H(\theta=\pi/2)>-1/e_1$. In Tab.(\ref{Tab-Param-Input}), we give the set of input parameters used for building three inflow/outflow solutions of the meridionally self-similar model.

\begin{table}
\begin{center}
\begin{tabular}{*{7}{c}}
    \hline
    \multicolumn{2}{c}{\rm Solution} & $a$ & $\Omega/\omega_{{H}}$ & $r_{\rm sta}/r_{\rm H}$ & $\gamma_{\rm max,ax}$ & $\xi_{\star}$   \\
    \hline\hline
    {M1} & {I1} & 0.5429 & 6.2167$\times 10^{-2}$ & 3.1777 & 15 & 3430  \\ 
    	& {O1} & 0.5410 & 6.2047$\times 10^{-2}$ & 3.1771 & 1.47 & 1.42 \\
    \hline   
    {M2} & {I2} & 0.4316 & 9.6912$\times 10^{-2}$ & 1.5031 & 11 & 1360  \\ 
    	& {O2} & 0.4316 & 9.6912 $\times 10^{-2}$ & 1.5031 & 4 & 1.5 \\
    \hline   
    {M3} & {I3} & 0.5189 & 0.5022 & 1.1755 & 12 & 1470 \\
    	& {O3} & 0.5189 & 0.5022 & 1.1750 & 10 & 19.6 \\
    \hline             
\end{tabular}
\caption{Output parameters for the three solutions. For each of them, the first line lists the parameters for the inflow solution and the second line the parameters for the outflow solutions. We give $a$ the dimensionless black hole spin, $\Omega/\omega_{H}$ the dimensionless isorotation frequency and $r_{\rm sta}/r_{H}$ the dimensionless stagnation radius using minimal matching conditions, for the three solutions M1, M2 and M3. The two last columns give the maximum Lorentz factor along the fluid axis and the effective enthalpy $\xi_\star$ at the Alfv\'{e}n point, on the polar axis. \label{Tab-Param-Match}}
\end{center}
\end{table}	

We show in Tab.(\ref{Tab-Param-Match}) the output results of our three inflow/outflow solutions, under minimal matching conditions given in Eq.(\ref{eq-mmc}). We give $a$ the usual dimensionless black hole spin in unit of the gravitational radius, $\Omega/\omega_{H}$ the isorotation frequency in unit of angular velocity on the black hole horizon, and $r_{\rm sta}/r_{H}$ the stagnation radius in unit of black hole horizon radius. We also give, for the three global solutions, the maximum Lorentz factor and the effective enthalpy $\xi_{\star}$ at the Alfv\'{e}n point, both along the polar axis.

We choose the input parameters, both for the inflow and the outflow, in order to satisfy specific conditions for the solutions and to match them under the minimal conditions Eq.(\ref{eq-mmc}). We require the final inflow Lorentz factor on the axis to be higher than $10$. The variation of the Lorentz factor with the magnetic flux in the inflow is negative or null on the north pole horizon. Thus, we use numerical gradient descent techniques in the parameter space (see Appendix \ref{an-gdt}). 

We start by building three inflow solutions with different kind of energy exchange with the black hole that satisfy our constraints. Then using a numerical gradient descent technique, we build three outflow solutions, each matching one of our inflow solutions. We get a discrepancy for $a$, $\Omega/\omega_{H}$ and $r_{\rm sta}/r_{H}$ between the inflow solution (I1, I2 or I3) and the outflow solution (O1,O2 and O3 respectively) lower than $10^{-2}$. The parameters are listed in Tab.(\ref{Tab-Param-Match}). The numerical value of $\gamma_{\rm max, ax}$ is not infinite on the black hole horizon, along the axis. It is numerically impossible to get an inflow solution with $\gamma=+\infty$ on the horizon so we choose to have $\gamma_{\rm max,ax}>10$. We could tune $\Pi_\star$ in the inflow to get this constrain for $\gamma$ close to the horizon. As explained in Appendix  \ref{an-sec-modelhorizonbehaviour}, getting $\gamma \underset{R\rightarrow R_H}{\sim} {1}/{h_z}$ implies that the $\Pi$ function on the horizon tends to $\ln(R-R_H)$, but if $\gamma$ is not infinite at $R=R_H$, $\Pi$ behaves as $1/(R-R_H)$. Instead of adapting the inflow $\Pi_\star$ in order to increase $\gamma_{\rm max,ax}$, we prefer to keep a degree of freedom to solve the difficulties of the matching conditions. Let us also remind that, in the outflow with high asymptotic Lorentz factor, this factor deviates slightly from its maximal value in the asymptotic regime. This is due to numerical reasons, because it is not possible to get a sufficiently precise value of $\Pi_\star$ to tune it to the non oscillating value.

To obtain the solution we also need to fix the value of the three parameters $M$, $B_\star$ and $\xi_\star$. The geometry and the velocity profiles of one solution are not depending on these free parameters.  We set the black hole mass to $6.6 \times 10^9 M_\odot$, which is the value mentioned in \cite{2011ApJ...729..119G} for the M87 black hole mass. This is within the range measured by the \citeauthor{2019ApJ...875L...1E}, $M=6.5\pm 0.7\times 10^9 M_\odot$. As discussed above, we use observational constraints to fix the magnitude of the second parameter, the magnetic field strength $B_\star$. Since the solution M1 has a stagnation radius larger than $5r_g$, we use $B(20 r_g)\approx 1{\rm G}$ and we take $B_{\star,{\rm out}}\approx G^2_{\rm out}(20r_g){\rm G}$. For solutions M2 and M3 the stagnation radius is lower than $5r_g$, then we use $B(5 r_g)\approx 4.9{\rm G}$ and we take $B_{\star,{\rm out}}\approx 4.9 G^2_{\rm out}(5r_g){\rm G}$. We put in Tab.(\ref{Tab-TypicalParam}) the values of $B_\star^{\rm in}$ obtained for each solution. 

Once the value of $B_\star$ is determined, the value of $\rho_\star \xi_\star$ is known from the definition of the Alfv\'{e}n surface. From Eqs.(73) and (74) of \cite{CC2018}, the effective enthalpy and the mass density fields are scaled by the factors $\xi_\star$ and $\rho_\star$, respectively. Once $\xi_\star$ is given, both the effective enthalpy and the mass density field scaling are fixed .

We choose $\xi_\star$ for the outflow solution as in \cite{CC2018}, such that the effective internal energy at infinity on the axis reduces to the internal energy of a gas at thermodynamic equilibrium. 

\cite{2014Natur.510..126Z} define a scaling-law $\Phi_{\rm jet}^Z=f_{Z}^{ }\sqrt{\dot{M}c \left(\dfrac{r_s}{2}\right)^2}$ between the magnetic flux of the jet and the total disk accretion rate $\dot{M}$. They assume that the jet power is the result of a pure Blandford-Znajek mechanism. Thus they deduce the black hole magnetic flux $\Phi_{\rm BH}$ and from observations a value of $f_Z^{ }\sim 50$.

We use a similar scaling law for the black hole magnetic flux $\Phi_{\rm BH} = f_{\rm inf}\sqrt{\dot{M}_{\rm inf}c \left(\dfrac{r_s}{2}\right)^2}$ and the inflow mass rate. Our scaling factor $f_{\rm inf}$ must be larger than $f_Z$ since the magnetic flux of the jet is only part of the one threading the black hole in the magnetosphere. We choose $f_{\rm inf} \sim 150$, a value three times higher than $ f_Z^{ }\sim 50$, because we choose $\dot{M}_{\rm inf}$ of order one tenth of $\dot{M}$ for the same magnetic flux $\Phi_{\rm BH}$. In our model the efficiency to create magnetic flux from the pair inflow is higher that the one deduced from standard Blandford-Znajek theory applied in the jet at 1 pc. This scaling law is used to define the value of $\xi_\star$ in the inflow.

Indeed, from the model we can derive,
\begin{equation}
    \Phi_{\rm BH}^2\approx\frac{\sqrt{\mu}\xi_\star\gamma_\star\left(1+\sqrt{1-\left(\frac{2l}{\mu}\right)^2}\right)}{|\nu|h_\star G^2_{H}}\dot{M}_{\rm inf}c\left(\frac{r_s}{2}\right)^2\,,
\end{equation}
which leads to,
\begin{equation}
    \xi_{\star} = \dfrac{ |\nu| h_\star G^2_H}{\sqrt{\mu}\gamma_\star \left(1+\sqrt{1-a^2}\right)}{f_{\rm inf}^2}\,,
\end{equation}
where all the quantities are evaluated in the inflow. 

	\subsection{Field line geometry with quasi-isotropic coordinates}

\begin{figure*}     
\centering
\includegraphics[width=0.49\linewidth]{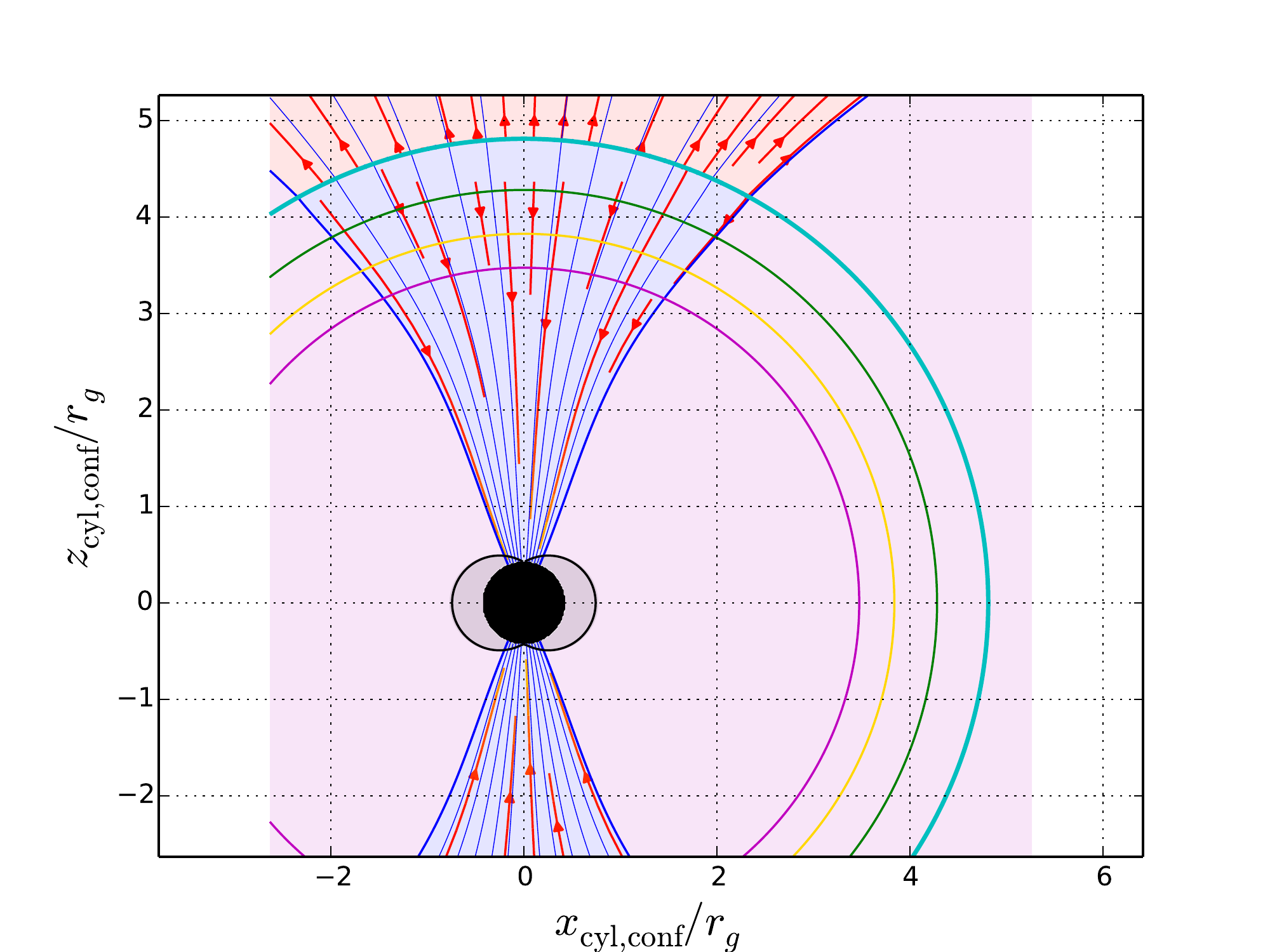}\includegraphics[width=0.49\linewidth]{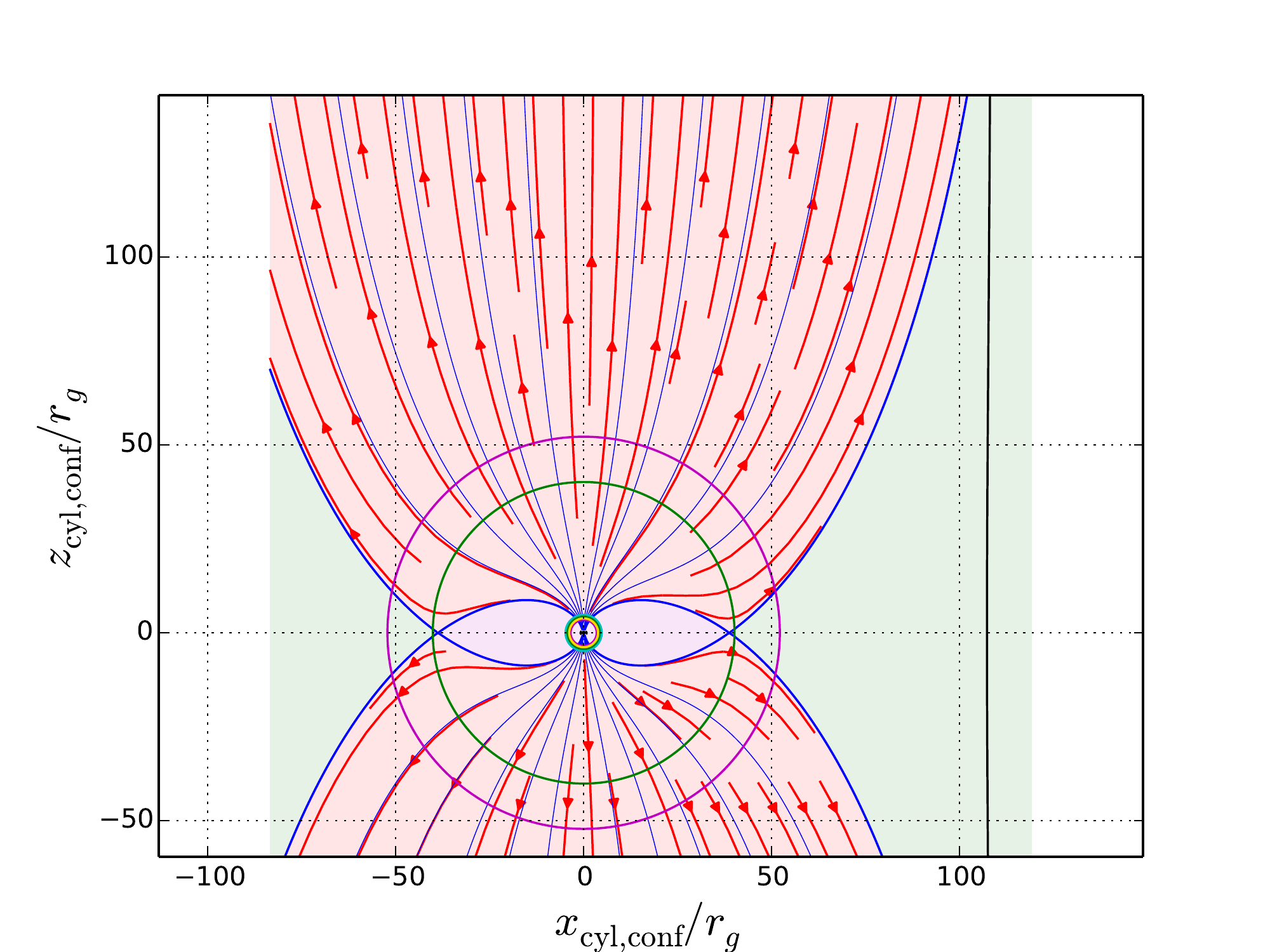}
\includegraphics[width=0.49\linewidth]{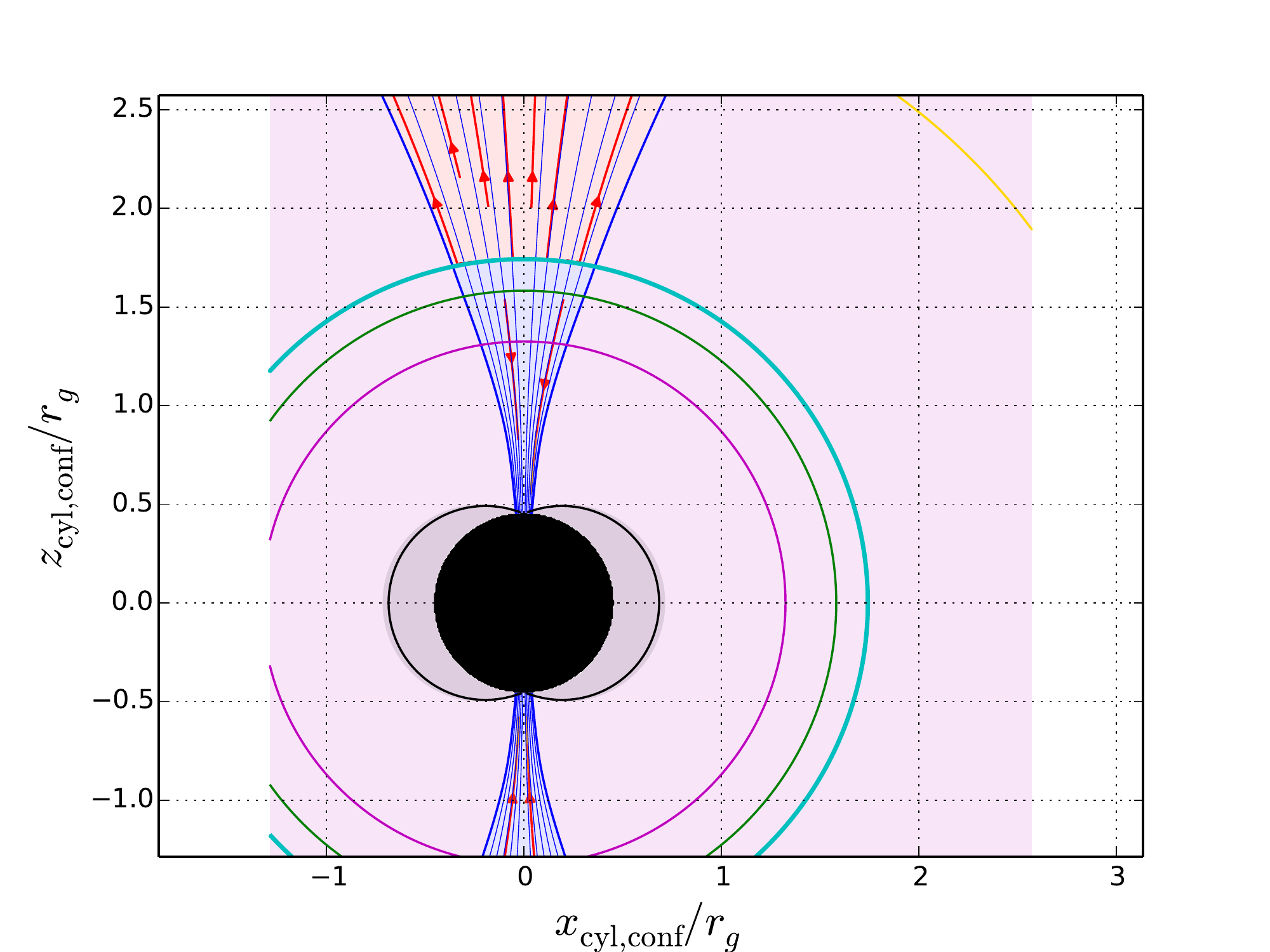}\includegraphics[width=0.49\linewidth]{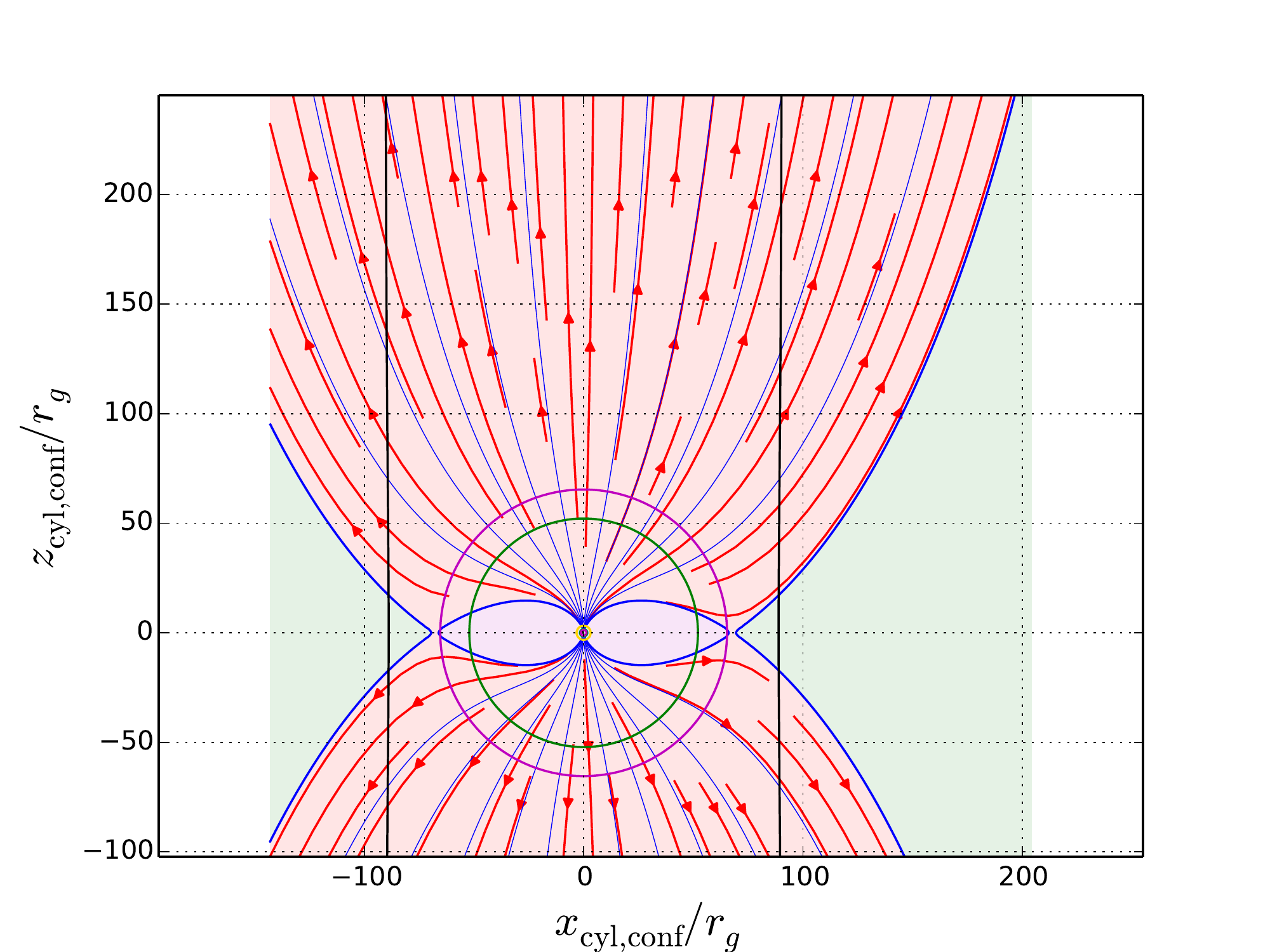}
\includegraphics[width=0.49\linewidth]{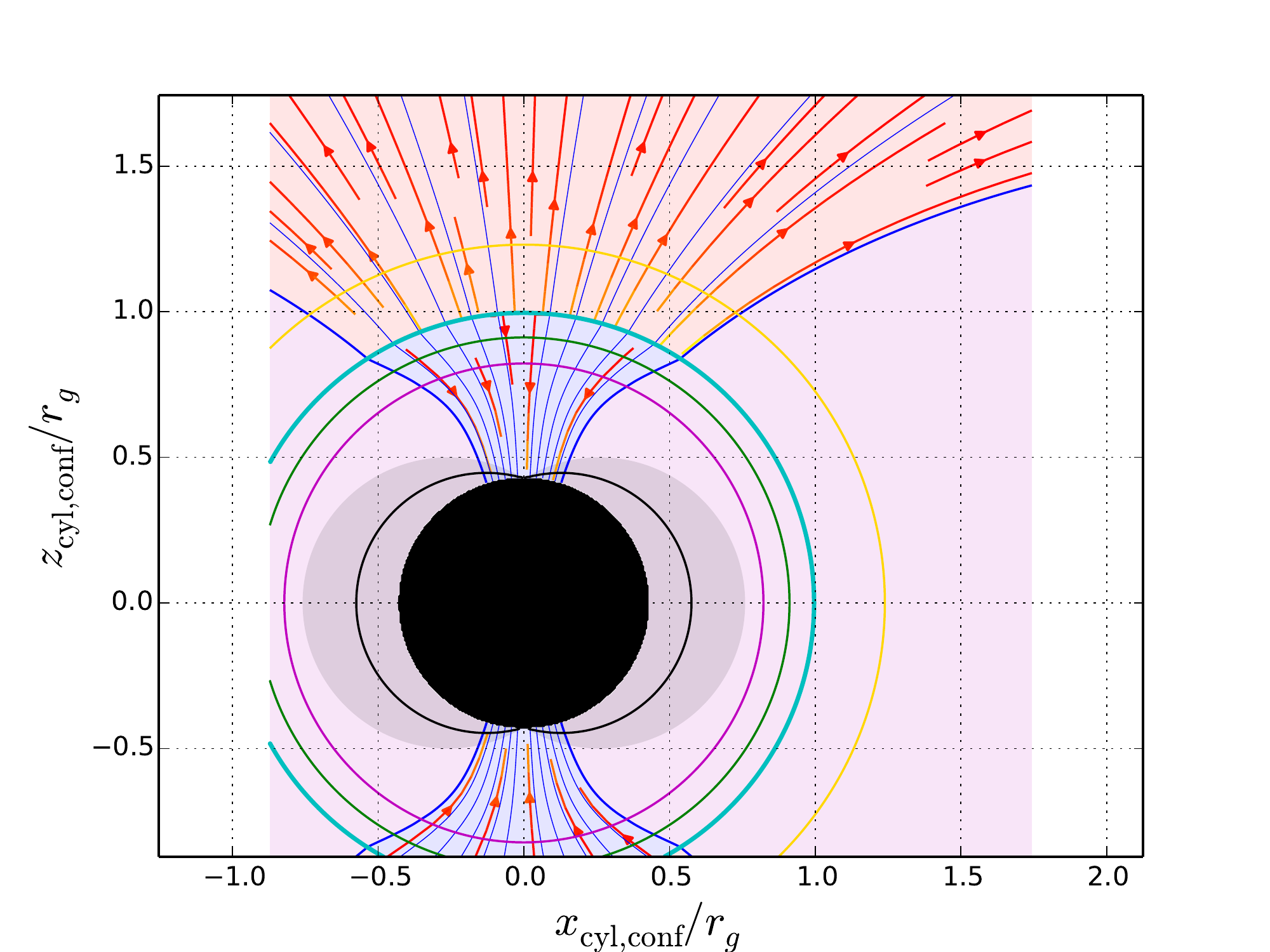}\includegraphics[width=0.49\linewidth]{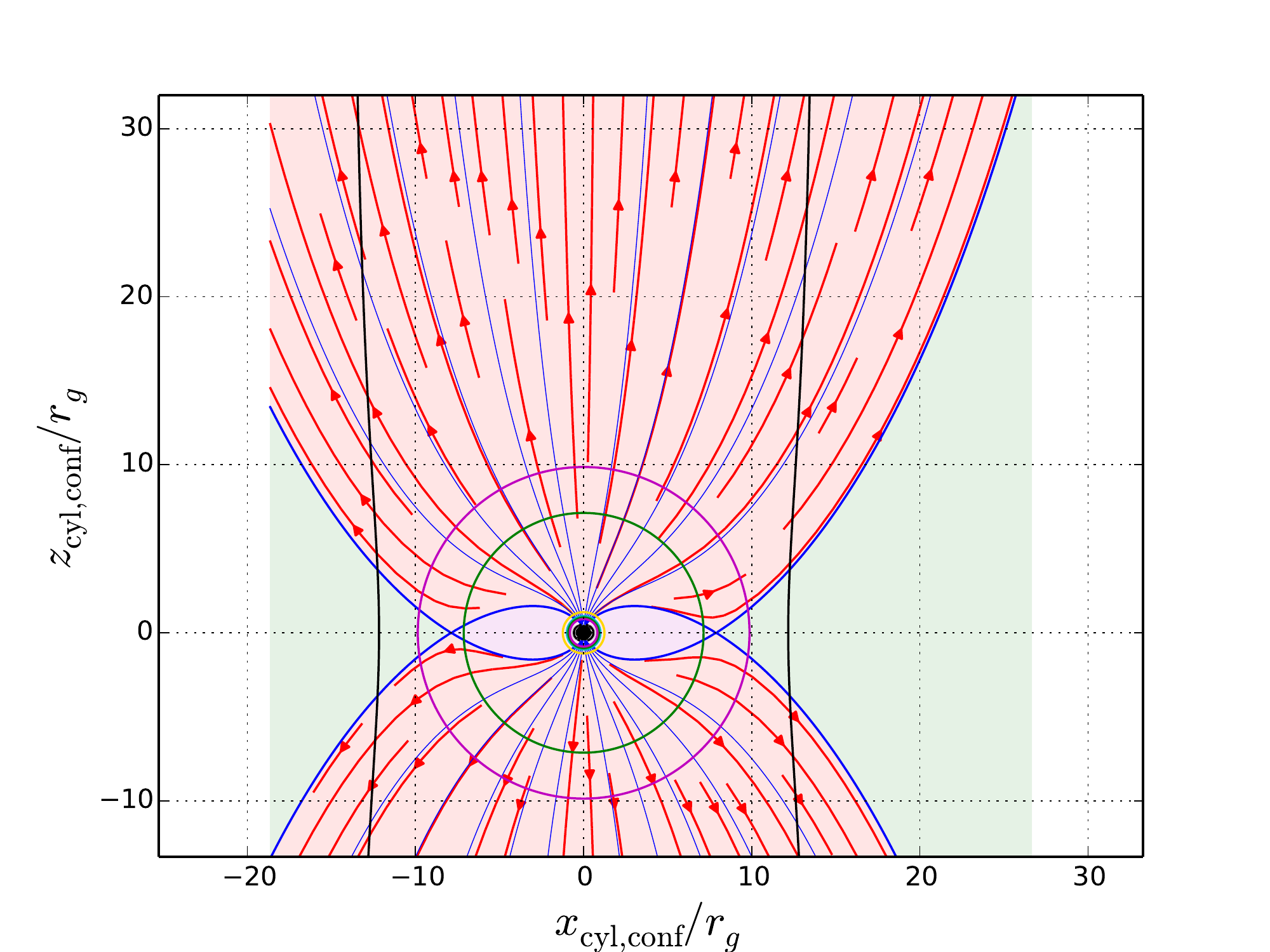}
\caption{ Poloidal field lines on two different scales for the three solutions (M1 top solution, M2 middle one and M3 bottom one). On the left we zoom inside the stagnation radius, and on the right the scale encloses the outer light cylinder radius. The red arrows represent the mass flux $\rho_0 h \gamma \mathbf{V}_{\rm p}$ and the thin blue lines, the poloidal magnetic field lines. The thick blue line marks the last open magnetic field line of the flow connected to the black hole. The yellow line represents the position of the corotation surface where $\Omega=\omega$. The cyan circle corresponds to the stagnation surface, the green ones correspond to the slow-magnetosonic surfaces and the magenta ones to the Alfv\'{e}n surfaces. The light cylinder surfaces are noted by the black solid lines. The magnetosphere is represented in purple, the open line flow region in light red. The region where we expect the disk wind is in green, the ergoregion in light gray and the inner horizon region in black. We used quasi-isotropic coordinates to plot this figure.}  
\label{Fig-Flow}
\end{figure*}

Fig.(\ref{Fig-Flow}) shows the field line geometry of the matching solutions. We plot for each solution a zooming view of the field lines close to the environment of the black hole for the inflow, and a larger view of the outflow, including the external light cylinder. Instead of using a simple Cartesian version of the Boyer-Lindquist coordinates, or what is called pseudo-Cartesian coordinates, we opted for the use of so-called quasi-isotropic coordinates. In \cite{Chantry_2020} we discussed in details their properties. This choice of coordinates presents two main advantages. First, it allows a conformal representation and therefore a correct representation of the angles. Thus the property of the orthogonal field line penetration into the horizon is correctly visualized. Secondly, these coordinates expand the representation in the black hole environment, which allows to show more details in this area. 

We plot in the left panel of Fig.(\ref{Fig-Flow}) the poloidal field geometry only for the open field lines linking the black hole horizon to infinity ($A<A_{\rm mag}$). First, our model (inflow and outflow) is deduced from an expansion to second order of the colatitude in Euler's equations. This explains why it is physically relevant close to the axis and in the region with small colatitudes. For a given second order expansion in colatitude we should quantify the deviation to the equilibrium of Euler's equations and normalize with the strongest volumic force to estimate the region of validity. This calculation is quite complicated, and we decide to examine the solutions inside of the region defined by the last open field line which contains the region of validity.

Second, we cannot use our model for the magnetospheric dead zone ($A>A_{\rm mag}$ and $r<r_{\rm mag}$). It would induce artificial source terms on the equatorial plane with $r_H<r<r_{\rm mag}$. For $r>r_{\rm mag}$ the source terms could be explained by the presence of the accretion disk but we prefer to avoid this region in our modelling.

As explained in \cite{1990ApJ...363..206T}, the stagnation surface and the injection are located between the two light cylinders. All the field lines are continuous but not $\mathcal{C}^1$ at the stagnation surface. On this surface there is a kink in the fieldlines related to the surface current density. Two different trends are observed for the the expansion factor of the streamlines $F$. In the matching solutions M1 and M2, the field lines are flaring more in the starting region of the inflow than at the base of the outflow $F_{\rm sta,in}\geq F_{\rm sta, out}$. The situation is the opposite for the M3 solution $F_{\rm sta,in}\leq F_{\rm sta, out}$. The corotation surface location appears below the stagnation surface for the solution M1 and above for the solutions M2 and M3. The larger is $\Omega$, the smaller is the mean radius of the corotation surface.

We also observe different sizes for the magnetosphere. The equatorial extent of the last open line in the outflow is significantly larger for the M2 solution. The size of the  magnetosphere of the M2 solution reaches on the equatorial plane an approximate value of $75r_g$. This is slightly larger than the Alfv\'{e}n radius of the outflow. While for the other solutions, the magnetosphere is located inside the Alfv\'{e}n surface and reaches approximately $40r_g$ for solution M1 and $8r_g$ for solution M3, on the equatorial plane. The open lines represent $14\%$ of the total magnetic flux passing through the black hole horizon for M1, less than $1\%$ for M2 and around $8,5\%$ for M3. 

    \subsection{Interface between inflow and outflow}

The interface corresponds to the region with poloidal velocities close to zero. This region, in our model, is the one where the flow is loaded via creation of pairs or any other mechanism. In fact, the matching of the inflow and the outflow solutions puts some constraints on the loading terms, as detailed in Subsec. (\ref{subsec-theoryloadingoninterface}). Once the matching of the two solutions is obtained the injection or loading terms can be calculated. First, we discuss the surface charge density and the toroidal current flux sources we obtain at this interface. Then we explain how we inject mass, angular momentum and energy. 

\subsubsection{Electromagnetic sources on the stagnation surface}\label{subsec-emagsource}
           
The electromagnetic sources on the stagnation surface are fully determined by $\sigma_e$ and $J_{\sigma}^{\hat{\theta}}$ because the ratio of $\sigma_e$ on  $J_{\sigma}^{\hat{\varphi}}$ is given by Eq.(\ref{Surfacic-DetermineByAngle}), 
\begin{equation}
    \dfrac{\sigma_e c}{J^{\hat{\varphi}}} = -\dfrac{\varpi \left(\Omega-\omega\right)}{hc} = -x  \, ,
\end{equation}
where $x$ is the dimensionless cylindrical radius.

\begin{figure}    
\centering
\includegraphics[width=\linewidth]{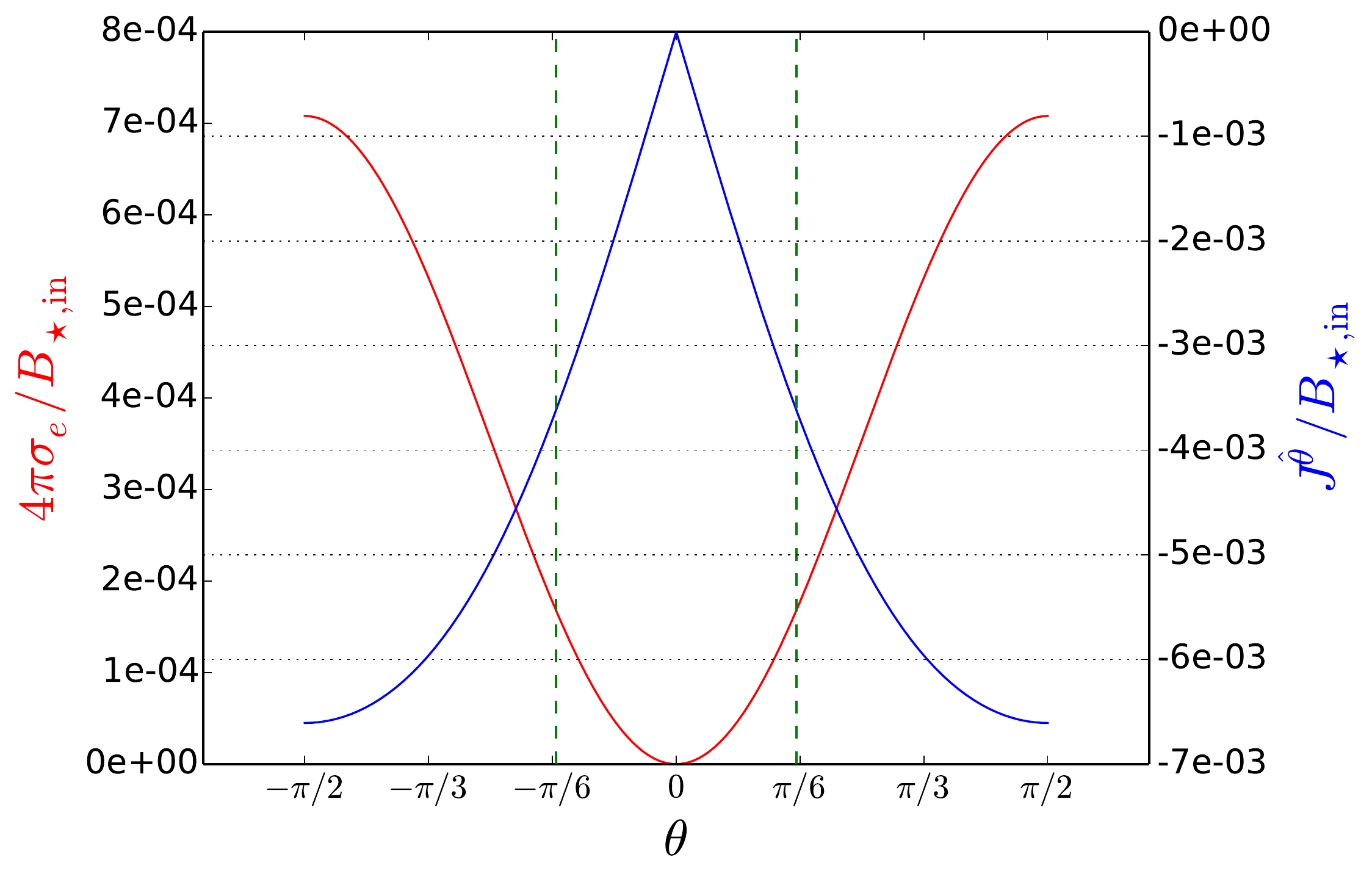}
\includegraphics[width=\linewidth]{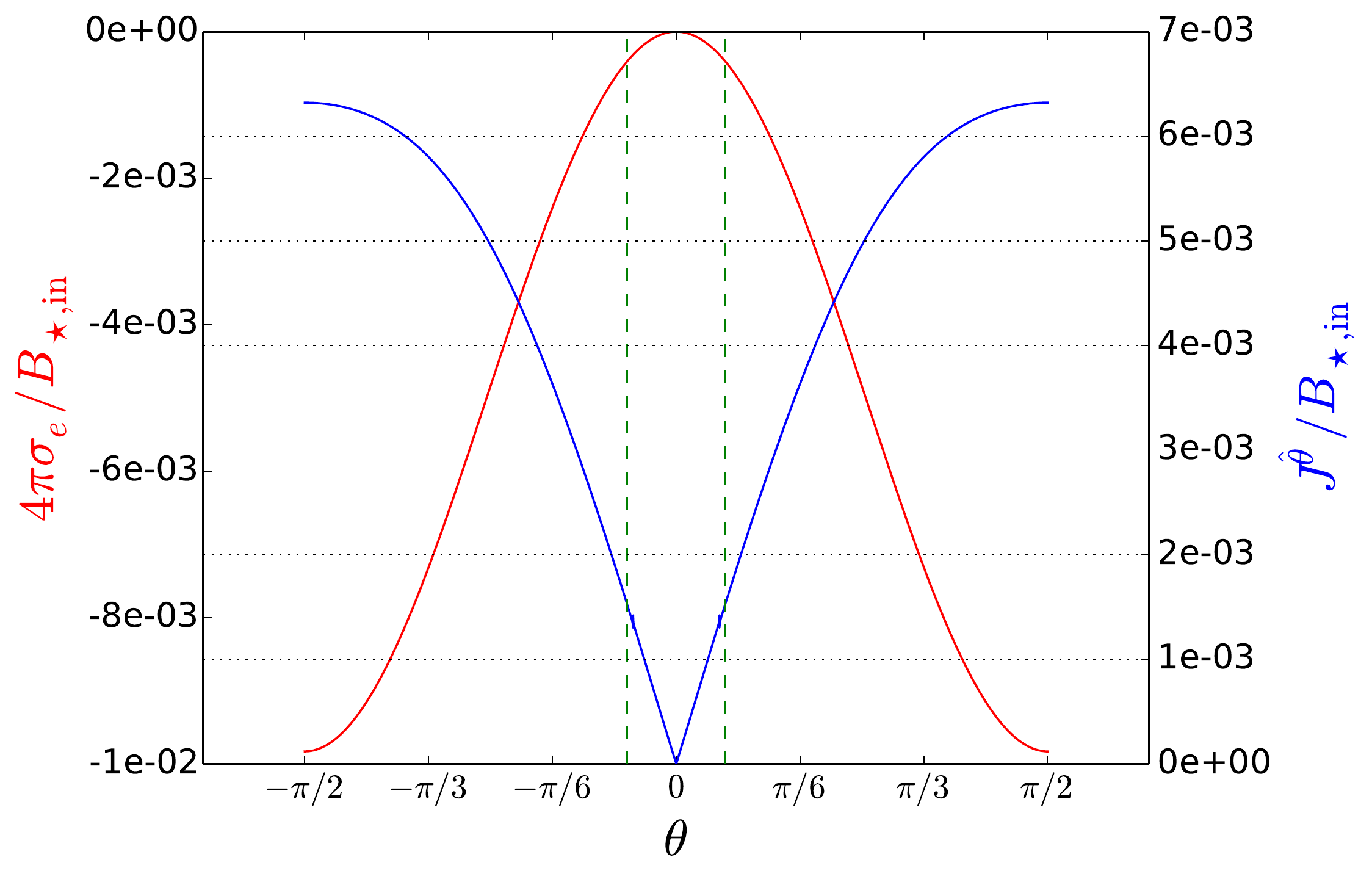}
\includegraphics[width=\linewidth]{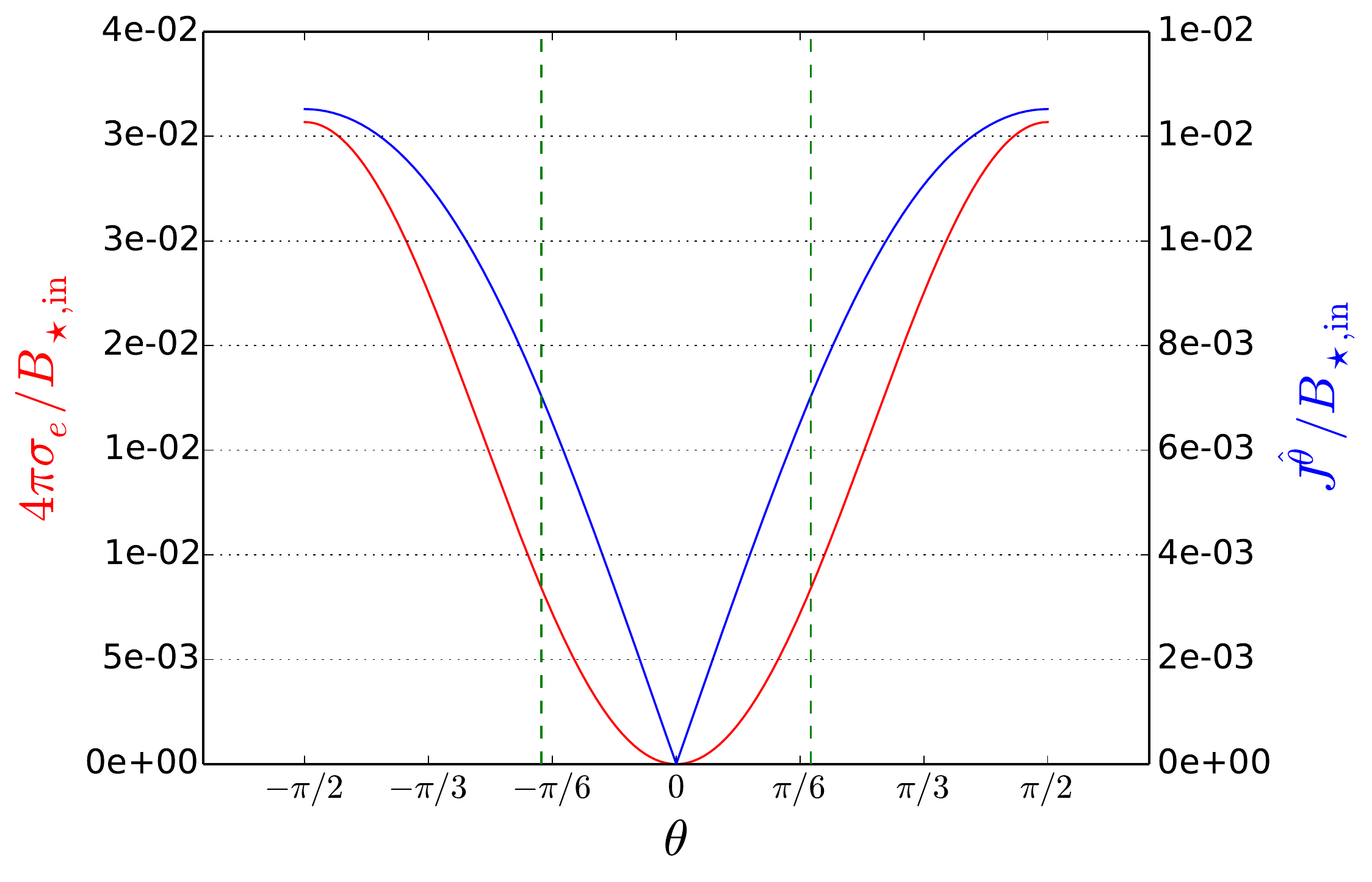}
\caption{ Electromagnetic surface sources as a function of the colatitude for the different solutions. The top solution is M1, the middle one M2 and  the bottom one is M3. In red, the surface density of charge is plotted as a function of the colatitude and, in blue, we plot the dimensionless toroidal surface current. The vertical dotted green lines represent the colatitude on the stagnation surface of the last open poloidal magnetic field lines.
\label{Fig-SurfacicEletromagSource}} 
\end{figure}

First, let us note that the sign of $\sigma_e$ is the same as the one of $-(F_{\rm sta,out}-F_{\rm sta, in})(\Omega-\omega)_{\rm sta}$. We plot $\sigma_e$ and $J^{\hat{\theta}}$ on Fig.(\ref{Fig-SurfacicEletromagSource}). For the M1 solution, the corotation surface is located below the stagnation surface where $\Omega>\omega$. We observe that $F_{\rm sta,out}<F_{\rm sta, in}$ and the flaring of the poloidal field lines increases where they cross the stagnation surface (see Fig.\ref{Fig-Flow}). These two facts explain the positive sign of $\sigma_e$ and the negative one of $J^{\hat{\varphi}}$. For M2, the corotation surface is above the stagnation surface. The same increase of the magnetic field line flaring occurs at the crossing of the stagnation surface implying $\sigma_e<0$. For the last solution M3, the corotation surface is also above the stagnation surface but the flaring of the magnetic poloidal lines decreases across the stagnation surface implying $F_{\rm sta,out}>F_{\rm sta, in}$ and $\sigma_e>0$.

As it can be seen on Fig.(\ref{Fig-SurfacicEletromagSource}), $J^{\hat{\theta}}_\sigma$ is negative for M1 and positive for M2 et M3. The sign of $J^{\hat{\theta}}_\sigma$ is determined by the direction of the shift at the stagnation surface of the current line with $I={\rm cst}$ (see Fig.(\ref{CurrentIntensity}) where $J^{\hat{\theta}}_\sigma<0$). A positive surface current $J^{\hat{\theta}}_\sigma$ implies a decreasing of the current $I$ across the stagnation surface (see Eq.(\ref{eq-increascurrent})) and then a increasing of the Poynting flux $\propto -I\Omega$. 
 
        \subsubsection{Loading terms }

The loading terms bring mass, angular momentum and energy to the MHD fields and to the black hole. Here these quantities are the result of the minimal matching conditions given by Eq.(\ref{eq-mmc}). 

\begin{table*}
\centering
\begin{tabular}{*{6}{c}}
   \hline
    {\rm Solution} & $B_\star^{\rm in}$ & $B_H$ & ${\dot{M}_{\rm Inj}}^\star$  & ${\dot{J}_{\rm Inj}}^\star$ & ${\dot{E}_{\rm Inj}}^\star$  \\
    &($\unit[]{cm^{-1/2}.g^{1/2}.s^{-1}}$)&($\unit[]{cm^{-1/2}.g^{1/2}.s^{-1}}$)& ($\unit[]{g.s^{-1}}$) & ($\unit[]{g.cm^2.s^{-1}}$) & ($\unit[]{erg.s^{-1}}$)\\
    \hline\hline
    M1 &$8,7\times 10^1$&$5,84\times 10^2$& $1.21\times 10^{23}$ & $3.56\times 10^{48}$ & $1.09\times 10^{44}$\\
    M2 &$8,2\times 10^{1}$&$4.7\times 10^{2}$& $1,1\times 10^{23}$ & $3.2\times 10^{48}$ & $9.8\times 10^{43}$\\
    M3 &$1,3\times 10^2$&$3,19\times10^2$& $2.62\times 10^{23}$ & $7.7\times 10^{48}$ & $2.36\times 10^{44}$\\
    \hline
\end{tabular}
\caption{In the first and second columns we give the estimated values of the magnetic field on the inflow Alfv\'{e}nic point and on the black hole horizon, respectively. In the three last columns we plot the constant values for the mass, angular momentum, and energy per infinitesimal intervals of time and of dimensionless magnetic flux for a black hole mass equal to the one of M87. Each line corresponds to one of the three inflow/outflow solutions of the meridional self-similar model. We use $B(20 r_g)\approx 1{\rm G}$ to calculate $B_\star^{\rm out}$ for the solution M1 since the stagnation radius is around $5 r_g$ and $B(5 r_g)\approx 4.9{\rm G}$ for the two other solutions, M2 and M3. \label{Tab-TypicalParam}}
\end{table*}

 The injection terms are proportional to the scaling factors of Eq.(\ref{eq-typicalscale}) determined by the value of $B_{\star}^{\rm in}$.
We already gave an estimation of the magnetic field along the polar axis at the Alfv\'{e}n radius for the M87 black hole (see Tab.(\ref{Tab-TypicalParam})). We found $B_{\star,{\rm out}}\approx G^2_{\rm out}(20r_g){\rm G}$ for M1 and $B_{\star,{\rm out}}\approx 4.9 G^2_{\rm out}(5r_g){\rm G}$ for M2 and M3. From this value, we can estimate, for the inflow solution, the alfv\'{e}nic magnetic field on the axis $B_{\star,{\rm in}}$ and the magnetic field on the black hole horizon $B_{H}$. For the three solutions we get a magnetic field on the black hole horizon between $300$ and $600$ Gauss.
Eq.(\ref{eq-typicalscale}) allows us to estimate, for the inflow solution, the constant values for the mass, energy and angular momentum injected per unit time and dimensionless magnetic flux, $\alpha$. These quantities have been calculated for the three global inflow/outflow solutions as shown in Tab.(\ref{Tab-TypicalParam}). 

Now, we can compare the physical quantities  for the three matching solutions with the ones obtained by other works. They depend on the considered phenomena and injection models.
For example, let us evaluate the amount of mass that can be injected via pair creation from hard photons emitted by the accretion disk. Following \cite{2017PhRvD..96l3006L}, the injection rate per unit volume is estimated as $\sigma_{\gamma\gamma}n_{\gamma}^2 c$, where $n_\gamma$ is the density of hard photons with an energy $\epsilon_\gamma$ ($\epsilon_\gamma>\unit[1]{MeV}\approx 2 m_e c^2$). We use the Thomson cross section $\sigma_{\gamma\gamma}\approx\unit[6,6\times 10^{-25}]{cm^2}$ for estimating the cross section of pair production. Now let introduce the dimensionless radius of the hottest part of the disk $R_\gamma=r_\gamma/r_g$, the dimensionless mass $m={M}/{M_{\odot}}$, and the dimensionless luminosity $\ell_\gamma=L_{\gamma} / L_{\rm Edd}$, where $L_\gamma$ is the luminosity of hard photons and $L_{\rm Edd}$ the Eddington luminosity. Then the luminosity, coming essentially from the disk, is related to the photon density $L_\gamma = 4\pi r_\gamma^2 n_\gamma c \epsilon_\gamma$, which leads to, 
\begin{equation}
    n_\gamma \approx \unit[10^{22}\dfrac{\ell_\gamma}{m R_\gamma^2}]{cm^{-3}}\,.
\end{equation}
The mass injection rate can be written as, 
\begin{equation}\label{eq-estimate}
    \dot{M}_{\rm Inj} \approx \unit[1,6\times 10^{20}\dfrac{\ell_\gamma^2 m}{R_\gamma^4}]{g\, s^{-1}} = 2.6\times 10^{-6} \unit[\dfrac{\ell_\gamma^2 m}{R_\gamma^4}]{M_\odot \, yr^{-1}}\,.
\end{equation}
Using values of luminosity mentioned in \cite{10.1093/mnras/stw166} for M87, we get $\ell_\gamma \sim 10^{-7}-10^{-4}$. The \cite{2021ApJ...910L..13E} have shown, using their library of disc models, that the inner radius of the disc lays within $\approx 10-20 r_g$. Taking $m \approx 6\times 10^9$ and $R_\gamma \sim 10-10^2$ we get $\dot{M}_{\rm Inj} = \unit[10^8 - 10^{20}]{g\,s^{-1}}$. The factor $R_\gamma^{-4}$ makes this estimate extremely sensitive to the value of $R_\gamma$. Calculations based on \cite{2015ApJ...801...56P}, with different matching conditions, lead to similar conclusions. Nevertheless, many works (\citealt{Levinson2011VARIABLETE}, \citealt{2016ApJ...833..142H}) show that this injection does not allow to reach the Goldreich-Julian density necessary for the screening of the transverse electric field. In this case, spark gap may form \citep{2017PhRvD..96l3006L} along the magnetic field. The electric acceleration combined with Compton and Inverse Compton processes allow an additional source of pair production. This mechanism leads to mass injection in the lower range, up to $10^{11}-10^{12}{\rm g\,s^{-1}}$. Indeed, the pair production in the gap cannot explain the injection required in our model. It is consistent with the infinite conduction assumption, which leads to a charge density equals to the Goldreich-Julian one everywhere. 

However, recent works with particle-in-cell simulations (e.g. \citealt{refId0}) explore the dynamics of the formation of such gaps and the role of magnetic reconnection allowing to visualise the location of pair formation. Their conclusions tend to suggest that gaps are intermittent. Thus, we use more recent radiative GRMHD simulations to give a more precise estimation of the mass injection rate \citep{10.1093/mnras/stab2462}. Fig.(10) of this publication gives an average in time of the pair production rate as a function of the radial distance in a region close of the axis. Assuming a spherical symmetry and the value mentioned in this plot for the MAD W18 disk model, we can estimate the total injected mass. We obtain a total amount of $\unit[2.5 \left(\dfrac{M_H}{10^9 M_\odot}\right)^3\times 10^{14} ]{g\, s^{-1}}$. For M87, it gives a total injected mass $\approx \unit[8\times 10^{16}]{g
\, s^{-1}}$. This amount is sufficient to screen the  electric field and avoid the formation of the spark gap along the axis region close to the black hole.

Typical values of a few solar masses per year are usually inferred for the accretion rate of hadrons from the disc. This is much larger than the total rate of pairs accreted in the inflow, which we have found to be a few $10^{-6}$ M$_\odot/$yr (from Eq.(\ref{eq-estimate})). Thus the inflow of pairs has a negligible contribution to the increase of the mass of the central black hole. Conversely, its contribution in removing angular momentum may be significant, especially if the disc wind is efficient in extracting angular momentum from the disc.

\cite{2013PhRvD..88h4046G} explored the injected critical mass for which no extraction occurs. For a cold flow and a magnetic flux $\approx \unit[10^{27}]{G\,cm^2} $, which crosses the black hole, the mass flux limit find by the authors depends from the value of $a$ and $\theta$ but is around $\unit[10^{25}-10^{30}]{g\,s^{-1}}$. 

\begin{table*}
\centering
\begin{tabular}{*{7}{c}}
   \hline
    {\rm Solution} & $\alpha_{\rm Mag}$ & $\left.\frac{1}{\dot{M}^\star_{\rm Inj}}\frac{{\rm d}^2 M_{\rm Inj}}{{\rm d}t {\rm d}\alpha}\right|_{\rm Min}$  & $\left.\frac{1}{\dot{M}^\star_{\rm Inj}}\frac{{\rm d}^2 M_{\rm Inj}}{{\rm d}t {\rm d}\alpha}\right|_{\rm Max}$ & $\langle\frac{1}{\dot{M}^\star_{\rm Inj}}\frac{{\rm d}^2 M_{\rm Inj}}{{\rm d}t {\rm d}\alpha}\rangle$ &  $\dot{M}_{\rm Inj}^{\rm tot}$ & $\dot{M}_{\rm Inj,in}^{\rm tot}$ \\
    \hline\hline
    M1 & $0.95$ & $8.2\times 10^{-3}$ & $1.12\times 10^{-2}$ & $9.7\times 10^{-3}$ & $ \unit[1.1 \times 10^{21}]{g.s^{-1}}$ & $ \unit[2.4 \times 10^{20}]{g.s^{-1}}$\\
    M2 & $0.99$ & $8.63\times 10^{-5}$ & $8.87\times 10^{-5}$ & $8.8\times 10^{-5}$ & $ \unit[9.5 \times 10^{18}]{g.s^{-1}}$  & $ \unit[8,95 \times 10^{18}]{g.s^{-1}}$ \\
    M3 & $0.98$ & $8.25\times 10^{-4}$ & $1.12\times 10^{-3}$ & $9.7\times 10^{-4}$ & $ \unit[1.3 \times 10^{20}]{g.s^{-1}}$  & $ \unit[4,3 \times 10^{19}]{g.s^{-1}}$ \\
    \hline
\end{tabular}
\caption{The first column indicates the dimensionless magnetic flux for the last open magnetic field line, which marks the limit of the magnetosphere in the outflow. In the second, third and fourth columns, we give the minimum, the maximum and the mean value of the mass injected per unit time and unit dimensionless magnetic flux, at the stagnation surface, for $A<A_{\rm mag}$, respectively. The fifth and sixth columns give the total mass injected and the mass injected in the inflow per unit time at the stagnation surface for $A<A_{\rm mag}$, respectively. Each line corresponds to one of the three inflow/outflow solutions. \label{tab-injmass}}
\end{table*} 

\begin{figure}       
\centering
\includegraphics[width=\linewidth]{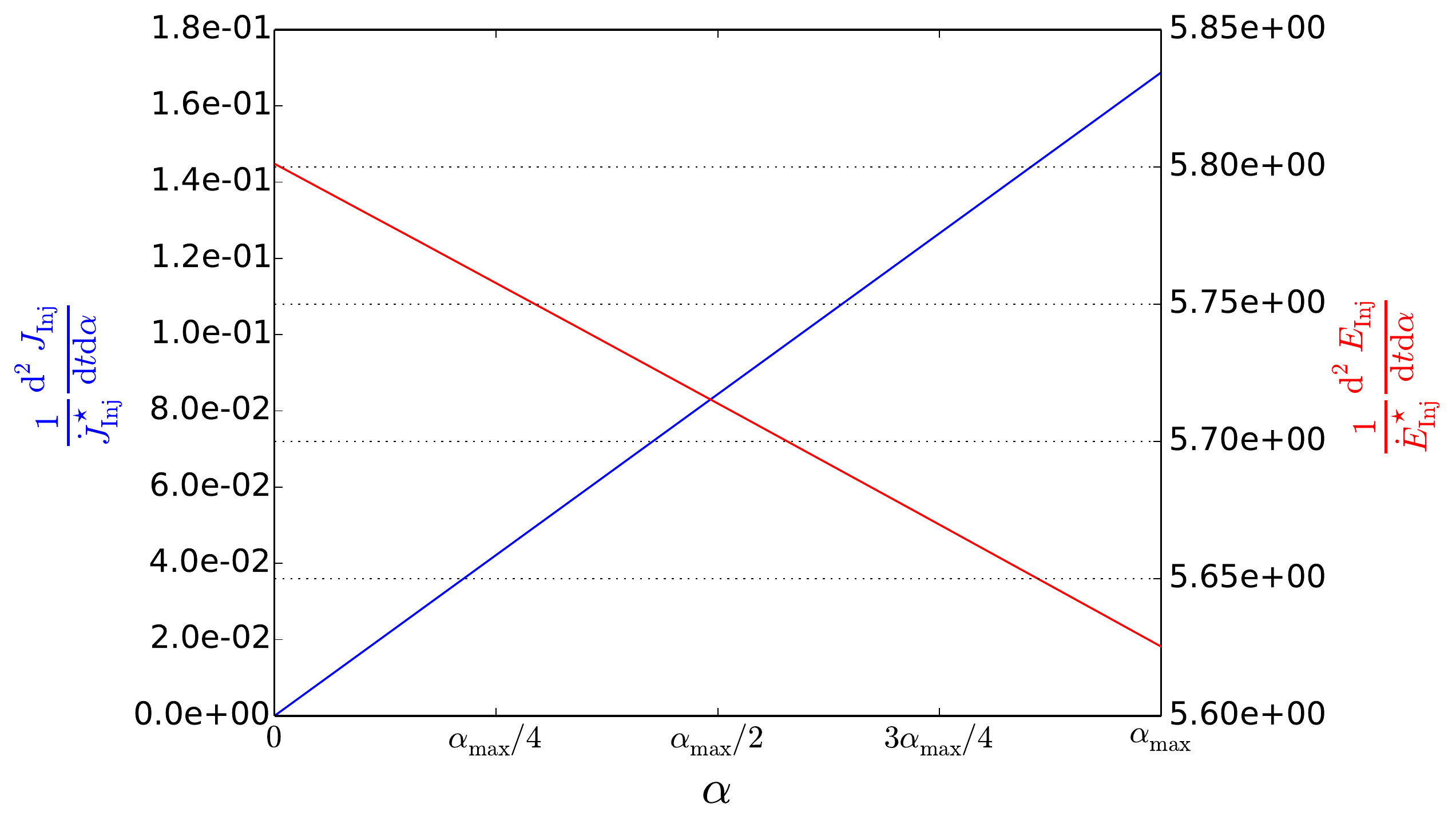}
\includegraphics[width=\linewidth]{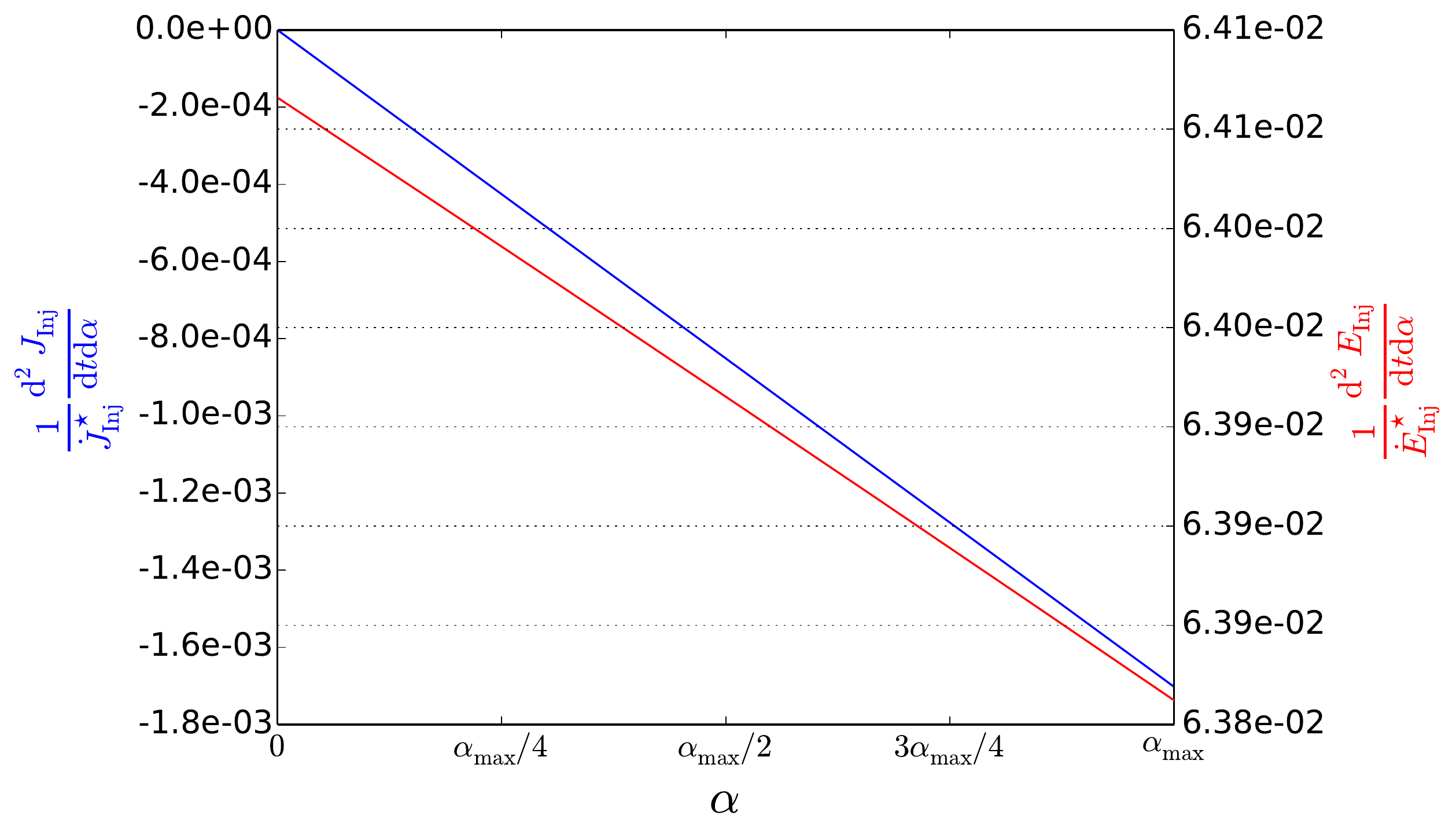}
\includegraphics[width=\linewidth]{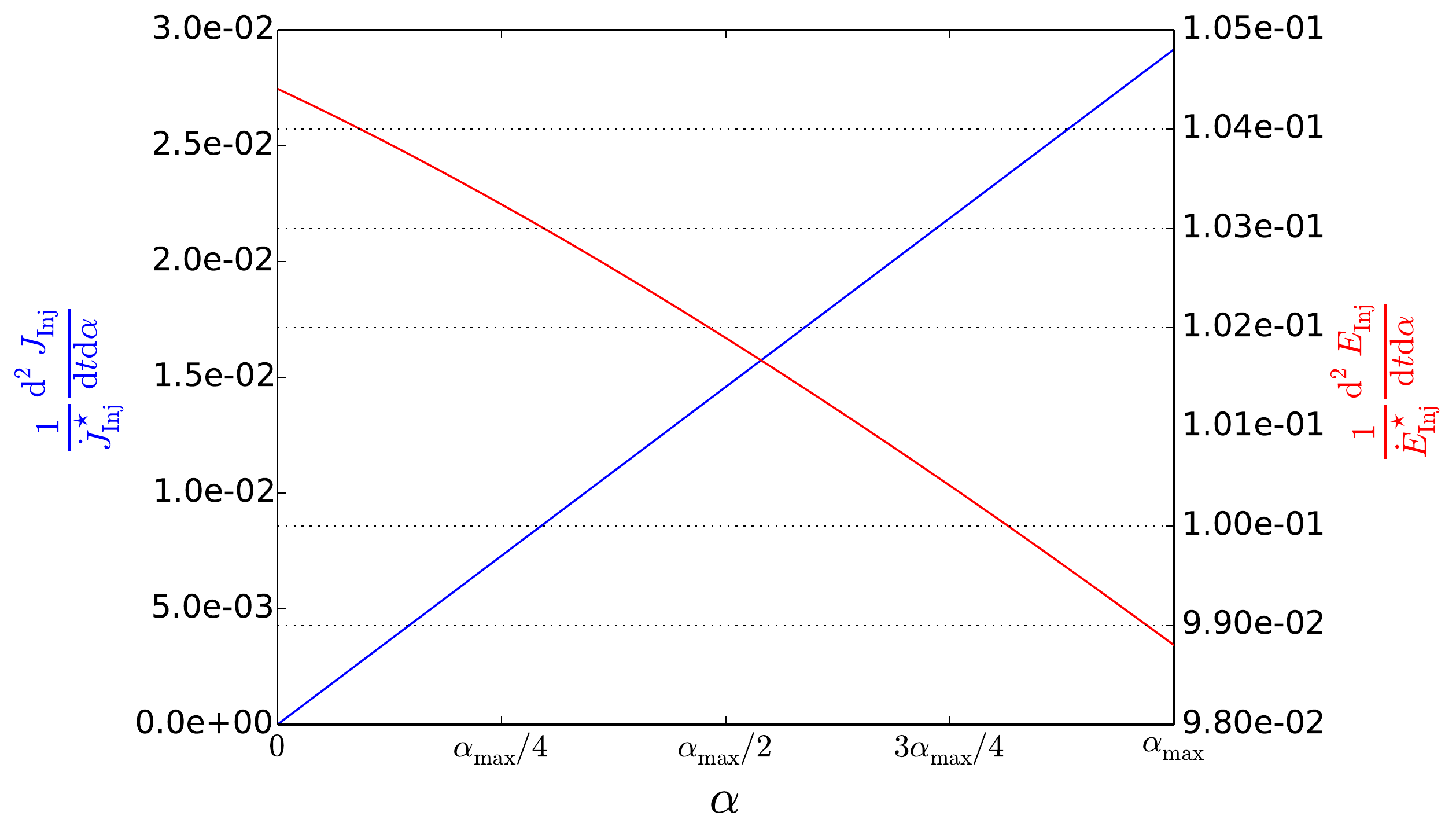}
\caption{Injected angular momentum and energy at the interface per unit time and dimensionless magnetic flux as a function of the magnetic flux for the different solutions. On top, we plot M1 ($\alpha_{\rm Mag}\approx0.95$), in the middle M2 ($\alpha_{\rm Mag}\approx0.99$) and at the bottom M3 ($\alpha_{\rm Mag}\approx0.98$). The angular momentum is plotted in blue. It is divided by its scaling value $J^\star_{\rm Inj}$. The energy is plotted in red and divided by its scaling value $E^\star_{\rm Inj}$.
\label{Fig-AngularEnergyInjection} }
\end{figure}

\begin{figure}       
\centering
\includegraphics[width=\linewidth]{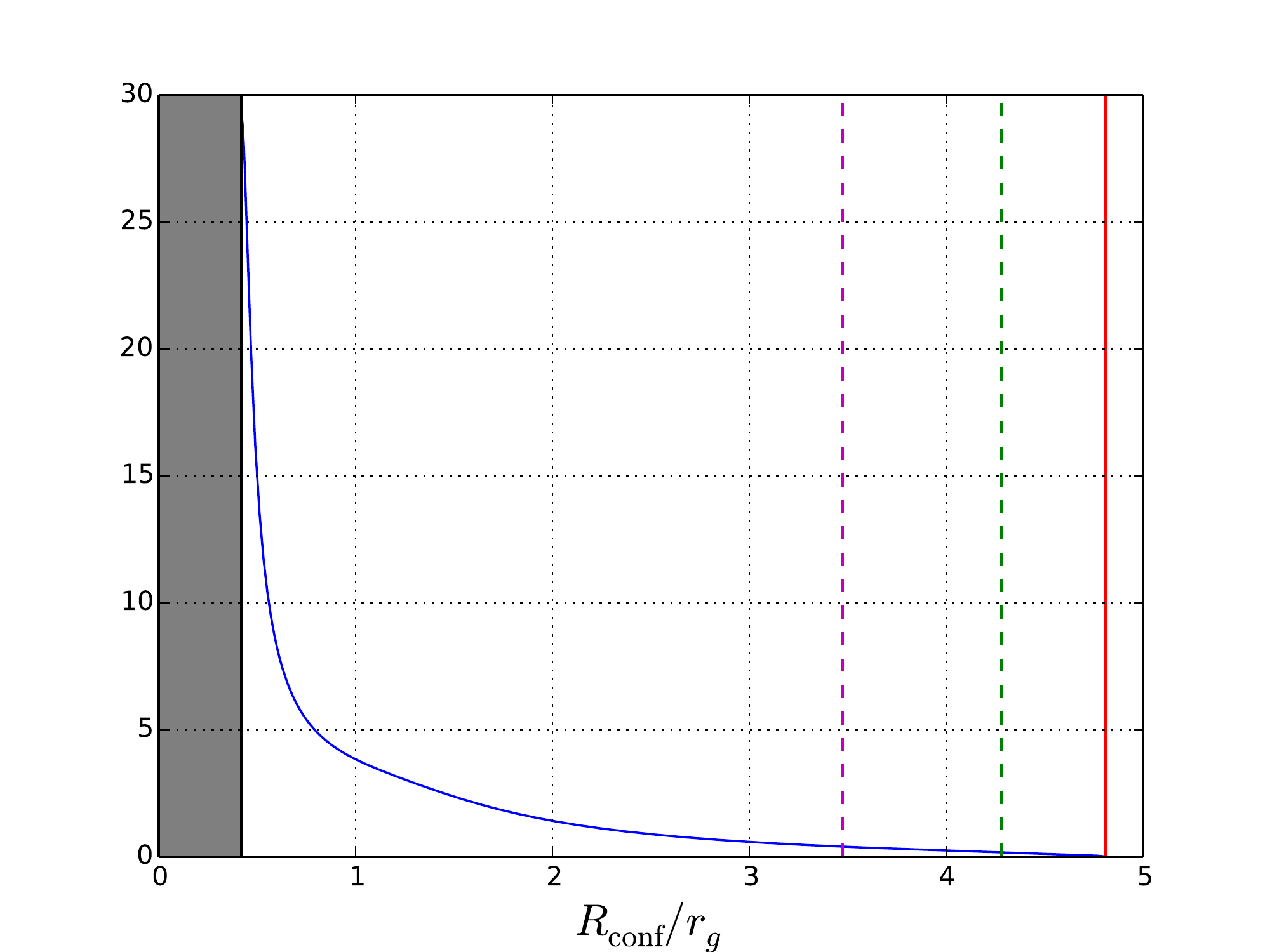}
\includegraphics[width=\linewidth]{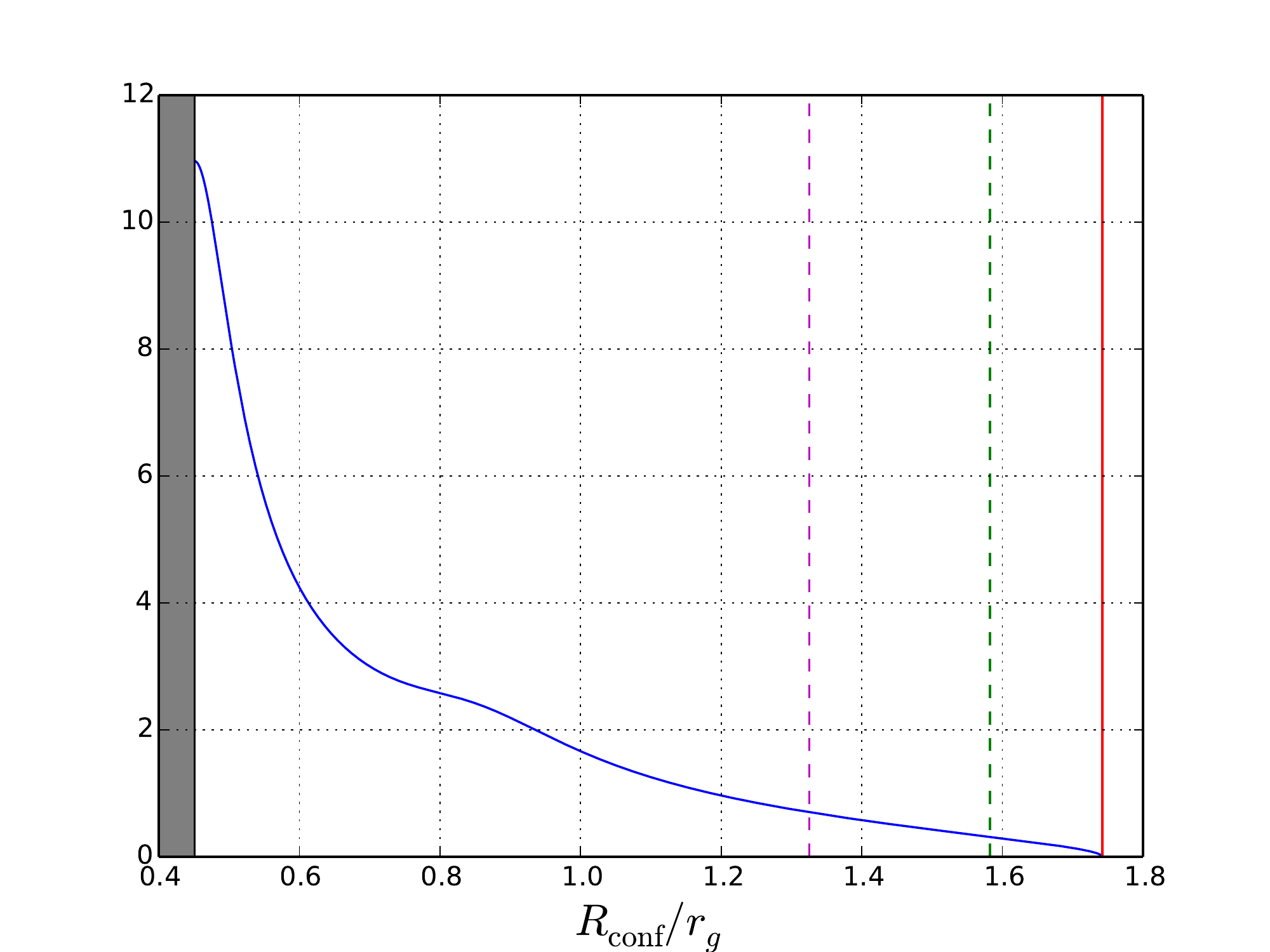}
\includegraphics[width=\linewidth]{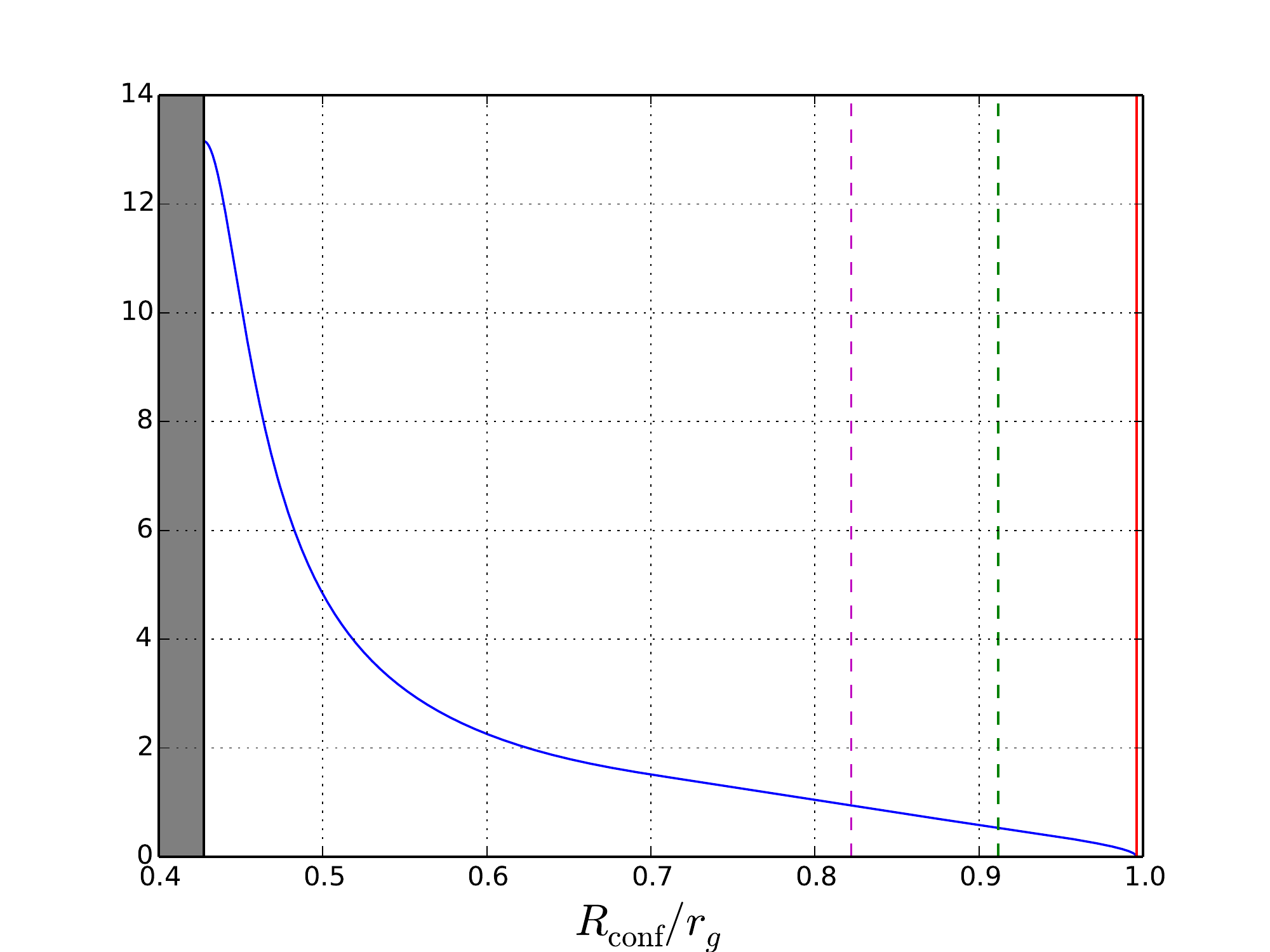}
\caption{ Plot of the celerity $\beta\gamma$ of the inflow plasma along the axis as a function of the quasi-isotropic radial distance for the different solutions, M1 on top, M2 in the middle and M3, bottom. The vertical lines correspond to the different critical surfaces, namely, the Alfv\'{e}n one in dotted magenta,the slow-magnetosonic one in dotted green and the stagnation surface in red. The grey shaded area corresponds to the inner part of the black hole horizon. 
\label{Fig-Celerity} }
\end{figure}

In Tab.(\ref{tab-injmass}), we put the minimum, the maximum and the mean value of the injected mass per unit time and per unit dimensionless magnetic flux. After integration for $A<A_{\rm mag}$, we obtained the total injected mass and the injected mass in the inflow per unit time. These values are put in the Tab.(\ref{tab-injmass}). In the first column we indicate the dimensionless magnetic flux for the last open magnetic field line of the outflow. Note that the values of $\alpha_{\rm Mag}$ are similar for the three global solutions, despite the large variation of the sizes of the magnetosphere on the equatorial axis.

In Tab.(\ref{tab-injmass}) the total injected mass is, for M2 and M3 in the upper range of the estimation based on the EHT emission ring size. We get a larger value only for solution M1. Globally, we obtain an injected mass which is two to four orders of magnitude higher than the one obtained by \cite{10.1093/mnras/stab2462}. For the M1 and M3 solutions, most of the injected mass per unit time is flowing outward, quantitatively 80\% for M1 and 66\% for M3. Conversely, for the M2 solution, only 6\% of the total injected mass per unit time is flowing outward.

In Fig.(\ref{Fig-AngularEnergyInjection}) we plot in blue the injected angular momentum rate per unit dimensionless magnetic flux. Its sign is positive for solutions M1 and M3 and negative for solution M2. The total amount of angular momentum rate per unit time is equal to $(10^{-3}-10^{-2}) J^\star_{\rm Inj}$. We also plot the injected power per unit magnetic flux. This quantity decreases slowly with the magnetic flux due to the negative value of $e_1$ for the inflow solutions. The order of magnitude of the total amount of injected energy is $5.4 E^\star_{\rm Inj}$ for M1, $6.3\times 10^{-2}E^\star_{\rm Inj}$ for M2 and $10^{-1} E^\star_{\rm Inj}$ for M3.

	\subsection{Kinetics and dynamics of the inflow}	
	
Fig.(\ref{Fig-Celerity}) shows the fluid celerity $\gamma\beta$ measured by the ZAMO observer along the polar axis. The Lorentz factor reaches relatively high values ($10-25$) as expected. In fact the parameters could be tuned in order to have $\gamma\longrightarrow +\infty$ and to smooth the pressure function $\Pi$ behaviour close to the black hole horizon.

\begin{figure}       
\centering
\includegraphics[width=\linewidth]{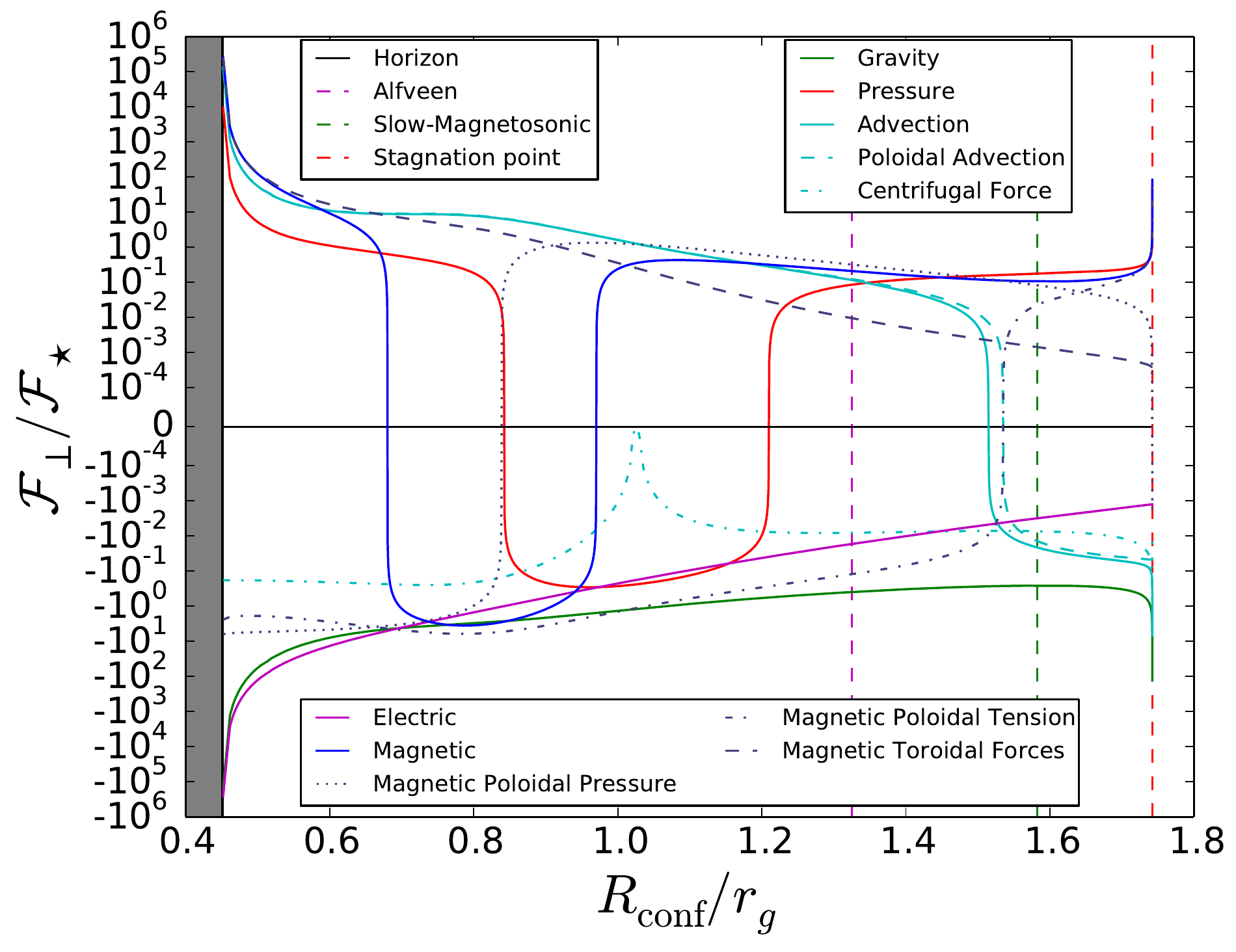}
\includegraphics[width=\linewidth]{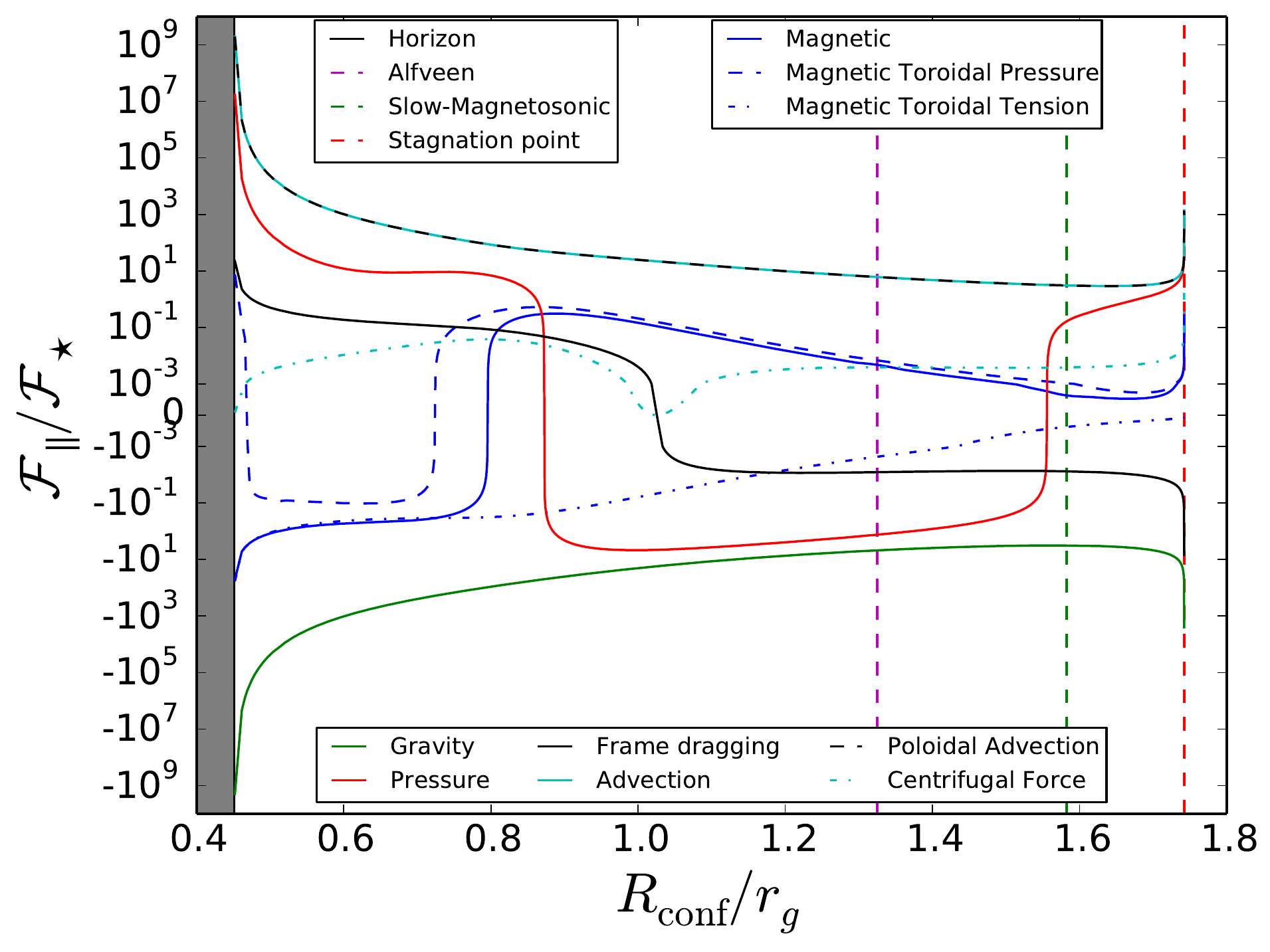}
\caption{ For the inflow of solution M2, we plot the transverse forces as function of the quasi-isotropic radial distance, on top, with a positive value for forces directed inward the flux tube. We plot the longitudinal forces, bottom, with a positive value for decelerating forces. The grey shaded area corresponds to the inner part of the black hole horizon. \label{Fig-I2-Forces} }
\end{figure}

Considering the forces acting on the inflow, the situation is quite similar for the different solutions. We plot in Fig.(\ref{Fig-I2-Forces}) the  transverse and longitudinal forces for a field line close to the axis for solution M2.

The upper part of Fig.(\ref{Fig-I2-Forces}) represents the transverse forces for the M2 inflow solution. Positive values correspond to the collimating forces. Near the stagnation surface the gravitational force is the main decollimating force, which is in quasi-equilibrium with the sum of the magnetic forces (i.e. mainly the magnetic poloidal pressure and the magnetic tension) and the pressure gradient. Near the black hole the gravitational and electrical forces (decollimating) are in equilibrium with the magnetic forces, i.e. essentially the magnetic poloidal tension, plus the poloidal advection force. 

The bottom part of Fig.(\ref{Fig-I2-Forces}) corresponds to the longitudinal forces. Negative (positive) values means that the forces are directed towards (outwards) the black hole center. The main force driving the flow is gravity. The pressure also plays an important role in the part of the flow where the acceleration decreases the pressure, i.e. by a cavitation effect. The main opposite force is due to the fluid inertia, the advection term, plus the pressure in areas near the stagnation layer and the black hole horizon.

	\subsection{Exchange between the black hole and the MHD fields}
	
\begin{figure}       
\centering
\includegraphics[width=\linewidth]{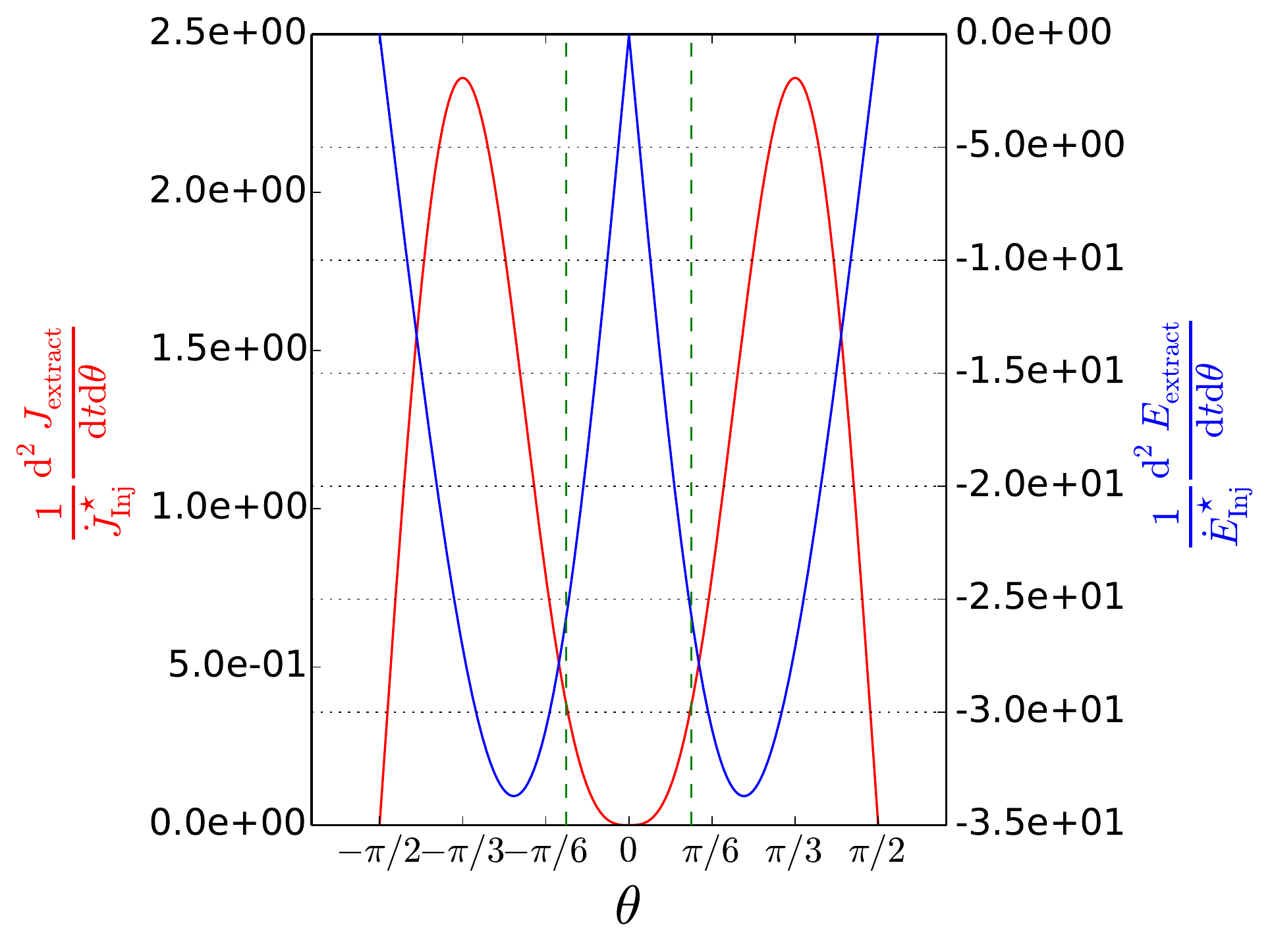}
\includegraphics[width=\linewidth]{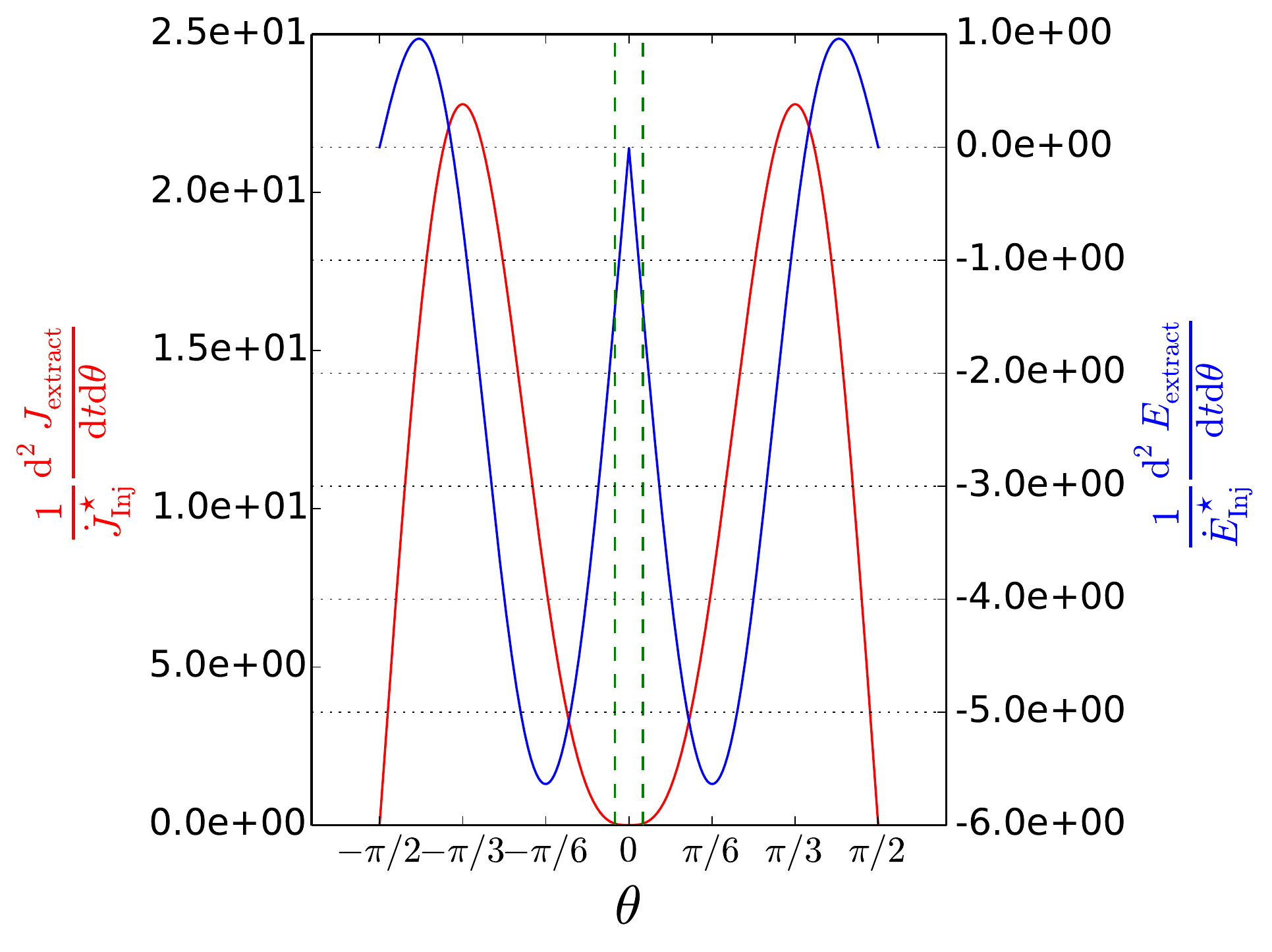}
\includegraphics[width=\linewidth]{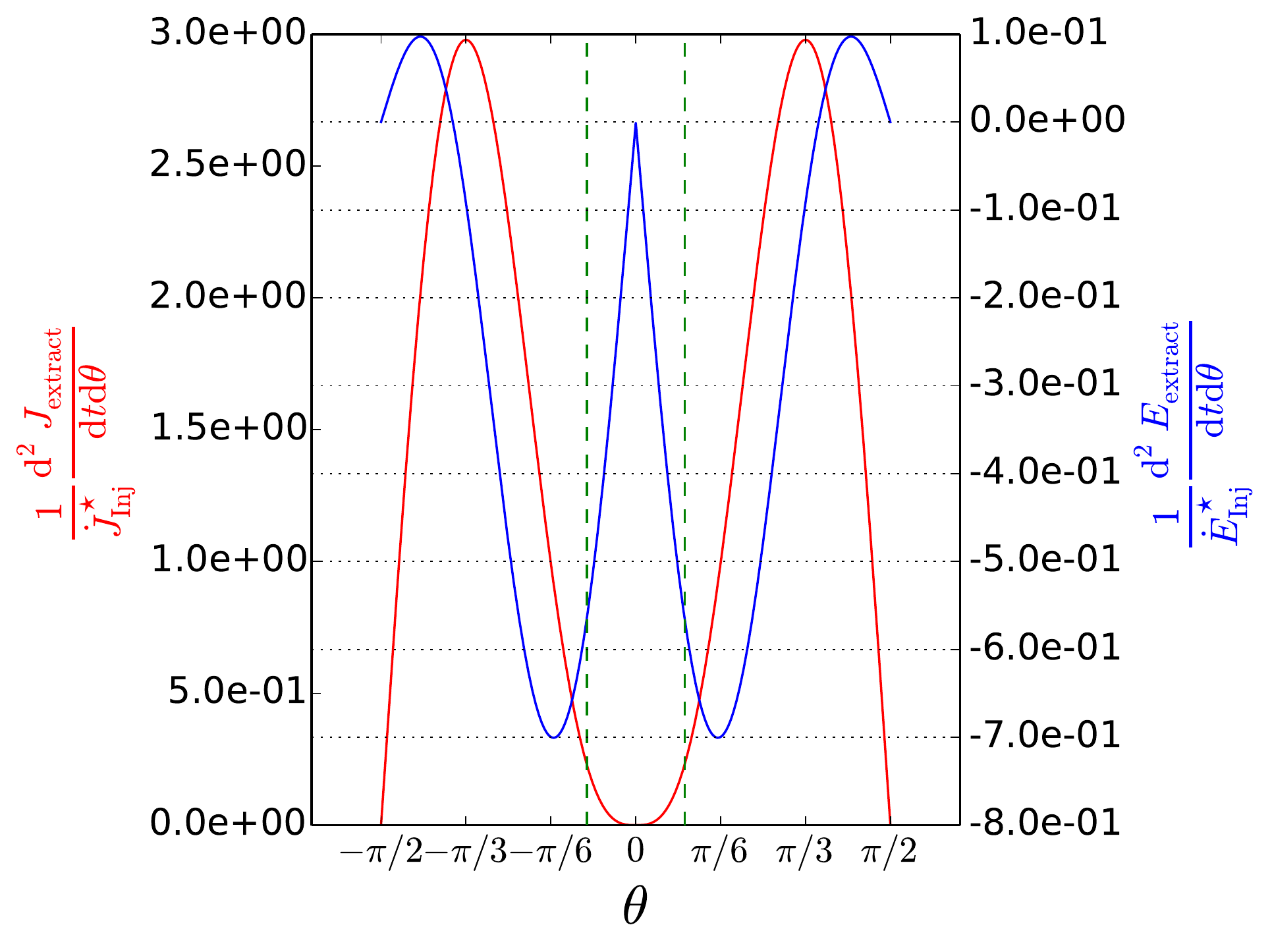}
\caption{Global balance on the black hole horizon of the Noether energy and angular momentum for the three solutions, M1 on top, M2 middle and M3 bottom. In red, the dimensionless extracted Noether angular momentum per unit colatitude and time. In blue, the dimensionless extracted Noether energy per infinite intervals of colatitude and time. The dotted green lines indicate the colatitude of the last magnetospheric field line at the horizon radius. \label{Fig-InflowEnergyAngularMomentumFlux}}
\end{figure}

We plot on Fig.(\ref{Fig-InflowEnergyAngularMomentumFlux}) the Noether's energy and angular momentum exchanges between the MHD inflow and the black hole. The colatitude of the open field lines at the black hole horizon is plotted by green dotted points. We have drawn the Noether's energy and angular momentum exchange beyond this angle because we were interested for getting these fluxes up to $\theta = \pi / 2$. The fluxes on the black hole are determined by the MHD fields on the horizon. 

Because we impose that the last inflow field line corresponds to the last open outflow field line, the colatitude of this last field line on the horizon depends also of the chosen outflow solution.

Angular momentum is extracted from the black hole for each inflow/outflow solution for the whole range of colatitude except on the axis. If we look at the Fig.(\ref{Fig-InflowEnergyAngularMomentumFlux}), the amount of extracted angular momentum in dimensionless units has a maximum which varies by one order of magnitude from solution to solution. Nevertheless taking into account the constant values in Tab.(\ref{Tab-TypicalParam}), we calculated the extracted angular momentum integrated over the open field lines and on the whole black hole horizon. These values are mentioned in columns two and three of Tab.(\ref{Tab-ExtractionAngularMomentum}). On the open field lines, the extracted angular momentum from the black hole horizon, ${\dot{J}_{\rm H,open}}$ is $40$ times lower for the M2 solution compared to the other ones. But on the whole black hole horizon, M2 extracts more efficiently angular momentum than the two other solutions. This result is due to the very small extension of the open field line region on the black hole horizon for the M2 solution and to some change of the stagnation radius. The colatitude of the last open magnetic field line on the black hole horizon is equal to, $\theta_{\rm open, H}=0.125\pi$ for M1, $0.03\pi$ for M2, and $0.10\pi$ for M3, respectively. Note that the ratio between the extracted angular momentum rate for the total and the inner regions is much higher for the M2 solution.

None of our solutions are capable of having a positive global extraction of Noether energy along the open magnetic field lines. This is due to the fact that, close to the axis, the inertial energy is dominant (see  Eq.(\ref{decompofluxenergy})). This energy is negative on the black hole horizon. For the global solutions M2 and M3, there is a positive extraction of energy but only at colatitudes close to the equatorial plane. We tried with our gradient descent method (see Appendix \ref{an-gdt}) to tune the inflow parameters in order to decrease the angle where the global extraction starts. We need further studies to see if solutions exist with a global extraction of Noether's energy occurring on some of the opening magnetic field lines. However, this may be an intrinsic limitation of our model, due to the self-similarity.

\begin{figure}[h]      
\centering
\includegraphics[width=\linewidth]{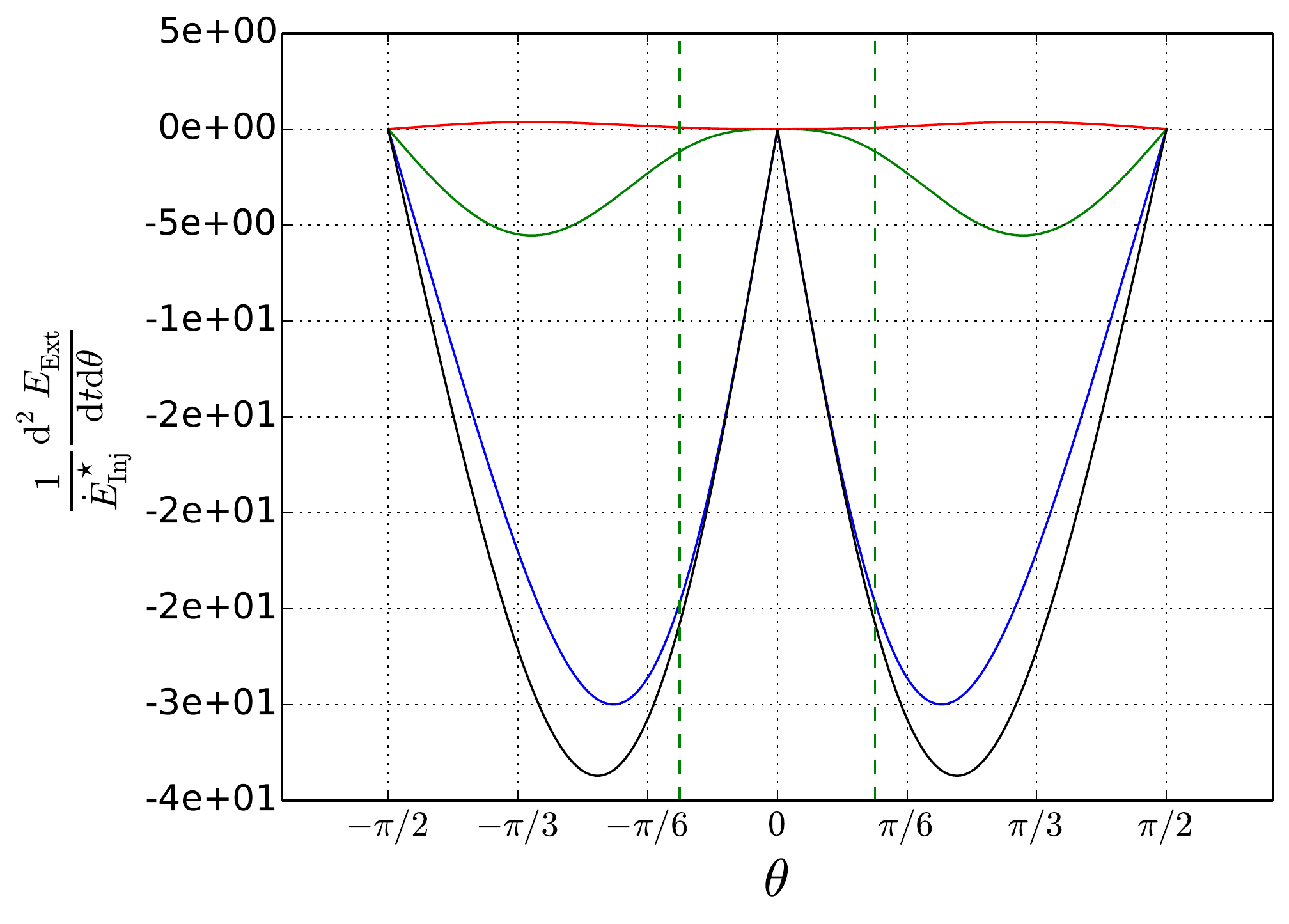}
\includegraphics[width=\linewidth]{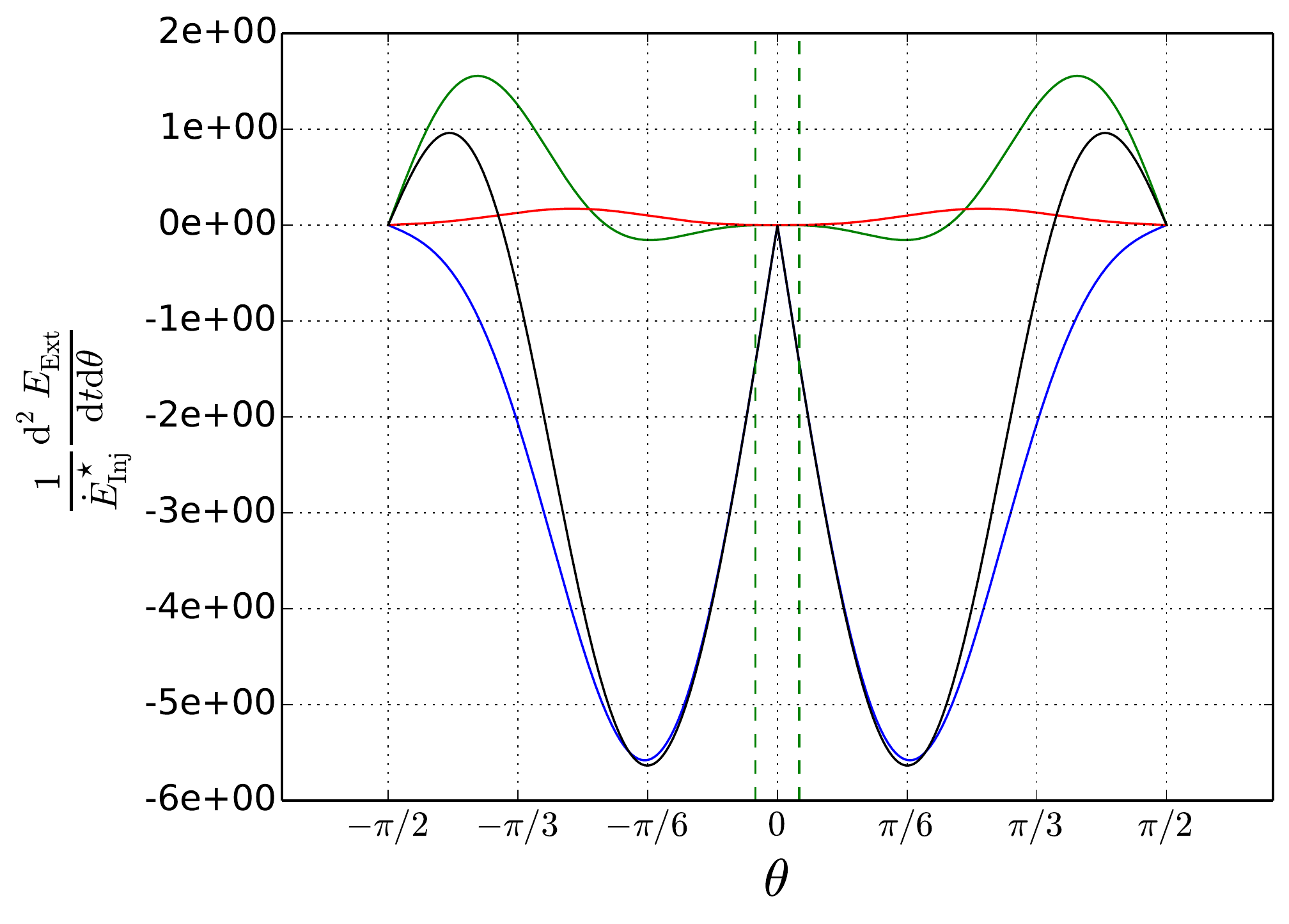}
\includegraphics[width=\linewidth]{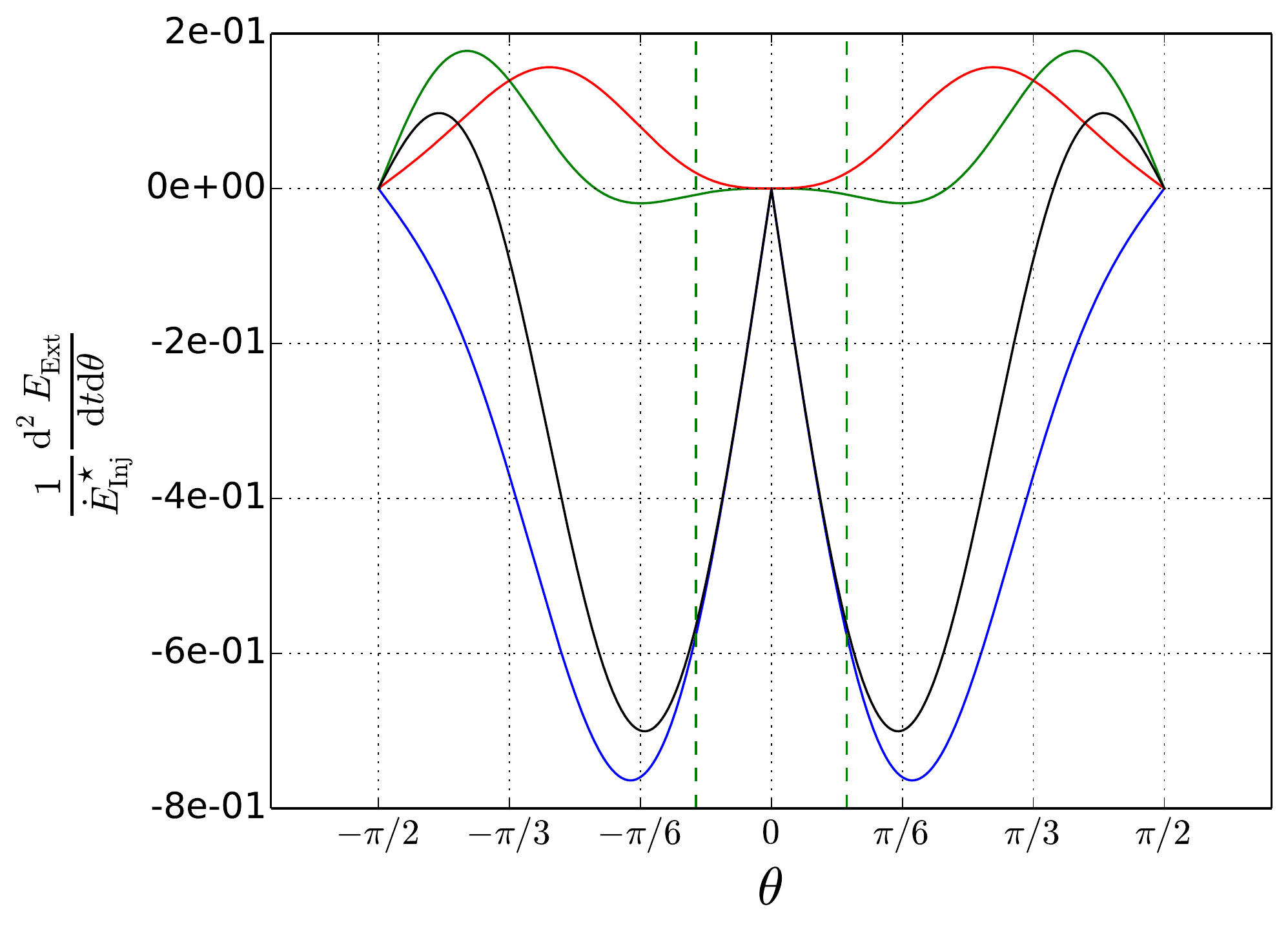}
\caption{ Noether energy flux component per unit colatitude flowing across black hole horizon in function of the colatitude for the three solutions (M1 on top, M2, middle and M3, bottom). In red the Poynting flux, in blue the inertial flux, in green the Lense-Thirring flux and in black the total MHD flux. As before the dotted green lines indicates the colatitude of magnetosphere field line at the horizon radius. \label{Fig-InflowEnergyFluxDecoomposition} }
\end{figure}

As presented above, the Noether energy flux of the MHD fields can be decomposed in three main terms, the inertial energy term, $\Phi_M=\Psi_A h \gamma \xi c^2$ strictly negative on the black hole horizon, the Lense-Thirring term, $\Phi_{\rm LT}=\Psi_A\gamma\xi\varpi \omega V^{\hat{\varphi}}$, and the Poynting flux, $\Phi_{\rm EM}=-h\varpi\Omega B^{\hat{\varphi}}$. In Fig.(\ref{Fig-InflowEnergyFluxDecoomposition}), we plot these fluxes per unit colatitude. In the first solution M1, the energy flux is fully dominated by the fluid one and the Poynting flux is extremely small. This could be explained by the value of $\Omega/\omega_H$ (see Tab.(\ref{Tab-Param-Match})). The Lense-Thirring flux is negative which means that the fluid falls into the hole with positive $V^{\hat{\varphi}}$. In the second solution M2, the Poynting flux is still small, but outside of the open field lines the Lense-Thirring flux turns positive. For biggest colatitudes, the Noether energy flux of the fluid, the sum of $\Phi_{\rm LT}$ and $\Phi_M$), become positive, which means that the Penrose fluid process is efficient. In the last solution M3, we get almost the maximum value of 0.5 for $\Omega/\omega_H$. The Poynting flux starts to increase even in the open field line region.

\cite{1977MNRAS.179..433B} derived the Poynting flux, using the boundary conditions given by \cite{1977MNRAS.179..457Z} close to the black hole horizon in Carter tetrad. In our model these boundaries conditions are satisfied because of the infinite conduction given by Eq. (\ref{idealconduction}) if the poloidal velocity in the ZAMO frame is equal to the speed of light. If we use this condition in ZAMO tetrad, we also get $B^{\hat{\varphi}}+E^{\hat{\theta}}=0$ on the horizon. In our model, at first order in colatitude, the fluid enter in the horizon with $V^{\hat{\theta}} = V^{\hat{\varphi}} = 0$ (see \citealt{CC2018}). If we tune the parameter such that $V^{\hat{r}} = -c$ on the horizon pole, this condition is then satisfied to the order one in colatitude, as a consequence of the infinite conductivity. We find for the Poynting power extracted from the black hole between the colatitudes $0$ and $\theta$,
\begin{align}\label{eq-bz}
    \dot{E}_{\rm H}^{\rm Poy}(\theta)=& \displaystyle\int_0^\theta h \varpi \Omega E^{\hat{\theta}} \dfrac{dA}{d\theta}d\theta\nonumber\\
    =& 4\dfrac{\Omega}{\omega_H}\left(1-\dfrac{\Omega}{\omega_H}\right)\dfrac{c \Phi_{\rm BH}^2}{128 \pi^2 r_g^2}\dfrac{a^2}{1+\sqrt{1-a^2}-a^2/2}\nonumber\\
    &\displaystyle\int_0^\theta \dfrac{\sin\theta }{1-\frac{a^2}{2\left(1+\sqrt{1-a^2}\right)}\sin^2\theta}\left(\dfrac{d\,}{d\theta} \dfrac{A}{A_H}\right)^2d\theta\,,
\end{align}
where the magnetic flux $A$ is evaluated on the horizon and is a function of $\theta$ only, and $\Phi_{\rm BH} = 2\pi A_H$. The Poynting power is then determined by $\Omega$ and the magnetic flux function on the black hole horizon. No fluid quantity is entering in this expression. It means that we obtain a Poynting power similar to the one obtained in the force-free assumption (\citealt{1977MNRAS.179..433B}. Eq.(26) of \citealt{2013PhRvD..88h4046G} gives the Poynting power in the force-free limit, a monopole geometry with $\Omega = \omega_H/2$. 

	\subsection{Angular momentum and energy sources}
	
Now we can compare the sources of injected angular momentum, in dimensionless units, by pair injection in the stagnation surface and by the black hole. In Tab.(\ref{Tab-ExtractionAngularMomentum}), we note $\dot{J}_{(\rm Inj,open)}$ the angular momentum flux injected by pair creation, limited to the inner region of open magnetic field lines. For the M1 and M3 solutions, the angular momentum injected by pair creation, $\dot{J}_{(\rm Inj,open)}$, is of the same order than the angular momentum extracted from the black hole, $\dot{J}_{(\rm H,open)}$. However note that for M2 the pair creation causes an absorption of angular momentum, $\dot{J}_{(\rm Inj,open)}<0$, two orders lower than the injected angular momentum in M1 and M3. This is equivalent to say that the total angular momentum flux transported in the open magnetic field region of the inflow is larger than the one in the outflow. 

\begin{table*}
\centering
\begin{tabular}{*{10}{c}}
   \hline
    {\rm Solution} & ${\dot{J}_{\rm (Inj,open)}}$ & ${\dot{J}_{\rm (H,open)}}$ & ${\dot{J}_{\rm (H,tot)}}$ & ${\dot{E}_{\rm (Inj,open)}}$ & ${\dot{E}_{\rm (H,open)}}$ & ${\dot{E}_{\rm (H,open)}^{\rm Poy}}$ & ${\dot{E}_{\rm (H,tot)}}$ & ${\dot{E}_{\rm (H,tot)}^{\rm Poy}}$ & $\dot{E}_{\rm (out,open)}$ \\
    & (${\rm g.cm^2.s^{-1}}$) & (${\rm g.cm^2.s^{-1}}$) & (${\rm g.cm^2.s^{-1}}$) & (${\rm erg.s^{-1}}$) & (${\rm erg.s^{-1}}$) & (${\rm erg.s^{-1}}$) & (${\rm erg.s^{-1}}$) & (${\rm erg.s^{-1}}$) & (${\rm erg.s^{-1}}$) \\
    \hline\hline
    M1 & $2,9\times 10^{47}$ &$1,4\times 10^{47}$& $6.5\times 10^{48}$ & $5.9\times 10^{44}$ & $-5.9\times 10^{44}$ & $8.7\times 10^{41}$ & $-3.6\times 10^{45}$ & $3.2\times 10^{43}$ & $1.3\times 10^{42}$\\
    M2 & $-2.7\times 10^{45}$ & $3.4\times 10^{45}$ & $5.6\times 10^{49}$ & $6.2\times 10^{42}$ & $-6.2\times 10^{42}$ & $1.6\times 10^{39}$ & $-3.4\times 10^{44}$ & $1.2\times 10^{43}$ & $ 1.1\times 10^{39}$ \\
    M3 & $1.1\times 10^{47}$ &$1.4\times 10^{47}$& $1.8\times 10^{49}$ & $2.4\times 10^{43}$ & $-2.2\times 10^{43}$ & $3.7\times 10^{41}$ & $-1.1\times 10^{44}$ & $2.8\times 10^{43}$ & $1.9\times 10^{42}$ \\
    \hline
\end{tabular}
\caption{Total angular momentum and power injected by pairs or extracted from the black hole for the three solutions in the region inside the last open field lines. ${\dot{E}_{(\rm H,open)}^{\rm Poy}}$ is the Poynting power extracted from the black hole inside the inner region. ${\dot{J}_{(\rm H,tot})}$,  ${\dot{E}_{(\rm H,tot)}}$ and ${\dot{E}_{(\rm H,tot)}^{\rm Poy}}$ are the angular momentum, the power and the Poynting power extracted from the whole black hole horizon. $\dot{E}_{(\rm out,open)}$ is the total power of the outflow spine jet in the inner region. \label{Tab-ExtractionAngularMomentum}.}
\end{table*} 

In Tab.(\ref{Tab-ExtractionAngularMomentum}) we put the injected power by the pairs into the inner region $\dot{E}_{(\rm Inj,open)}$, and the power extracted from the black hole into the inner region $\dot{E}_{(\rm H,open)}$. In order to quantify the weight of the Poynting flux in the energetic balance, we include the extracted Poynting power into the inner region $\dot{E}_{(\rm H,open)}^{\rm Poy}$. We also give the total power $\dot{E}_{\rm H, tot}$ and the total Poynting power extracted from the black hole $\dot{E}_{(\rm H,tot)}^{\rm Poy}$. In Tab.(\ref{Tab-ExtractionAngularMomentum}) the last column is the outflow total power $\dot{E}_{\rm out,open}$.

If most of the injected energy flux is going down in the inflow, we already remarked that most of the mass is moving outwards for the M1 and M3 solutions. Nevertheless the analysis in terms of mass is depending on the enthalpy value $\xi_\star$. This is a free parameter which does not change the analytical solution. This value for the inflow depends on the choice of the ratio between the inflow mass rate and the black hole magnetic flux. $\xi_\star$ of the outflow is given by the choice of $P_0$ (see Eqs. 54 and the discussion in sub-section 5.2 of \citealt{CC2018}). 

In the inner region, the solutions present a negligible Poynting power $\dot{E}_{(\rm H,open)}^{\rm Poy}$ in comparison to the power extracted from the black hole $\dot{E}_{(\rm H,open)}$. I3 inflow solution has been optimised to reach the canonical value of $\Omega/\omega_{\rm H}= 1/2$ and the ratio between the Poynting power $\dot{E}_{(\rm H,open)}^{\rm Poy}$ and $\dot{E}_{(\rm H,open)}$ is at least one order of magnitude higher than for the two other solutions. The size of the M2 magnetosphere leads to a small opening angle of the last open field lines at the black hole horizon and then we get a small value of extracted Poynting power on the inner region for the M2 solution. On the whole black hole horizon, the ratio between the extracted Poynting power and the total power is again higher for the M3 solution.  

As explained above in subsec.(\ref{subsec-emagsource}), the pairs are contributing to the Poynting flux of the outflow via the surface current $J^{\hat{\theta}}_\sigma$. For the M3 solution, inside the open field lines, the Poynting power at the base of the outflow is around $\unit[5.4\times 10^{41}]{erg.s^{-1}}$ while it is around $\unit[1.2\times 10^{41}]{erg.s^{-1}}$ below the stagnation radius in the inflow. Thus, the contribution to the Poynting power from $J^{\hat{\theta}}_\sigma$, created by pairs, is of the same order as the the Poynting power extracted from the black hole at the horizon. 

\section{Discussion}\label{sec-sec}

Near the system axis, the energy fluxes have different behaviour. The Poynting flux $\Phi_{\rm EM}$ is proportional to $\varpi^2$ and the fluid energy flux $\Phi_{\rm FL}$ has a non-zero constant term in its $\varpi$ expansion. Then around the axis the fluid energy flux will be the dominant term. Furthermore, it is difficult to increase the Poynting flux relatively to the fluid energy flux. This can be explained by the equality at the Alfv\'{e}n point on the axis,
\begin{equation}\label{eq-equipartmodl}
    \dfrac{B_\star^2}{8\pi}=\dfrac{1}{2}\rho_\star\xi_\star \gamma_\star^2 V_\star^2 ,
\end{equation}
which links the typical volumetric magnetic energy to the typical kinetic energy of the fluid.

In our model in order to increase the ratio of the Poynting and fluid energy fluxes, one way is to increase the flow speed at  the Alfv\'{e}n point and the factor $\gamma_\star^2 \dfrac{V_\star^2}{c^2}=\dfrac{\mu/\nu^2}{1-\mu/\nu^2}$. For the inflow, the parametric study seems to lead to a decreasing of the inflow Alfv\'{e}n radius and then to the stagnation surface. However for small values of the stagnation radius, it is more difficult to match an outflow solution because the escape speed is higher.

Eq. (\ref{eq-bz}) shows that the extracted Poynting flux depends strongly on $a$. Increasing the Poynting flux of the M3 solution can be achieved by an increase of the black hole spin, keeping the ratio $\Omega / \omega_H$ equal to $1/2$.

For the three solutions, most of the injected energy flux $\dot{E}_{\rm Inj,\ open}$ goes to the inflow $\dot{E}_{(\rm H,open)}$ and only about ten percent goes to the outflow $\dot{E}_{(\rm out,\ open)}$, as it can be seen in Table  \ref{Tab-ExtractionAngularMomentum}. The inflow energy is absorbed by the black hole since $\dot{E}_{(\rm H,\ open)}$ is negative. Part of this energy flux is given back by the black hole into a positive Poynting flux (see Fig. \ref{Fig-InflowEnergyFluxDecoomposition}). Poynting flux transfer occurs between the black hole and the inflow on the horizon. In a further study, we aim at searching for solutions that could extract energy from the black hole. There may be a way to optimize the eight parameters of the inflow. The first step would be to get less flaring of the inflow flux tubes in order to increase the size of opening magnetic field line region on the horizon. It is equivalent to have the last open magnetic line connected to the black hole starting at a larger colatitude. Second, we should increase the parameter ratio involved in Eq. (\ref{eq-equipartmodl}) without reducing to much the radius of the stagnation, in order to increase the magnetisation and then the ratio of the Poynting flux to the fluid power. 
 
Solving the Bernoulli equation,  \citet{2013PhRvD..88h4046G,2014ApJ...796...26G} and also \cite{2015ApJ...801...56P} have developed models with a fixed geometry, obtaining double flow. The pair creation zone is either a thin layer as in our model either a volumetric injection. Despite the assumption of a radial geometry, which is a limitation of the model, \citet{2013PhRvD..88h4046G} use the particle source $q_n$ (that we note $k_n$ in the present publication) as a parameter. Above a given threshold for particle injection, the total energy flux cannot be extracted any longer. This threshold depends on the field line colatitude and on the black hole spin. The higher the spin, the easier the extraction. \citet{2014ApJ...796...26G} define the source term as a radial power law. In both publications, they find a low total energy extraction around the axis, which our results confirm. \cite{2015ApJ...801...56P} use a geometry obtained from a parabolic force free field solution. They inject pairs on a stagnation surface. They impose in addition two matching conditions at the stagnation surface, the continuity of Poynting flux and the equality of inflow and outflow pair fluxes. Their double flow solution is electromagnetically dominated and similar to the results of the GRMHD simulations of \cite{2007MNRAS.375..531M}. 

In our case, once the inflow solution has been fixed the three matching conditions leaves us with four degrees of freedom for the outflow solution. These degrees could be used in a further study for different aims. First, in order to get the total current continuity condition and second, in order to obtain a smaller magnetosphere and to increase the horizon colatitude angle of the last opened field lines. Differently by selecting some radial solution for the outflow as the K4 solution presented in \cite{CC2018}, we could obtain a larger colatitude extension on the stagnation surface. It will allow to match an inflow with a larger colatitude extend at the horizon. Finally, since for our three solutions, the total MHD power of the outflows corresponds to the range of the power transition between FR1 and FR2 galaxies as mentioned in \cite{2019A&A...621A.132M}, we could search for different outflows, which match the same inflow to increase or decrease the outflow jet power.

Using an iterative procedure on the magnetic flux, the current and isorotation integrals, \cite{2014ApJ...788..186N} solved the Grad-Shafranov equation in a force free configuration. It allows, in the force free assumption, to recover the field geometry starting from a radial or a paraboloidal configuration. To pursue this theoretical approach, \cite{2019ApJ...880...93H} solved the Grad-Shafranov equation in the same way without neglecting the fluid forces, and simultaneously solved the Bernouilli equation. With different matching conditions compared to our model, they obtained double flow solutions with injection on the stagnation layer. Their 4-force $\mathbf{k}$ of the radiation field on the pair fluid is assumed to be equal to the product of the source term $k_m$ and the 4-velocity $\mathbf{u}$. \cite{2020ApJ...894...45H} applied their model to produce double flow solutions for the stratified M87 jet. In their Cases V and VI, the outflow fluid energy is equal to 43\% of the total energy, lower than the ratio we obtained for our M3 solution, which is $\approx 71\%$.

Several authors studied electrical gaps of charge separation in black hole magnetospheres, \citep{2017PhRvD..96l3006L,2016ApJ...818...50H} showed that the formation of a gap occurs for small accretion rates. In these works, the flow geometry is fixed and mainly radial. The kinetic and the dynamics of the electron fluid, the positron one and the radiation are treated separately, leading to a self-consistent model of pair formation. If the electrical gap is spherical and the black hole rotates maximally in \cite{2017PhRvD..96l3006L}, we can estimate the mass creation rate of pairs inside the gap. Take as an example the case of M87. Using the maximal value of the magneto-spheric current of Fig.(2) in \cite{2017PhRvD..96l3006L}, it gives a rate $\approx 10^{11}\mathrm{g\,s^{-1}}$. This value is $7$ orders of magnitude less than the rates we obtained in our solutions. We calculate the pair multiplicity $n/n_{GJ}$ on the Alfv\'en surface on the polar axis both for inflow and outflow, and we obtain values between $10^{10}$ and $10^{12}$. This amount is not consistent with pair production in potential gaps near the polar axis in the magnetosphere, since the current theory and simulations of potential gap process predict values not higher than few thousands.

In fact, electrical gaps are thought to be intermittent phenomena. The gap induces pair creation. The pairs fills the gap and produce high energy emission via the inverse Compton mechanism. Then the gap reforms that is the reason for an intermittent emission at high energy. Particle-in-cell (\citealt{refId0}) and radiative GRMHD (\citealt{10.1093/mnras/stab2462}) simulations have shown that the gap size is difficult to estimate and could be as small as $0.05r_g$. Moreover, on the axis the gap could disappear as in the MAD W18 disk model of \cite{10.1093/mnras/stab2462}. Our solutions without gap are relevant to describe the double-flow near the polar axis. The value of the electron temperature mentioned in \cite{10.1093/mnras/stab2462} is similar to the one we obtained for our three solutions (around $\approx 0.5 {\rm MeV}$). 

\section{Conclusions}\label{sec-sec4}

We presented here a meridionally self-similar MHD model of a relativistic double flow in Kerr geometry that incorporates mass, angular momentum and energy loading on magnetic field lines threading the black hole horizon. The semi-analytical model of \cite{CC2018} allows us to produce double flow solutions above the black hole pole, passing critical surfaces with a spherical stagnation surface, where the loading of pairs occurs. The pair production is supposed to be due to two-photon interaction via the gamma-ray-emission from the accretion disk. The goal of this modelling is to describe MHD fields close to the polar axis, where the total power is matter dominated resulting in a fast spine-jet propagating at large distances and a polar pair accretion onto the horizon. 

To perform the loading we introduced two source terms, the mass injection rate and the four-force exerted by the radiation at the origin of pairs. It implies that mass, angular-momentum and energy fluxes along the field lines are not conserved. We derived the corresponding particle number continuity and energy-momentum conservation equations. We fixed the following working assumptions at the stagnation surface in order to match the inflow and outflow solutions, the continuity of the radial magnetic field component and the magnetic flux, the continuity of the isorotation frequency and a meridional surface current, which creates a discontinuity of the current intensity along the polar axis. This meridional surface current, accompanied by a charge density on the stagnation surface, causes a discontinuity of the toroidal magnetic field component and the field line opening. It implies a contribution of the injected pairs to the electric current and then to the Poynting flux. 

With those assumptions we got three matching conditions for the inflow and outflow parameters. Since we have to fix eight parameters for the inflow component and seven for the outflow component, the three matching conditions reduce the number of free parameters in the outflow to four. Since we model a flow without neglecting fluid forces, we are able to discuss the possibilities of extracting energy from the black hole with a fluid alone and in a non force-free MHD configuration. Once the inflow/outflow solution is obtained by solving a system of first order differential equations, this model allows to quantify the exchange between the black hole and the MHD fields, and the sources of energy and angular momentum due to pair loading.

In this paper we present three inflow/outflow solutions (M1,M2 and M3) and we apply them to the case of the black hole mass of M87. We use for the magnetic field magnitude, the values observed by the EHT at a distance of a few Schwarzschild radii. The inflow model respects a scaling law similar to that mentioned in \cite{2014Natur.510..126Z} between the magnetic flux and the mass flux. In our publication, we use this scaling-law, assuming that the mass inflow rate is a fraction of the mass accretion rate. The efficiency is three times higher to create magnetic flux. The magnitude of the specific enthalpy in our model is fixed thanks to this law.

We estimate a range of mass injection rates by pair formation from hard disk photons. Our solutions M2 and M3 require a large amount of injected mass and are located close to the upper boundary of this estimated range. The M1 solution requires an injection rate higher than the maximum value by an order of magnitude. It could involve other mechanisms of injection to be physically relevant. The number of injected pairs in our three models is too high to be associated  with pair production in spark-gaps seen as unscreen parallel electric field regions.

As expected the inflow acceleration is dominated by gravity. Pressure also plays a role, decelerating the inflow close to the stagnation layer and close to the black hole horizon and accelerating it in between. Due to the flaring of the flux tubes, the gravity works for opening the tubes where the radius increases. This force is counter-balanced by the magnetic forces, mainly composed of magnetic poloidal pressure. For the transverse equilibrium, the pressure also counter-balances gravity at the beginning and at the end of the inflow.

Inside the open field lines, the injected angular momentum on the stagnation surface can be of both signs and is of the same order of magnitude as the extracted one from the black hole. In terms of Noether's energy, most of the injected energy falls down into the black hole. Inside of the open magnetic field lines the energetic distribution is dominated by the inertial term. The black hole is fed by mass, kinetic and internal energy. At larger colatitudes, the energetic budget is dominated by the Lense-Thirring effect for the inflow of M2 and by the Poynting flux for the inflow of M3. For this last solution M3, the total Poynting power on the horizon, even if it is less than the inertial energy, represents a quarter of the total energy absorbed by the black hole. The pair injection contribution to the Poynting flux has been calculated for the M3 solution and is comparable to the Poynting flux extracted from the black hole.

To conclude, the solution M3 is the most interesting solution, having an isorotation frequency equals to one half of the black hole one. It has a quite high but still reasonable injection mass rate and also a reasonable value for total outflow power. The final Lorentz factor of the outflow is around $\gamma \approx 10$. The extracted Poynting power from the black hole is comparable to the one given by a force-free model in a monopole geometry with the same isorotation frequency and the same total magnetic flux. The outflow of the M3 solution has a quite high magnetisation just above the stagnation surface and the total power is inside the expected range for extragalactic jets.

\section*{acknowledgements}

We thank the referee for valuable comments and suggestions. LC thanks the Observatoire de Paris - PSL University for an ATER position and the Dipartimento di Fisica - Università degli Studi di Torino for the post-doctoral position which enabled him to complete this work. CS thanks LUPM for hosting him during his CNRS delegations and his "CRCT". 

We acknowledge financial support from "Programme National des Hautes Energies" (PNHE) and from ``Programme National de Physique Stellaire'' (PNPS), CNRS/INSU, France.

\section*{Data availibility}
A catalogue of the three complete inflow/outflow solutions is available on request to the main author.



\bibliographystyle{mnras}
\bibliography{Biblio} 



\appendix

\section{Behaviour of the inflow solutions near the black hole horizon}\label{an-sec-modelhorizonbehaviour}
    
Let discuss how the model equations and the four functions $M^2,G^2,F,\Pi$ behave close to the black hole horizon (see Appendix C of \citealt{CC2018}). We adopt here all notations coming from \cite{CC2018}. $M$ is the poloidal Alfv\'{e}n Mach number on the polar axis, $G$ the dimensionless cylindrical radius in unit of the Alfv\'{e}n radius, $F$  is the expansion factor of the streamlines and $\Pi$ is the dimensionless pressure along the polar axis. The equations of the model are determined by the functions $\mathcal{D},\mathcal{N}_{M^2},\mathcal{N}_{F}$ and $\mathcal{N}_{\Pi}$ depending of $R,M^2,G^2,F$ and $\Pi$. To determine the behaviour of our solutions near the black hole horizon, we need to express these functions at the radius $R=R_H=\dfrac{\mu}{2}\left(1+\sqrt{1-\left(\dfrac{2l}{\mu}\right)^2}\right)$ expressed in Alfv\'{e}n radius unit. As explained in \cite{CC2018} and previous meridional self-similar models, we build from this model a constant, $\epsilon$, which measures the efficiency of magnetic collimation (see Appendix C of \citealt{CC2018}). This parameter writes as,
\begin{align}
\epsilon=&\dfrac{1}{h_z^2}\left[2\lambda^2\left(\dfrac{\Lambda^2N_B}{D} +\dfrac{\overline{\omega}_z}
 {\lambda}\right)+\dfrac{\nu^2(2m_1-2e_1+\kappa-\delta)R}{(R^2+l^2)} \right. \nonumber \\
 &\left. -\dfrac{M^4}{G^2 h_\star^4 (R^2+l^2)}\left(1-\dfrac{l^2}{(R^2+l^2)}+(\kappa-2m_1)\dfrac{(R^2+l^2)}{G^2}\right)\right. \nonumber\\
 &\left. -\dfrac{\nu^2l^2RG^2}{(R^2+l^2)^3} + \dfrac{M^4 h_z^2 F^2}{h_\star^4 G^2} \right] +\lambda^2\left(\dfrac{\Lambda N_V}{h_\star G D}\right)^2 \,.
\end{align}
Close to the black hole horizon, we have $h_z^2\underset{R\rightarrow R_{H}}{\sim}\frac{2}{\mu}\frac{\sqrt{1-a^2}}{1+\sqrt{1-a^2}} (R-R_H)$. Numerically, we find that the function $h_z F \underset{R\rightarrow R_H}{\longrightarrow} 0$ near the horizon. The functions $M^2$ and $G^2$ remain finite and do not reach $0$ on $R=R_H$. Since $\epsilon$ is constant, it implies that there exist a constant $\epsilon '$ such that,
 \begin{align}
&-\dfrac{M^4}{G^2h_\star^4(R^2+l^2)}\left(1+(\kappa-2m_1)\frac{X_+}{G^2}-\dfrac{l^2}{(R^2+l^2)}\right)-\dfrac{\nu^2 l^2 R G^2}{(R^2+l^2)^3}\nonumber\\
 &-\dfrac{\nu^2}{\mu}\left(2e^1-2m^1+\delta-\kappa\right)+2\lambda^2\left(\dfrac{\Lambda^2 N_B}{D}+\dfrac{\bar{\omega_z}}{\lambda}\right)\underset{R\rightarrow R_H}{\sim} \epsilon' h_z^2 \, ,
 \label{eq-epsiprim}
 \end{align}
 where we used $(R^2+l^2) = \mu R_H$ on the black hole horizon. The function $\mathcal{D}$ can be singular only for $R=0$. The function $\mathcal{N}_{M^2}$ could also have a singularity if $G^2 =0$ or $M^2=0$. Another possibility of singularity could happen for $R=R_H$. The function $\mathcal{N}_{M^2}$ can be written as:
\begin{align}
    \mathcal{N}_{M^2} = & \dfrac{\mu h_\star^4 D R G^2}{2 h_z^2 X_+ M^2}\dfrac{X_-}{X_+}\left[ -\frac{M^4}{G^2h_\star^4}\left(1+(\kappa-2m_1)\dfrac{X_+}{G^2}-\dfrac{l^2}{X_+}\right) -\dfrac{\nu^2 l^2 G^2}{\mu^3 R_{H}^2} \right. \nonumber \\
    & \left.-\dfrac{\nu^2}{\mu}\left(2e^1-2m^1+\delta-\kappa\right)+2\lambda^2\left(\dfrac{\Lambda^2 N_B}{D}+\dfrac{\bar{\omega_z}}{\lambda}\right)\right] +R_{M^2} \, .
\end{align}
$R_{M^2}$ does not have any singularity for $R=R_H$. The equation Eq.(\ref{eq-epsiprim}) insures the regularity of $\mathcal{N}_{M^2}\left(R,M^2,G^2,F,\Pi\right)$ where $M^2,G^2,F,\Pi$ are solutions of model equations. It explains the numerical regularity of the $M^2$ function close to the horizon. Indeed, we found numerically that $M^2$ reaches a limit value different of $0$. 
Such arguments no longer hold for the $F$ function. In this case, $\mathcal{N}_{F}\left(R,M^2,G^2,F,\Pi\right) \underset{R\rightarrow R_H}{\sim} \dfrac{{\rm Cst}}{h_z^2}$ (with a non vanishing constant), then we expect a behaviour such as $F \underset{R\rightarrow R_H}{\propto} \ln(R-R_H)$, which is the observed behaviour of the numerical solution. Nevertheless, the angle of the magnetic field line with the radial direction $\chi$ is linked to the colatitude with,
\begin{equation}
    \tan \chi = \dfrac{1}{2}\dfrac{\sqrt{X_+}h_z F}{R}\tan\theta\,.
\end{equation}
${\sqrt{X_+}h_z F}/{R}$ reaches $0$ on the horizon even if $F$ diverges because $F$ varies logarithmically. This ensures that the magnetic field lines are always perpendicular to the even horizon, as expected. The behaviour of $F$ implies the convergence of $G^2$ to a finite value on the horizon. 

The solutions of the model verify on the axis $\gamma \xi \underset{R\rightarrow R_H}{\longrightarrow} +\infty$ where $\xi$ is the specific enthalpy. In order to avoid $\xi \underset{R\rightarrow R_H}{\longrightarrow} +\infty$, we may tune the inflow parameters to get $\gamma \underset{R\rightarrow R_H}{\sim} {1}/{h_z}$ on the horizon. This requirement also induces the $\Pi$ function behaviour on the horizon. It implies that $\nu^2h_\star^4-\frac{\mu M^4}{G^4} \underset{R\rightarrow R_H}{\sim} h_z^2$. Using the model equations, we get $\Pi \underset{R\rightarrow R_H}{\sim} \ln(R-R_H)$ on the horizon. If $\gamma$ instead does not behave like ${1}/{h_z}$, then $\Pi \underset{R\rightarrow R_H}{\sim}(R-R_H)^{-1}$. Nevertheless, this requirement is difficult to obtain in pratical terms, during the matching procedure between inflow and outflow solutions.

\section{Gradient descent techniques}
\label{an-gdt}

To adapt the input parameters of the outflow to the inflow, we do not use a simple technique of optimisation of a residual function. Instead we decided to follow the direction in parameter space, obtained by a technique able to calculate dual-like basis of the space generated by linear combination of gradients in the parameter space. Indeed, the minimisation of a residual function often leads to difficulties of different kinds. It also leads to regions where the automation of the crossing of the slow magnetosonic point undergoes a discontinuity due to the non-linearity of the equations or the crossing is impossible. Then, we need to explore the parameter space using different possibilities for the chosen direction.    

For the matching, we need to find the outflow parameter $(\lambda,\kappa,\mu,\nu,l,\mu,e_1)$ such that the quantities,
\begin{equation}
\begin{array}{cc}
    f_1 =& \dfrac{R_{\rm sta}}{\mu} , \\
    f_2 =&\dfrac{l}{\mu} ,\\
    f_3 =& \dfrac{\mu^3}{2\left(1+l^2\right)^2}+\dfrac{ \lambda\mu^{3/2} }{\nu}\sqrt{1-\dfrac{\mu}{1+l^2}}
\end{array}\,,
\end{equation}
are equal to some specific values (the corresponding values given in  Eq.(\ref{eq-mmc}) and calculated for the inflow). In what follows, we discuss a more general procedure where we have $n$ (with $n\leq 7$) function $\left(f_k\right)_{k=1,...n}$ of the solution parameter $(\lambda,\kappa,\mu,\nu,l,\mu,e_1)$ to adapt. We will refer to them as control functions.

If we note,
\begin{equation}
    u_k = \nabla f_k \in \mathbb{R}^7 \quad \text{ for } \quad k\in \llbracket 1,n\rrbracket\,.
\end{equation}
Assume that $\left(u_k\right)_{k=1,...n}$ are linearly dependent, and call $E_n$ the subspace generated by linear combination of $\left(u_k\right)_{k=1,...n}$ and $E_{n,j}=\text{Span}\left\{u_k\,|\, k=1...n \text{ and }k\neq j\right\}$ for each $j=1...n$. Then,
\begin{equation}
    \exists ! d_j\in E_n \text{ such that }\left\{\begin{array}{c}
            ||d_j||=1 \\
         d_j\cdot u_j>0  \\
         d_j  \perp E_{n,j} 
    \end{array}\right.\,,
\end{equation}
$d_j$ is the normalised projection of $u_j$, orthogonal to $E_{n,j}$. We calculate $d_j$ using a recurrence formula. Note $p_n\left(u_j;\left(u_k\right)_{\substack{k=1...n\\ k\neq j}}\right)$ the unit vector embedded in $E_n$, orthogonal to $E_{n,j}$, and such that $d_j\cdot u_j>0$. Then for all $i\neq j$, it follows that,
\begin{equation}
    p_n\left(u_j;\left(u_k\right)_{\substack{k=1...n\\ k\neq j}}\right)=p_{n-1}\left(p_2(u_j;u_i);\left(p_2(u_k,u_i)\right)_{\substack{k=1...n\\ k\neq j \\ k\neq i}}\right)
\end{equation}
 which allows to explicitly calculate $d_j$, considering that for each non-colinear vector $u,v$,
\begin{equation}
    p_2\left(u;v\right)=\dfrac{u-\dfrac{u\cdot v}{||v||^2}v}{\left|\left|u-\dfrac{u\cdot v}{||v||^2}v\right|\right|}\,.
\end{equation}

If we note $s=(\lambda,\kappa,\mu,\nu,l,\mu,e_1)$, then for all $j$ and for a small displacement $\varepsilon d_j$, we expect that for all $i\in\llbracket 1,n\rrbracket$ a typical behaviour,
\begin{equation*}
    f_i(s+\varepsilon d_j)=f_i(s)+\delta_{ij}\varepsilon \left(u_i \cdot d_j\right) + O(\varepsilon^2)\,.
\end{equation*}
which makes it possible to deal with the control functions one after the other. Nevertheless due to the strong non-linear behaviour of the system of equations, mentioned in the appendix of \cite{CC2018}, the control functions suffer of a lot of discontinuities. It implies that this method can be used only locally. The closer the family $\left(u_k\right)_{k=1,...n}$ is to the orthogonal family, the more efficient this method. This method can be used to fit the solutions of the self-similar model to the observational constraint.


\bsp	
\label{lastpage}
\end{document}